\documentclass[final]{elsarticle}

\usepackage{amsmath,amssymb}
\usepackage{graphicx}
\usepackage{footnote}
\usepackage{algorithmic}

\graphicspath{{./images/}}

\title{An analysis of the periodically forced PP04 climate model, using the theory of non-smooth dynamical systems.}

\author[bath]{Kgomotso S. Morupisi}
\ead{K.Morupisi@bath.ac.uk}
\author[bath]{Chris~Budd*}
\ead{mascjb@bath.ac.uk}
\cortext[cor1]{Corresponding author}
\address[bath]{Department of Mathematical Sciences, University of Bath, BA2 7AY, UK}

\begin{document}

\begin{abstract}
In this paper we perform a careful analysis of the forced PP04 model for climate change, in particular the behaviour of the ice-ages. This system models
the transition from a glacial  to an inter-glacial state through a sudden release of oceanic Carbon Dioxide into the atmosphere. This process can be cast in terms of a Filippov dynamical system, with a discontinuous change in its dynamics related to the Carbon Dioxide release. By using techniques from the theory of non-smooth dynamical systems, we give an analysis of this model in the cases of both no insolation forcing and also periodic insolation forcing. This reveals a rich, and novel, dynamical structure to the solutions of the PP04 model. In particular we see synchronised periodic solutions with subtle regions of existence which depend on the amplitude and frequency of the forcing. The orbits can be created/destroyed in both smooth and discontinuity induced bifurcations. We study both the orbits and the transitions between them and make comparisons with actual climate dynamics. 
\end{abstract}


\maketitle

\section{Introduction} 
\label{sec:Intro}

\subsection{Overview}
\label{sub:Intro:Overview}

\noindent Reduced climate models (RCMs), see for example \cite{KaperVo}, \cite{saltzman1990first}\cite{saltzman1991first}, \cite{paillard2001glacial}, \cite{ashwin2015middle}, \cite{EW}, \cite{crucifix2012oscillators},\cite{kaper2013mathematics},\cite{dijkstra2013nonlinear}, have been used extensively to study various forms of climate dynamics. Whilst not in any way a substitute for general climate models (GCMs) for an accurate simulation of climate dynamics from which predictions can be made, they are nonetheless very useful for investigating certain types of qualitative climate phenomena, particularly those that occur over time scales which are too long for a realistic calculation on a GCM.  RCMs can be used both to give insights into the macroscopic behaviour of certain types of climate phenomena, and also as ways of testing the predictions of the GCMs. In this paper we will in particular perform a careful analysis of the RCM usually called the PP04 model \cite{paillard2004antarctic} which has been used both to gain insight into the past behaviour of the Earth's glacial cycles (ice-ages) and also to predict future glacial events  \cite{crucifix2013}, \cite{ashwin2018chaotic}. Glacial cycles themselves show very subtle dynamics, with an interplay of variations of ice, temperature and of Carbon Dioxide, all coupled together by various feedback loops, and with external forcing from the Sun. These cycles have led to periodic, and significant, variations in the temperature, ice cover and Carbon Dioxide levels of the Earth. In the most recent glacial periods (over the last half a million years) these cycles have a roughly 100 kyr periodicicty, whereas before that period the cycles has a shorter period of around 40 kyr. It is generally believed \cite{imbrie1993structure}, \cite{kaper2013mathematics} that these cycles are either directly driven by the (quasi-)periodic variations in the insolation forcing received from the Sun through the Milankovitch cycles, or are a result of internal processes on the Earth with comparable time-scales, which are in turn synchronised by the Milankovitch cycles \cite{crucifix2012oscillators}. In this paper we will look at the PP04 model described in \cite{paillard2004antarctic} which is based on the latter assumption.

\vspace{0.1in}

\noindent Climate models are also very interesting examples of dynamical systems. Indeed dynamical systems theory has been used extensively to study them. Because of the huge disparity in time-scales for climate driven events, it is natural to approximate some as being near instantaneous when compared to others. From this perspective we expect to see climate models containing discontinuities and switches. The PP04 model which We will study in this paper has exactly this structure. To study it we can then make use of the relatively new theory of non-smooth dynamical systems \cite{bernardo2008piecewise}, to both determine possible climate states and to find the transitions between them. A similar approach has also been considered in \cite{EW} in the study of ice line dynamics in the Budyko-Sellars model for climate change.

\subsection{Results}

\noindent The emphasis of this paper will the study of the solutions of the PP04 model when it is driven by (quasi-)periodic insolation forcing. 

\vspace{0.1in}

\noindent If the magnitude of the insolation forcing is zero we will show that the model admits periodic solutions, associated with natural internal time-scales for the growth and retreat of ice sheets coupled to Carbon Dioxide levels in the atmosphere. These solutions have a period of 147 kyr $\equiv 1/\omega_0$. In the non-smooth PP04 model these orbits are created and destroyed in border collision bifurcations, arising when certain fixed points intersect a discontinuity surface. In a smoothed version of the they arise instead through Hopf bifurcations from the steady state followed by cyclic saddle-nodes. 

\vspace{0.1in}

\noindent If the insolation forcing is purely periodic, of frequency $\omega$ and amplitude $\mu$, we find that if $\mu$, $|\omega - n \omega_0/m|$ (with $n,m=1,2, ..$) are both small, then a {\em mode-locked periodic solution} exists which is a perturbation of the unforced solution. This solution is {\em synchronised} to the insolation forcing, but in general has a different phase. For small $\mu$, the regions of existence in the $(\mu,\omega)$ space are {\em linear tongues} for all values of $n$ if $m=1$, and are bounded by saddle-node bifurcations.  Outside of these tongues we see quasi-periodic motion which combines the insolation forcing frequency and the natural frequencies of the system. As the amplitude $\mu$ increases, the mode locked solution persists until it typically loses stability at a grazing bifurcation. At this point we see complex and multi-modal behaviour. As $\mu$ increases, the tongues for different values of $n$ can overlap and expand, leading to the co-existence of different period states. Close to the boundaries of these regions we see a variety of different types of behaviours (with subtle domains of attraction), leading to interesting transitions between the states as parameters are varied, with some behaviour having a qualitative resemblance to that at the Mid-Pleistocene Transition. When the periodic insolation forcing is extended to being quasi-periodic we then see the stable periodic solutions perturbing to invariant tori.

\vspace{0.1in}

\noindent The layout of the reminder of this paper is as follows. In Section 2 we will briefly review some of the observed features of the glacial cycles which we seek to reproduce in our models. In Section 3 we will review some of the existing models and will motivate the PP04 model for glacial dynamics. In Section 4 we will explain the basic ideas, and necessity, of using non-smooth dynamical systems theory in the analysis of the PP04 model, showing that it takes the form of a Filippov system without sliding. In Section 5 we will study the unforced PP04 model. We will analyse the changes in the dynamics as parameters vary, looking at both the existence of fixed points, and of periodic solutions, and the transitions between them as a result of border collision bifurcations in the non-smooth system. In Section 6 we will give a detailed mathematical analysis  of the existence, and stability, of the periodic solutions of the PP04 model under the effects of both small, and large, periodic insolation forcing. In Section 7 we will support these results through a series of numerical computations of the $\Omega-$ limit set of the solutions using a Mont\'e-Carlo method, and also the transitions between different states. Finally in Section 8 we will discuss the implications of these results to our understanding of the dynamics of the climate.

\section{Observed climate dynamics and glacial cycles}
	\noindent It is a feature of observed climate dynamics over the last few million years, that the Earth experiences glacial cycles, which are roughly periodic variations between hot and cold states. The hot (or interglacial states) tend to last for relatively short periods compared to the longer (or glacial) states. The period of these in the past 800,000 years has been roughly 100k years. Before then the period was closer to 40k years. The change between these types of behaviour is called the Mid-Pleistocene Transition (MPT), and has been studied by many authors \cite{paillard2004antarctic,saltzman1991first,saltzman1990first,paillard1998timing}. We see this in the following two figures. The first shows the changes in ice volume and temperature over the last million years. The second shows the MPT.
	
	\begin{figure}[htbp]
  \centering
  \includegraphics[width=\textwidth]{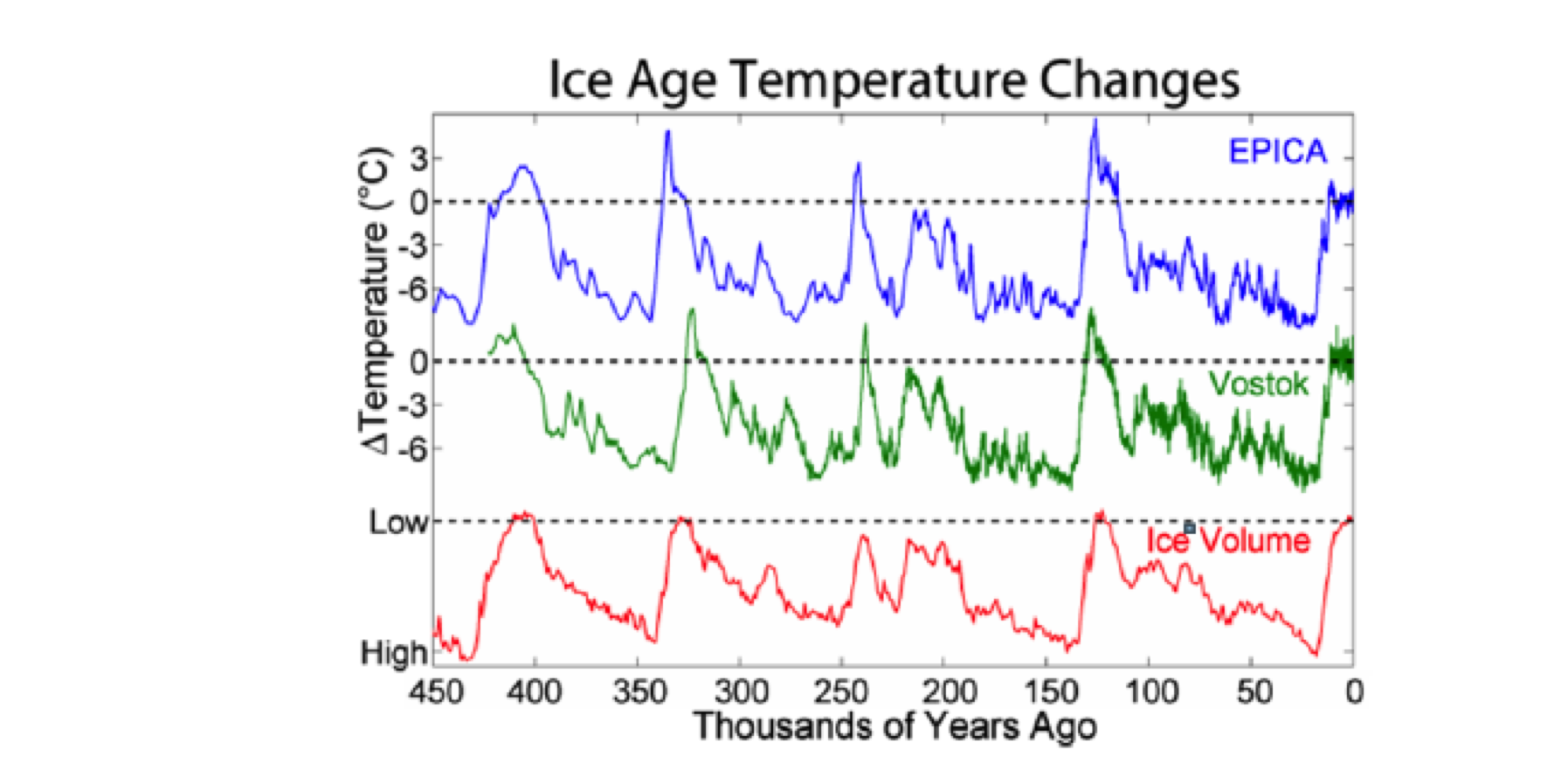}
  \caption{Ices age temperatures and ice volume taken from the Vostok and EPICA ice cores. (Image from {\tt en.wikipedia.org}.)}
\label{figc:1}
\end{figure}
	
	\vspace{0.1in}	
	
	\begin{figure}[htbp]
  \centering
  \includegraphics[width=\textwidth]{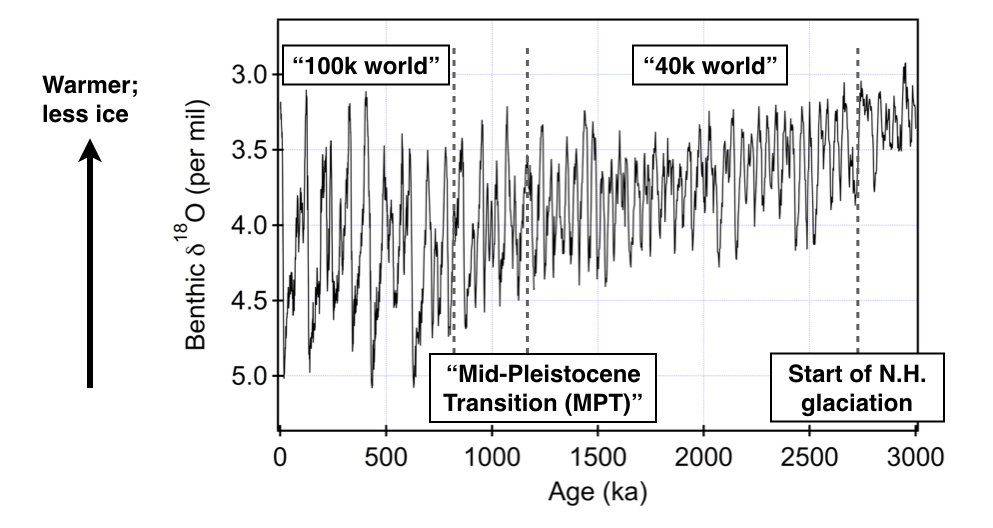}
  \caption{The oscillation cycles obtained from an oxygen isotope $\delta^ {18} O$. (Image from Lisiecki and Raymo 2005).}
\label{figc:mpt}
\end{figure}
	
	\vspace{0.1in}

	\noindent The changes to the past Earth climate, shown in these figures, can be studied through paleo-data sources such as coral reefs, deep sea sediments,continental deposits of flora and fauna and ice cores \cite{jouzel1987vostok}. When studying the reconstructed data from the Vostok ice core, a correlation between temperature and  concentration of Carbon Dioxide and methane has been identified which suggested that greenhouse gases are  causes or drivers of glacial cycles. Moreover, the temperature record shows that temperature decrease (slightly) leads the Carbon Dioxide decrease, and that at the end of every glacial period, global ice volume  changes have lagged changes to both the Antarctica air temperature and atmospheric Carbon Dioxide concentrations\cite{petit1999climate}. These observations implied that temperature changes partly drove Carbon Dioxide changes, and also led to a proposal that some mechanisms  that occur in the  Southern Ocean play a significant role  in long term changes of atmospheric Carbon Dioxide \cite{petit1999climate}.
	
	\vspace{0.1in}
	
	\noindent There has been much speculation about the causes of glacial cycles. A common explanation is that it is related to the changes in the Solar insolation forcing due to the  Milankovitch cycles in the Earth's orbit. The link between  atmospheric temperature  and such astronomical forcing was  established, for example, through the ice core \cite{jouzel1987vostok} particularly through the similarities between the mid-June insolation forcing at $65^0 N$ and data from the $\delta^{18}O$ isotope \cite{petit1999climate}. Hence  orbital forcing is viewed as the cause of the initial temperature changes at the  beginning of glacial cycles. 
Moreover, Hays et al \cite{hays1976variations}, when studying $\delta^{18}O$ isotope, showed that the fluctuations of volume of ice had experienced periods  of 23 kyr and 41 kyr which supports the contribution of  orbital forcing, more precisely the variations in precession and  obliquity, as the cause of these oscillations \cite{held1982climate}, \cite{paillard2017climate}. The oscillations of the glacial cycles with periods of 23 kyr and 41 kyr have been successfully reproduced  by the inclusion of astronomical forcing. However, the dominant asymmetrical 100 kyr period oscillations seen in the above figures observed for the last 450 kyrs \cite{de2013astronomical}
 have been difficult to explain through astronomical variation theory alone. It seems clear that a full explanation of this periodicity must involve considerations of internal processes occurring on the Earth, such as the time-scales for the advance and retreat of the ice sheets. Hence in an attempt to address this problem, different conceptual models using the physics of ice sheets and the ocean-atmosphere feedback have been developed. We now briefly review some of these.

\section{Dynamical models for climate change}
\subsection{Conceptual models}

\noindent Whilst many sophisticated models for climate change exist, such as the Global Climate Models, for example the HadGem3 model \cite{hadgem3}, these cannot be run to simulate the long periods associated with the glacial cycles. Hence, in order to obtain insight into the glacial cycles it is often useful to make use of simpler conceptual models, which study variations of ice sheets and effects of astronomical forcing on these ice sheets. These are usually expressed in terms of low-dimensional dynamical systems. Various such dynamical systems models have been proposed to explain the glacial cycles, and a good review of these is given in the papers \cite{crucifix2012oscillators,de2013astronomical} and books \cite{kaper2013mathematics,dijkstra2013nonlinear}. Many of these models make use of some form of relaxation oscillator to explain the observed behaviour of the ice ages, and in particular the mechanism  of repeated slow growth of ice sheets  followed by rapid decay of ice sheets. In these models phase locking is often observed between the  astronomical variations and the internal variability of the Earth's climate \cite{dijkstra2013nonlinear}.

\vspace{0.1in}

\noindent {\em Smooth models:} Many such models use the concept of smooth excitable systems \cite{saltzman1991first,saltzman1990first,crucifix2012oscillators} and explain the origin of the glacial cycles through mechanisms such as the Hopf bifurcation. In particular, Saltzman and Maasch  \cite{saltzman1990first,saltzman1991first}, advocated that the glacial cycles could be viewed as limit cycles synchronized by the insolation forcing.  The Saltzman and Maasch (SM90) \cite{saltzman1990first} and 1991 (SM91) \cite{saltzman1991first} models adopt the hypothesis that the increase in insolation causes a decrease in ice sheet mass and that the change in atmospheric carbon is driven by tectonic forcing. The SM90 and SM91 models are smooth  dynamical systems with non-linearity on the Carbon Dioxide equations playing an important role in inducing  the existence of limit cycle. These models interpret the Mid-Pleistocene Transition (MPT) as a bifurcation from a quasi-linear response to a nonlinear resonance \cite{crucifix2012oscillators} with the SM90 model experiencing a Hopf bifurcation \cite{ashwin2015middle}.

\vspace{0.1in}

\noindent {\em Non-smooth threshold models:} A second type of models, described  in \cite{paillard1998timing,paillard2004antarctic,ashwin2015middle, EW}, use the concept of thresholds to link the Carbon cycle  to the retreat of the ice sheets. The justification  for this choice being the presence of abrupt transitions in the paleo-climate geological records \cite{paillard2001glacial}. In 1998 Paillard (P98) suggested that the climate system can be represented as three quasi-stable states that are driven by astronomical forcing. In P98, threshold criteria are used to bring about the instability into the systems so that it can switch between different climatic states. The presence of such thresholds in the P98 model give it the form of a {\em hybrid dynamical system} \cite{bernardo2008piecewise}. Gildor and Tziperman \cite{gildor2000sea} proposed a (higher complexity) box model of the climate with thresholds to explain the MPT. This  model (GT2000) coupled ocean, atmosphere, sea ice and land ice behaviour, with the ocean divided into  eight boxes, the atmosphere into four vertically averaged boxes and with the sea ice responding to the energy balance equations. In this model the sea ice was considered to have a hysteretic response to the variations in the land ice volume with thresholds in the land ice volume (due to growth and melting) bringing about the switching mechanism. This mechanism suggested that the MPT could be as a result of climate cooling which in turn allowed sea ice cover to expand, hence activating the sea ice switch. Therefore implying that the glacial cycles of 100 kyr timescale did not rely on the astronomical forcing. Another model of glacial cycles is given in \cite{ashwin2015middle} (AD15). This model proposes that  global ice volume relaxes to an equilibrium state depending on a climatic state and that melting of the ice sheet is governed by astronomical variations of the insolation forcing. In this model the climatic states are  governed by a drift function  which describes a nonlinear relationship between the state and the ice volume, and the transitions from the 40 kyr to the 100 kyr period states are described  as a trans-critical bifurcation on the slow manifold. Non-smooth effects, and an analysis based on non-smooth dynamics, are also considered in the paper \cite{EW} which looks at a glacial ice-line model formulated as a Filippov system.

\subsection{The PP04 threshold model for the ice ages}

\noindent The model that we will study in this paper was introduced by Paillard and Parrenin in 2004 \cite{paillard2004antarctic} (PP04) and describes the evolution and feedback mechanisms associated with the global ice volume $V$, the extent of Antarctic ice sheet $A$ and atmospheric Carbon Dioxide content $C$. PP04 is a piece-wise smooth model of the glacial cycles that incorporates  physical mechanisms involving the influence of the Antarctic ice-sheet extent on the bottom water formation.  These in turn cause dramatic changes in the amount of atmospheric Carbon Dioxide  during the glacial-interglacial transitions when Carbon Dioxide is thought to be released from the deep ocean.  Fuller details of the physical motivation of this model can be found in \cite{paillard2004antarctic}

\vspace{0.1in}

\noindent In the PP04 model the equations for the change in  $V$  depend on the amount of atmospheric Carbon Dioxide $C$ and involve the astronomical forcing, and the extent of the Antarctic ice sheets $A$ is then coupled with global ice volume, with full details of the model given in \cite{paillard2004antarctic}. The amount of Carbon Dioxide in the atmosphere is considered to depend on the reduction of the amount of global ice volume, the insolation forcing, and the state of the Southern Ocean. In particular the ocean contribution is represented by the (discontinuous) Heaviside function $H(-F)$ and is dependent on the the 'salty bottoms efficiency' parameter $F(V,A,C)$ which is positive when the climate is in a glacial state, and negative when it is in an inter-glacial state.  In this model the atmospheric Carbon Dioxide rapidly increases when the Southern Ocean suddenly ventilates. The ventilation process is described in \cite{paillard2004antarctic}, and occurs when the deep ocean stratification (which usually prevents the water mixing) ceases due to 
a difficulty in salty bottom water formation, leading to a release of Carbon Dioxide. This release of Carbon Dioxide then leads to a warming of the Earth which drives a rapid deglaciation process. After this event the ice sheets accumulate slowly, until the threshold value of $F$ is again reached and another ventilation is triggered. Consequently, during the glacial periods, there is  no ocean contribution in the system until another release of Carbon Dioxide from the ocean is initiated. 

\vspace{0.1in}

\noindent Perhaps the most important part of this model is the inclusion of the function $F$, which acts as a switch in the system between the glacial and inter-glacial states. According to \cite{paillard2004antarctic} $F$ should increase when changes in $V$ lead to global cooling, and decrease when continental shelf areas are reduced. The function $F(V,A,C)$ is then defined by
\begin{equation}
F = a V - b A - c I_{60}(t) + d. 
\label{cnov1}
\end{equation}
In this model $F$ increases with global ice volume and decreases with the Antarctic ice sheet and Southern Hemisphere insolation forcing. The constant parameter $d$  controls the threshold crossing for the model from glacial to interglacial states, and $I_{60}$, is the daily insolation forcing at $60^0 S$. The value of $c$ is taken in \cite{paillard2004antarctic} to be very small. The full PP04 model for the glacial cycle model is then defined by
the piece-wise smooth, non-autonomous dynamical system given by:

\begin{align}
\frac{dV}{dt}& = \frac{(-x C - y I_{65}(t) + z - V)}{\tau_V},
\label{pp41}\\
\frac{dA}{dt}& =
\frac{(V - A)}{\tau_A},\\
\frac{dC}{dt}& = \frac{(\alpha I_{65}(t) - \beta V + \gamma H(-F)+\delta-C)}{\tau_C},
\label{pp4}
\end{align}

with $H$ being the Heavyside function defined by
\begin{align*}
   H(-F) =
        \begin{cases}
              1&\quad \text{if}\quad   F<0,\\
              0&\quad \text{ if} \quad F>0.
         \end{cases}
\end{align*}
Here  $x,y,z,\alpha,\beta,\gamma,\delta$ are physical constants, $\tau_V,\tau_A,\tau_C$ are the time-scales associated with ice formation and Carbon Dioxide growth, and $I_{65}(t)$ is the insolation at $65^{\circ}$ North. According to Garcia-Olivares et al \cite{garcia2013simulation}, the parameter $\delta$ can be interpreted as the Carbon Dioxide reference level and $\beta$ represents the positive feedback between the temperature and the Carbon Dioxide levels through the ice volume $V$. Values for $I_{65}(t)$ are provided in  Mitsui et. al. \cite{mitsui2014dynamics}, \cite{de2013astronomical,mitsui2014dynamics}. As a result of the introduction of the ocean contribution into the equation for Carbon Dioxide, the PP04 system has a {\em derivative discontinuity} when $F= 0$. This is described in \cite{paillard2004antarctic} as a {\em reflection of the nonlinearity of the interactions between deep stratification, bottom water formation and thermohalide circulation}. 

\subsection{Model parameters}

 \noindent In the paper \cite{paillard2004antarctic}, Paillard et. al. considered the values given in a table \ref{par} to produce their figures. These parameter values were considered from first principles and were obtained experimentally.  We will use the same values for our analysis, except that following a discussion with Prof. Paillard at the July 2017 CliMathNet Conference, we take $\gamma = 0.7$.

\begin{table}[htp]
\centering

\label{tab. table1.}
\begin{tabular}{|c|c|c|}
\hline 
variables  & values& Range \\
\hline 
	$ a$ & $0.3$&0.26-0.39\\
			$b$ & $0.7$&0.63-0.74\\
			$c$ &$ 0.01$&0-0.15 \\
			$d$ &$ 0.27 $&0.253-0.302\\
			$x $ & $1.3$&1.23-1.44\\
			$y$ & $0.5$&0.4-0.64\\
			$z $ & $0.8$&0.77-0.82\\
			$\alpha$ & $0.15$&0-0.35\\
			$\beta$ & $0.5$&0.46-0.54\\
			$\gamma$& $0.5$&0.37-0.7\\
			$\delta$& $0.4$&0.39-0.42\\	
			$\tau_V$ & $15(kyr)$&13.2-18.1\\
			$\tau_C$& $5(kyr)$&3.1-15\\
			$\tau_A$& $12(kyr)$&9.5-26\\
			\hline
		\end{tabular}
		\label{par}
		\caption{The model parameter values used in the original PP04 model. In this paper we take $\gamma = 0.7$.}
	\end{table}

\subsection{Realistic values for the insolation forcing. }

\noindent In \cite{de2013astronomical,mitsui2014dynamics,ashwin2015middle} a Fourier series representation was given for the astronomical forcing at the Northern hemisphere summer solstice at $65^0$latitude. The resulting expression is given by
\begin{equation}
 I_{65}(t) =\frac{1}{e}\sum _{i=1}^{35}[s_i \sin  (\omega_i t)+ c_i \cos (\omega_i t)].
 \end{equation}
Here the values of $e$, $w_i$, $s_i$ and $c_i$ are given in \cite{de2013astronomical,mitsui2014dynamics,mitsui2015bifurcations} and are found through through linear regression over the past one million years to the present. The parameter $e$ is  a scale factor used to make the function for astronomical forcing dimensionless. Different researchers take different values of $e$  for different models. For instance  Ashwin and Dietlevsen \cite{ashwin2015middle}  considered $e=1$ for the AD15 model. However  Mitsui and Aihara \cite{mitsui2014dynamics} considered $e = 11.77 \; Wm^{-2}$ for the  Crucifix-De Saedeleer model and $e = 18.3 \; Wm^{-2}$ for the SM90,SM91 and PP04 models, basing the value on the three frequency components of astronomical forcing that they considered to be significant. In contrary Mitsui et. al. \cite{mitsui2015bifurcations} considered the parameter $e = 23.58 \; Wm^{-2}$ for the PP04 model. We will take the latter value for this paper. 

\vspace{0.1in}

\noindent According to Mitsui et al \cite{mitsui2014dynamics}, the three astronomical forcing components: the precession terms at $i = 1$ (23.7 kyr) and $ i = 3$ (19.1 kyr) as well as  obliquity term at $i = 4$ (41.0 kyr) constitute $78$ percent of original insolation forcing.  It is thus reasonable to consider the simplified astronomical forcing as a quasi-periodic function comprising three harmonics that includes precession at 19 or 23 kyrs and the (dominant) obliquity forcing at $41 kyr$ ($\omega_4 = 0.1532$).  We observe that the amplitude of the obliquity forcing, obtained by taking the coefficient $s_4 \approx -11$ reported in \cite{mitsui2014dynamics} and setting all other coefficients to zero, and dividing by $e = 23.58$, is approximately $\mu = 0.467.$    
Accordingly, for the remainder of this paper we will use
the frequency and forcing amplitude  
\begin{equation}
    \omega = 0.1532, \quad \mbox{and} \quad \mu = 0.467
\label{muandomega}
\end{equation}
as parameters of a physically realistic single mode insolation forcing \cite{mitsui2014dynamics,de2013astronomical}. However, to understand the general behaviour of the model, we will explore the dynamics which results from taking other values of these parameters.

\section{The PP04 model as a Filippov system.}

\subsection{Overview} 

\noindent Although the threshold models described above, and in particular the PP04 model, are non-smooth in nature, they have been studied so far as smooth systems, See for example \cite{mitsui2014dynamics,mitsui2015bifurcations}. In this analysis only the types of dynamics peculiar to smooth dynamical systems were observed, and Mitsui remarked that this was a limitation of the smooth analysis. Indeed, hybrid dynamical systems are canonical examples non-smooth systems and can be studied best by using the theory of non-smooth dynamical systems \cite{bernardo2008piecewise} in order to find all the dynamics present in the system. This is the motivation for the approach used in this paper. 

\vspace{0.1in}

\noindent Non-smooth dynamical systems arise in a large number of  applications and as models of a number of phenomena. They are used  in mechanical engineering in vibro-impacting systems, or in switches in electronic circuits such as thermostats and also in climate models \cite{di2010discontinuity,guardia2008topological}. Discontinuous dynamical systems  are systems where the vector field is piece-wise smooth (discontinuous). Therefore the dynamical system is known to be  non-smooth as its trajectories may not be differentiable everywhere. Non-smooth dynamical systems  are  characterized by some discontinuity in their right hand side, which can be due to the discontinuities in evolution with respect to time, or the system state reaching a discontinuous boundary \cite{di2008bifurcations,colombo2010discontinuity}. Consequently they can be used to represent numerous physical processes which are characterized by periods of smooth evolution being interrupted by an instantaneous event or the systems whereby the physical states switches between two or more different states \cite{bernardo2008piecewise}. Therefore, when modelling such  physical states, each state is given by a different set of differential equations \cite{awrejcewicz2005continuous}. That is, in each region, the evolution of trajectories are defined by the smooth dynamical system which changes to a different defining system  across the discontinuity boundary \cite{glendinning2016classification}. Thus the  behaviour is that of a piece-wise smooth dynamical system \cite{bernardo2008piecewise}.

\subsection{Filippov systems}

\noindent A Filippov system is a general piece-wise smooth dynamical system comprising a finite set of ordinary  differential equations, which can be expressed as
\begin{equation}
\dot{\textbf{x}}={\mathbf N}_{i}(\textbf{x})\:\:\:\:\:\:\:\:\: \textbf{x}\in S_i\subset R^n.
\end{equation}
Here  each subspace or region $S_i$ has a non-empty interior, and the vector field ${\mathbf N}_i$ is smooth and defined on the disjoint open regions $S_{i}$. The intersection   $\Sigma_{ij}$ of $S_i$ and $S_j$ is either an $R^{n-1}$ dimensional manifold included in the boundaries of the two regions or it is an empty set. A non empty border between any two or more regions $S_{i}$ is called a discontinuity boundary or switching manifold \cite{bernardo2008piecewise,colombo2012bifurcations}.
A piece-wise smooth system with a single discontinuity boundary (such as the PP04 model) can be defined by:
\begin{equation}
	\begin{aligned}
	\dot{\textbf{x}} =
	\begin{cases}
	{\mathbf N}_{1}(\textbf{x})&\quad \text{if}\quad   \textbf{x}\in S_1\\
	{\mathbf N}_{2}(\textbf{x})&\quad \text{ if} \quad \textbf{x}\in S_2.
	\end{cases}
	\end{aligned}
	\end{equation}
and we call $\Sigma_{12} \equiv \Sigma$
According to Cort\'es \cite{cortes2008discontinuous} and di Bernardo et al \cite{bernardo2008piecewise}, the {\em degree of smoothness}  of the piece-wise smooth system depends on whether the system exposes jumps and or switches on its state, vector field or its Jacobian. The degree of smoothness at the point $x_0$ on  the discontinuity boundary set is given by the highest order $r$ such that the Taylor expansions of the flows either side of $\Sigma$ (assumed to be at time $t = 0$) agree up to terms of $\mathcal{O}(t^{r-1})$. This informs us about the behaviour of the flow as it crosses the boundary\cite{bernardo2008piecewise}.
Systems have  degree of smoothness one  if $F_{i}(\textbf{x},\mu)-F_{j}(\textbf{x},\mu)\neq 0$ for $\textbf{x}\in \Sigma_{i j}\cap  D$ \cite{bernardo2008piecewise} and are called {\em Filippov} systems.
According to Piiroinen et al \cite{piiroinen2008event}, an important feature of a general Filippov systems is the possibility of  motion to be  constrained to the discontinuity  boundary where the orbit can slide. We will show that this does not arise in the PP04 model, which is an important aspect of its dynamical behaviour. 

\subsection{Features of the PP04 model as a Filippov system}
 \noindent We can formulate the PP04 model as a forced Filippov System. (A similar formulation of an ice-line model for the glacial dynamics is given in \cite{EW}). To do this we introduce a state vector
 $${\mathbf X} = (V,A,C)^T.$$
 
 \vspace{0.1in}
 
 \noindent According to Paillard et. al. \cite{paillard2004antarctic}, the inclusion of the  $I_{60}(t)$ term in the definition of the function $F$ does not affect the times when the glacial cycles terminates or the qualitative form of the overall dynamics. The proportionality coefficient $c=0.01$ in their model is very small (in comparison to $a,b$ and $d$) and their range of values for $c$ includes $c=0$. Setting $c$ equal to zero significantly simplifies the theoretical analysis of the PP04 model, without changing the observed dynamics in any significant way. Accordingly we set $c=0$ for the remainder of this paper.
 
 \vspace{0.1in}
 
\noindent With this simplification, it then follows from (\ref{cnov1}) that in all regions 
\begin{equation}
    F({\mathbf X}) = (a,-b,0)^T \; {\mathbf X} + d \equiv {\mathbf c}^T {\mathbf X} + d. 
    \label{cnov3}
\end{equation}
The discontinuity surface $\Sigma$ is then given by the linear relation
\begin{equation}
{
\Sigma = \{ {\mathbf X}:  {\mathbf c}^T {\mathbf X} + d = 0. \}
}
\label{c1}
\end{equation}

\noindent We define the following two states corresponding to the glacial and inter-glacial states
\begin{equation}
{
S_1 \equiv S^{+} = \{ {\mathbf X} : F({\mathbf X}) > 0, \quad S_2 \equiv S^{-} = \{ {\mathbf X} : F({\mathbf X}) < 0. \}.
}
\label{c2}
\end{equation}
The PP04 model in $S^{\pm}$ can then be written as:
\begin{equation}
{
\dot{{\mathbf X}} = L {\mathbf X} + {\mathbf b}^{\pm} + I_{65}(t) \; {\mathbf e}
}
\label{c3}
\end{equation}
Here the linear operator $L$ and the vector ${\mathbf e}$ are defined by
\begin{equation}
L = \left(
\begin{array}{c c c}
-1/\tau_V & 0 & -x/\tau_V \\
1/\tau_A  & -1/\tau_A & 0 \\
-\beta/\tau_C & 0 & -1/\tau_C
\end{array}
\right),
\quad 
{\mathbf e} = \left(
\begin{array}{c}
-y/\tau_V \\
0 \\
\alpha/\tau_C
\end{array}
\right).
\label{cnov5}
\end{equation}
\noindent It follows from a direct calculation that the linear operator $L$ has negative eigenvalues $-\lambda_1 < -\lambda_2 < -\lambda_3$, with corresponding eigenvectors ${\mathbf e}_i, i=1,2,3$. These values do not depend upon the system state. 

\vspace{0.1in}

\noindent In contrast the vectors
${\mathbf b}^{\pm}$ depend upon which region ${\mathbf X}$ lies in and 
are given by:
\begin{equation}
    {\mathbf b}^+ = \left(
    \begin{array}{c}
    z/\tau_V \\
    0 \\
    \delta/\tau_C
    \end{array}
    \right),
    \quad
   {\mathbf b}^- = \left(
    \begin{array}{c}
    z/\tau_V \\
    0 \\
    (\gamma + \delta)/\tau_C
    \end{array}
    \right).
    \label{cnov6}
\end{equation}

\vspace{0.1in}

\noindent It is clear from this formulation that the PP04 model has a piece-wise linear Filippov structure.  We can thus expect it to have similar 
dynamics to a typical Filippov problem and to show both 'smooth' and 'discontinuity induced' bifurcations (for example grazing bifurcations \cite{book,simpson2010bifurcations}) as parameters are varied. Indeed this is exactly what we will see in this paper. In the paper \cite{EW} a Filippov system of a similar form to the above was analysed for the ice-line model, and some of the ideas used in studying that system can be applied in the PP04 model. 

\vspace{0.1in}

\noindent We now look at the structure of the Fillipov formulation of the PP04 model. 

\vspace{0.2in}

\noindent {\bf Lemma 4.1}  {\em (i) The solutions of the PP04 system remain bounded for all time. 

\noindent (ii) There is an attracting region $B$ in the ${\mathbf X}$-phase space into which all trajectories enter.}

\vspace{0.1in}

\noindent {\bf Proof} As $L$ has all negative real eigenvalues, it can be written as $L = U\Lambda U^{-1}$ where $\Lambda = diag(-\lambda_1,-\lambda_2,-\lambda_3).$ If we set ${\mathbf Y} = U^{-1} {\mathbf X}$, ${\mathbf p}^{\pm} = U^{-1} {\mathbf b}^{\pm}$ and ${\mathbf q} = U^{-1} {\mathbf e}$ then 
$$\dot{\mathbf Y} = \Lambda {\mathbf Y} + {\mathbf p}^{\pm} + {\mathbf q} \; I_{65}(t).$$
Now consider $N = {\mathbf Y}^T {\mathbf Y}/2$ then it is immediate that
if $N$ is sufficiently large then
$$\dot{N} = {\mathbf Y}^T \Lambda {\mathbf Y} + {\mathbf Y}^T\left({\mathbf p}^{\pm} + {\mathbf q} \; I_{65}\right) < -\min({\lambda_i}) N + {\mathbf Y}^T\left({\mathbf p}^{\pm} + {\mathbf q} \; I_{65} \right) < 0.$$
Hence $N$, and thus $|{\mathbf X}|$, is bounded. To prove (ii) we note (from inspection of the actual matrix) that the matrix $U^TU$ is positive definite. It follows that bounded sets in $Y$ correspond to bounded sets in $X$ and vice-versa. Hence the $N-$ball in the $Y$ space corresponds to a bounded set $B$ in the $X$ space. \qed

\vspace{0.2in}

\noindent {\bf Lemma 4.2}  {\em The degree of discontinuity of the PP04 model is one. }

\vspace{0.1in}

\noindent {\bf Proof} It is clear from the formulation that ${\mathbf X}$ is continuous on $\Sigma$, but that $\dot{\mathbf {X}}$ has a jump discontinuity. The result then follows.  \qed

\vspace{0.1in}

\noindent The following results describe the change of $F$ across $\Sigma$ and show that we do not have sliding solutions.

\vspace{0.2in}

\noindent {\bf Lemma 4.3} {\em (i) $F$ and $dF/dt$ are continuous across $\Sigma$.

\noindent (ii) At any point on $\Sigma$ we have $(d^2 F/dt^2)^+ =  (d^2F/dt^2)^- + \alpha$
where $\alpha > 0$ is a positive constant.}

\vspace{0.1in}

\noindent {\bf Proof}  (i) If  $$F({\mathbf X}) = {\mathbf c}^T{\mathbf X} + d.$$
The continuity of $F$ is immediate. It also follows immediately that 
\begin{equation}
\begin{split}
\frac {dF}{dt} = {\mathbf c}^T \frac {d}{dt}{\mathbf X}\;
={\mathbf c}^T L {\mathbf X} +{\mathbf c}^T {\mathbf b}^{\pm} + {\mathbf c}^T {\mathbf e } I_{65}(t).
\end{split}
\end{equation}
Then if we define 
 \begin{equation}
 {
 {\mathbf h} = {\mathbf c}^T L , \quad r^{\pm} = {\mathbf c}^T {\mathbf b}^{\pm}, \quad g(t) = {\mathbf c}^T{\mathbf e} \; I_{65}(t),
 }
 \end{equation}
 we have 
 \begin{equation}
 \frac {dF}{dt} = {\mathbf h}^T{\mathbf X} +  r^{\pm} + g(t).
 \end{equation}
 However, it follows directly from the definition of ${\mathbf c}$ in (\ref{cnov3}) and of ${\mathbf b}^{\pm}$ in (\ref{cnov6}) that
 $$ r^- = r^+ \equiv  r.$$ 
 So  
 \begin{equation}
 \frac {d}{dt}F = {\mathbf h}^T{\mathbf X}+  r + g(t).
 \end{equation}
 It is clear that $dF/dt$ is then continuous across the discontinuity surface. 
 
 \vspace{0.1in}
 
 Similarly we have 
 \begin{equation}
     \frac{d^2 F}{dt^2} = {\mathbf h}^T (L{\mathbf X} +  {\mathbf b} ^{\pm}) + \dot{g}(t).
 \end{equation}
 Thus 
 $$[\ddot{F}]^+_- = {\mathbf h}^T({\mathbf b} ^{+} - {\mathbf b}^-) \equiv \alpha  = 0.00364.$$
 \qed
 
 \vspace{0.1in}
 
 \noindent If we approach $\Sigma$ from $S^+$ it follows that $dF/dt \le 0$. In particular if $dF/dt < 0$ on $\Sigma$ then from Lemma 4.2 it follows immediately that the corresponding trajectory must immediately enter the region $S^-$ and does not slide on $\Sigma$.
 
 \vspace{0.1in}
 
 \noindent It is possible for {\em grazing} to occur on $\Sigma$. This arises when $F = 0$ {\bf and} $dF/dt = 0$. In the case of an unforced system this will arise when
$${\mathbf c}^T{\mathbf X} + d = 0 \quad \mbox{and} \quad {\mathbf h}^T{\mathbf X }+ r = 0.$$
It follows immediately that in this case grazing on $\Sigma$ occurs along a straight line, the {\em grazing set} ${\cal G}$, which is parallel to the
vector ${\mathbf c} \times {\mathbf h}.$

\vspace{0.1in}

\noindent We note further that in this case we have
$$\frac{d^2 F}{dt^2} = {\mathbf h}. (L {\mathbf X} + {\mathbf b^{\pm}}).$$
Hence the surface $d^2 F/dt^2 = 0$ is another plane in each region $S^{\pm}$. This can intersect ${\cal G}$ at at most one point. This rules out the possibility of sliding.    

\vspace{0.2in}

\noindent Following this result, we can, without ambiguity make the following definitions:

\begin{equation}
    \Sigma^+ = \{{\mathbf X} \in \Sigma: dF/dt > 0. \} \quad \Sigma^- = \{{\mathbf X} \in \Sigma: dF/dt < 0. \} 
\end{equation}

\section{The dynamics of the unforced PP04 model}

\noindent We now study the unforced PP04 model which arises when there is zero insolation forcing, and consequently $\mu = 0$. In this study we show that for certain parameter values this (non-smooth) model has periodic solutions, which arise at border collision bifurcations between the fixed points and $\Sigma$
as parameters in the model change. This form of the periodic solutions are similar to that observed in \cite{EW}.

\subsection{Fixed Points}

\noindent It is easy to see that the PP04 model has two fixed points given by
\begin{equation}
    {\mathbf Z}^{\pm} =-L^{-1} {\mathbf b}^{\pm}.
    \label{cfp1}
\end{equation}
As $L$ has negative eigenvalues, these are both attracting nodes. We define
\begin{equation}
K^{\pm} = F({\mathbf Z}^{\pm}) = -{\mathbf c} ^T L^{-1}{\mathbf b}^{\pm}+d.
\label{cfp2}
\end{equation}

\vspace{0.1in}

\noindent If $K^+ > 0$ then ${\mathbf Z}^+$ lies in $S^+$ and is a {\em physical} fixed point. Any orbit which remains in $S^+$ for all time will evolve towards it. 

\vspace{0.1in}

\noindent If in contrast  $K^+ < 0$, then ${\mathbf Z}^+$ lies in $S^-$, and is a {\em virtual} fixed point. It has a stable manifold in $S^+$ and attracts trajectories in $S^+$ towards it. Such trajectories ultimately cross $\Sigma$ and enter $S^-$.
An exactly similar situation arises for the fixed point ${\mathbf Z}^-.$
A {\em border collision bifurcation} (BCB) occurs when either of the two fixed points crosses $\Sigma$ as a parameter varies. 

\vspace{0.1in}

\subsection{The dynamics of the unforced system as parameters vary.}

\noindent We now establish the following result which describes the changing dynamics of the unforced system as parameters vary.

\vspace{0.1in}

\noindent {\bf Theorem 5.1} {\em Let ${\mathbf Z}^{\pm}$ be defined as above

\vspace{0.1in}

\noindent (i) If ${\mathbf c}^T{\mathbf Z}^{+} + d \equiv d -L^{-1}{\mathbf b}^+ > 0$ then ${\mathbf Z}^+$ is a unique globally attracting fixed point.

\vspace{0.1in}

\noindent (ii) If ${\mathbf c}^T{\mathbf Z}^{-} + d \equiv d -L^{-1}{\mathbf b}^- < 0$ then  ${\mathbf Z}^-$ is a unique globally attracting fixed point.

\vspace{0.1in}

\noindent (iii) If ${\mathbf c}^T{\mathbf Z}^{-} + d \equiv d -L^{-1}{\mathbf b}^- > 0$ and if ${\mathbf c}.{\mathbf Z}^{+} + d \equiv d -L^{-1}{\mathbf b}^+ < 0$ then the system has a periodic solution $P(t)$ and no fixed points. 
}

\vspace{0.2in}

\noindent NOTE  We see from this lemma that the unforced system has either a fixed point or a period orbit, but not at the same time. This is in contrast to the Saltzman and Marsh models \cite{saltzman1991first,saltzman1991first}, but it is identical to the situation described in \cite{EW} where the periodic orbit is called a 'flip-flop' orbit. 

\vspace{0.1in}

\noindent {\em Proof} (i) Let ${\mathbf X}_0 \in S^+$ then provided that ${\mathbf X}(t) \in S^+$ we have
\begin{equation}
    {
    {\mathbf X}(t) = e^{Lt}({\mathbf x}_0 - {\mathbf Z}^+) + {\mathbf Z}^+. 
    }
    \label{cjuna1}
\end{equation}
Hence, if ${\mathbf X}$ remains in $S^+$ for all time, then (as $L$ has negative eigenvalues) it must asymptotically tend towards ${\mathbf Z}^+$. 

\vspace{0.1in}

\noindent Now, suppose that ${\mathbf X}(t)$ enters $S^-$. In this region we have
\begin{equation}
    {
    {\mathbf X}(t) = e^{Lt}({\mathbf x}_0 - {\mathbf Z}^-) + {\mathbf Z}^-. 
    }
    \label{cjuna2}
\end{equation}
Hence it is attracted towards the fixed point ${\mathbf Z}^-$ which lies within the region $S^+$. Thus $X$ must reenter the region $S^+$ at some later time. We claim that the trajectory either remains in $S^+$ for all time following this, and converges to ${\mathbf Z}^+$, or has a finite number of further 'visits' to $S^-$ before remaining $S^+$ and then converging to ${\mathbf Z}^+$

\vspace{0.1in}

\noindent To establish this result we suppose first that the trajectory enters $S^+$ at time $t_0$, leaves at time $t_1$, renters at time $t_2$ etc. so that $F(t_k) = 0, \dot{F}(t_{2j}) > 0, \dot{F}(t_{2j+1} < 0$. The function $F$ is defined by
$F = {\mathbf c}^T {\mathbf X} + d$, and hence in each regions $S^{\pm}$ it has the general form
\begin{equation}
F(t) = a_j e^{-\lambda_1 (t-t_j)} + b_j e^{-\lambda_2 (t-t_j)} + c_j e^{-\lambda_3 (t-t_j)} + K^{\pm}, \quad t_j < t < t_{j+1},
\label{Fapreqn}
\end{equation}
where, in this case, $K^+ >0$ and $K^- > 0.$  We firstly establish the following

\vspace{0.2in}

\noindent {\bf Lemma 5.2} {\em   If $c_{2j} > 0$ then the trajectory remains in $S^+$ for all $t > t_{2j}.$}

\vspace{0.1in}

\noindent {\bf Proof}. Suppose the converse. There must be a later time $t_{2j+1}$ for which $F(t_{2j+1}) = 0$. Now consider the globally defined function
$$F^*(t) = a_{2j} e^{-\lambda_1 (t-t_{2j}} + b_{2j} e^{-\lambda_2 (t-t_{2j})} + c_{2j} e^{-\lambda_3 (t-t_{2j})} + K^{+}.$$
As $c_{2j} > 0$ we must have that $F^*(t)$ tends to $K^+ > 0$ from above for large $t$, but $F*(t_{2j+1}) = 0$. By considering the shape of the curve $F^*$ we deduce that there are times $t_{2j} < t_a < t_{2j+1} < t_b < t_c$ so that $F^*(t_a) > 0, F^*(t_b) < 0, F^*(t_c) > K^+ > 0$ and $\dot{F^*}(t_a) = \dot{F^*}(t_b) = \dot{F^*}(t_c) = 0.$  However, it is immediate that $\dot{F^*}(t)$ is a sum of three different exponential functions. It is well known that a function which is the sum of $n$ different exponential functions can have at most $(n-1)$ zeros. Thus we have a contradiction. \qed

\vspace{0.1in}

\noindent Now consider the case of $c_{2j} < 0$ and assume that the trajectory crosses into $S^-$ at a time $t_{2j+1}$ and then back into $S^+$ at a time $t_{2j+2}$. We consider the map $G(c_{2j}) \to c_{2j+2}.$

\vspace{0.1in}

\noindent {\bf Lemma 5.3} {\em $G(z) = \alpha_j z + \beta_j$ where $0 < \alpha_j < 1$ and $0 < \beta_j < 1.$}

\vspace{0.1in}

\noindent {\bf Proof} At the time $t_{2j+1}$ we have (from Lemma 4.3) that $F(t_{2j+1}^-) =  F(t_{2j+1}^+)$, $\dot{F}(t_{2j+1}^-)=\dot{F}(t_{2j+1}^+)$ and $\ddot{F}(t_{2j+1}^-)=\ddot{F}(t_{2j+1}^+)+\alpha$. It follows, after some manipulation,  that the coefficients $a_{2j}, a_{2j+1}$ etc. obey the linear Vandermonde equation
\begin{equation} 
    \left(
    \begin{array}{r r r}
    1 & 1 & 1 \\
    \lambda_1 & \lambda_2 & \lambda_3 \\
    \lambda_1^2 & \lambda_2^2 & \lambda_3^2 \\
    \end{array}
    \right)
    \left(
    \begin{array}{r}
    e^{-\lambda_1 \Delta_{2j}} \; a_{2j}  - a_{2j+1}\\
    e^{-\lambda_2 \Delta_{2j}} \; b_{2j}  - b_{2j+1} \\
    e^{-\lambda_3 \Delta_{2j}} \; c_{2j}  - c_{2j+1} \\
    \end{array}
    \right)
= 
    \left(
    \begin{array}{c}
    K^- - K^+ \\
         0    \\
         \alpha \\
         \end{array}
         \right),
\end{equation}
where $\Delta_{2j} = t_{2j+1} - t_{2j}.$

\vspace{0.1in}

\noindent From the data given, 
$K^- -K^- = {\mathbf c}^T ({\mathbf Z}^+ - {\mathbf Z}^-) = 1.04 \quad \mbox{and} \quad \alpha = 0.00364.$
We deduce, on inverting the Vandermonde matrix, that
\begin{equation}
\left(
    \begin{array}{r}
    e^{-\lambda_1 \Delta_{2j}} \; a_{2j}  \\
    e^{-\lambda_2 \Delta_{2j}} \; b_{2j}   \\
    e^{-\lambda_3 \Delta_{2j}} \; c_{2j}   \\
    \end{array}
    \right)
    =
    \left(
    \begin{array}{r}
     a_{2j+1}\\
     b_{2j+1} \\
     c_{2j+1} \\
    \end{array}
    \right)
    +
    \left(
    \begin{array}{r}
    0.1399 \\
    -0.8004 \\
    1.7005 \\
    \end{array}
    \right).
    \end{equation}
\vspace{0.1in}

\noindent Applying the same result at the time $t_{2j+2}$ with $\Delta_{2j+1} = t_{2j+2}-t_{2j+1}$ we have
\begin{equation}
    c_{2j+2} = e^{-\lambda_3 \Delta_{2j+1}}\left(e^{-\lambda_3 \Delta_{2j}} c_{2j} - 1.7005 \right) + 1.7005.
    \label{cmay1}
\end{equation}

\noindent The form of $G$ given in the Lemma follows immediately.

\vspace{0.1in}

\noindent Now suppose that the trajectory always re enters $S^-$. It follows from Lemma 5.2 that $c_{2j}$ must always be negative. However, from Lemma 5.3 we have that
$$c_{2j+2} - c_{2j}  = (\alpha_j-1) c_{2j} + \beta_{2j} > 0.$$
Thus the sequence $c_{2j}$ bounded above (by zero) and is monotone increasing. It must therefore tend to a limit $c$, which in the limit satisfies $c = \alpha_j c + \beta_j$. As $\alpha_j$ and $\beta_j$ are always positive, this is a contradiction.
We deduce that $c_{2j}$ is eventually positive, at which point the trajectory remains in $S^+$ and hence tends to the fixed point ${\mathbf Z}^+$.   
This proves part (i).

\vspace{0.1in}

\noindent The proof of (ii) is identical that that given above.

\vspace{0.1in}

\begin{figure}[htbp]
    \centering
	\includegraphics[width=0.8\linewidth]{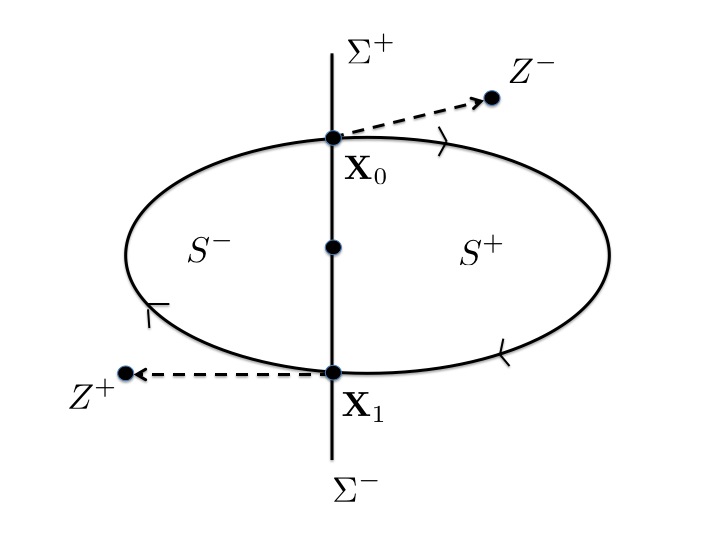}	\caption{Schematic showing the period orbit $P(t)$ and the virtual fixed points $Z^{\pm}$. }
	\label{fig:sketch}	
\end{figure}

\noindent To prove part (iii) we use the following argument, which is illustrated in Figure \ref{fig:sketch}.

\noindent Consider a trajectory which starts at the point ${\mathbf X}_0$ on $\Sigma^+$ and initially enters $S^+$ so that $dF/dt > 0$. As we are in case (iii), it follows that
\begin{equation}
F = a^+ e^{-\lambda_1 t} + b^+ e^{-\lambda_2 t} + c^+ e^{-\lambda_3 t} + K^+,
\label{cF1}
\end{equation}
where $K^+ < 0.$ For large $t > 0$, we must have $F < 0$. Then there must be a first time $t = t_1$ at which $F = 0$ and $dF/dt < 0$. 
At this point the trajectory intersects $\Sigma^-$ at the point ${\mathbf X} = {\mathbf X}_1.$
The flow now crosses over into $S^-$ with $F < 0$. The resulting flow is then given by
$$F = a^- e^{-\lambda_1 t} + b^- e^{-\lambda_2 t} + c^- e^{-\lambda_3 t} + K^-$$
with $K^- > 0$.
By the same argument, it follows that there is a first time $t_2 > t_1$ such that $F = 0$ and the trajectory intersects $\Sigma^+$ at the point ${\mathbf X}_2$.
The condition for this trajectory to be a periodic solution is that
\begin{equation}
    {
    {\mathbf X}_0 = {\mathbf X}_2 \equiv {\mathbf M}({\mathbf X}_0)
    }
    \label{cjuna3}
\end{equation}
\noindent This can be considered to be a fixed point condition for the nonlinear map ${\mathbf M}: \Sigma^+ \to \Sigma^+$ defined above. 

\vspace{0.1in}

\noindent It follows immediately from Lemma 4.1 that the function ${\mathbf M}$ maps the finite dimensional and bounded region $B \cap \Sigma^+$ into itself. 

\vspace{0.1in}

\noindent we now show that ${\mathbf M}$  is continuous. The trajectories in $S^+$ depend smoothly upon the initial value ${\mathbf X}_0$, hence $F(t)$ is a smooth function of ${\mathbf X}_0$. It therefore follows from the implicit function theorem that, provided $dF/dt < 0$ at ${\mathbf X}_1$, then the time $t_1$, and the point ${\mathbf X}_1$, are continuous (indeed differentiable) functions of ${\mathbf X}_0$. Similarly, the point ${\mathbf X}_2$ will also be a continuous function of ${\mathbf X}_1$. The continuity of the map ${\mathbf M}$ then follows provided that $\dot{F}({\mathbf X}_1) < 0.$ To prove this we establish a contradiction. Consider the function given (\ref{cF1}).
Suppose that at times $t_0$ and $t_1$ we have $F = 0$ and that also $\dot{F}(t_1) = 0$ so that $F(t) > 0$ if $t$ is close to $t_1$ and $t > t_1$. As $K^+ < 0$ there must be a later time $t_2$ such that $F(t_2) = 0.$ It follows from Roll\'e's Theorem that there must be times $t_a$ and $t_b$ with $t_0 < t_a < t_1 < t_b < t_2$ such that
$$\dot{F}(t_a) = \dot{F}(t_1) = \dot{F}(t_b) = 0.$$
Now, as before, $\dot{F}$ is a sum of three exponential functions. Such a function cannot have three zeros. Thus we have established the desired contradiction.

\vspace{0.1in}

\noindent We have thus established that the function ${\mathbf M}$ maps a bounded finite dimensional region into itself, and is continuous. The existence of a fixed point, and hence of a periodic orbit, then follows immediately from the Brouwer Fixed Point Theorem.

\qed

\vspace{0.1in}

\noindent NOTE A similar system was studied in \cite{EW} by using a contraction mapping argument which could be applied directly to their problem and using which they could also prove uniqueness of their periodic orbit for certain parameter values.  

\vspace{0.2in}

\begin{figure}[htbp]
\centering
  \includegraphics[width=0.7\textwidth]{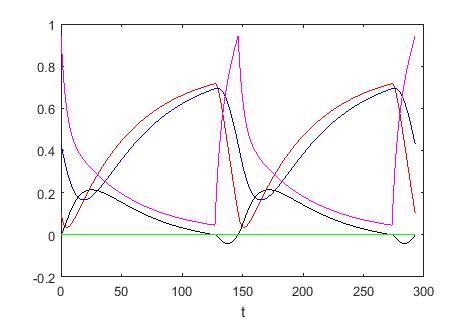}
  \caption{Unforced periodic solution showing $V$ (red), $A$ (blue) and $C$ (magenta), as well as $F$ (black).}
\label{fig:nc2}
\end{figure}

\noindent If we take the tabulated values for the PP04 model, with $\gamma = 0.7$ and $d = 0.27$ then we have fixed points at ${\mathbf Z}^+ = (0.8,0.8,0)$ and ${\mathbf Z}^- = (-1.8,-1.8,2)$ At these points we have $K^+ = c^T Z^+ + d = -0.05$ and
$K^- = c^T Z^- + d = 0.99$ so that the condition for a periodic solution is satisfied. The time series of the components of the resulting periodic solution (which appears from these calculations to be unique), is then illustrated in Figure \ref{fig:nc2}. These show a saw-tooth like structures similar to those evident in the geological reconstructed data. Such an oscillation was observed in the original PP04 model (see \cite{crucifix2012oscillators}). The relaxation oscillator obtained for the parameters we use has a period of about $147 kyr$. Note that the period of this unforced oscillation is higher than the observed period of $100 kyr$, but is not dissimilar. This model therefore suggests that the natural timescales of the Earth do play a role in determining the frequency of the ice ages.

\subsection{Border Collision and smooth bifurcations of the fixed points and periodic solution.}

\noindent If we vary one of the parameters of the system, say $d$,
then the periodic solution can lose existence at a {\em border collision bifurcation} (BCB), when either one of the two (virtual) fixed points ${\mathbf Z}^{\pm}$ intersects $\Sigma$. We then see a change from a periodic solution to a fixed point. 

\vspace{0.1in}

\noindent Qualitatively, the behaviour close to the BCB is illustrated by the representative phase-plane diagram in Figure \ref{fig:sketch} given earlier. In this we show the periodic solution when the two 
fixed points are virtual. The solid lines show the true dynamics in $S^+$ and $S^-$, and the dotted lines the 'virtual' dynamics if, for example the dynamics in $S^+$ is extended into 
$S^-$ so that it approaches the virtual fixed point. Even if the fixed point is close to $\Sigma$ this periodic orbit has a non-vanishing amplitude, indeed the amplitude tends
to a non-zero limit as one of the fixed points, say ${\mathbf Z}^+$ approaches $\Sigma$. 
However, as the BCB is approached the period of the periodic solution increases as it takes longer to approach $\Sigma$. The period rises to infinity when the fixed point ${\mathbf Z}^+$ lies on $\Sigma$.

\vspace{0.1in}

\begin{figure}[htbp] 
\centering
  \includegraphics[width=0.7\textwidth]{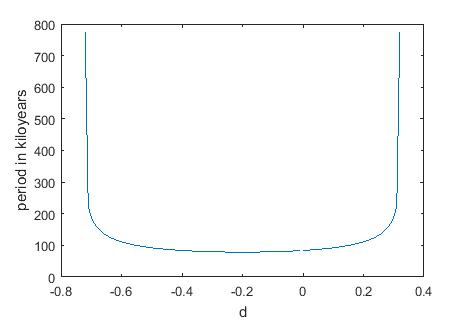}
  \caption{The change in the period of the periodic solution as the parameter $d$ is varied. In this we can see the Border Collision Bifurcations at $d = -0.72$ and $d = 0.32$ where the period becomes infinite.}
\label{fig:n3}
\end{figure}

\noindent The values of $d^{\pm}$ at which we have a BCB occur when either $K^+ = 0$ or $K^- = 0$. These cases arise when
\begin{equation}
d^{\pm} = -{\mathbf c}^T{\mathbf Z^\pm}
\end{equation}
For the tabulated values we obtain 
$$ d^- = -0.72, \quad d^+ = 0.32,$$ 
and hence a periodic solution exists when  $-0.72 < d < 0.32$.  
In Figure \ref{fig:n3} we show the period of the periodic solution as a function of $d$. In which we can see the two BCBs at which the period tends to infinity. 
 
\vspace{0.1in}

\noindent It is of interest, both theoretically, and also from the need to do computations, to consider how this bifurcation structure arises if we replace the non-smooth system by a smooth one. A convenient way to do this (see for example \cite{crucifix2012oscillators}) is to replace the (non-smooth) Heaviside function, by the regularized function
\begin{equation}
H_{\eta}(z) = \frac{1}{2} (1 + \tanh(\eta z) ).
\label{capr1}
\end{equation}
For large values of $\eta$ this closely approximates the Heaviside function. Using this approximation we integrate the system (\ref{pp41}-\ref{pp4}) forward in time numerically by using the Matlab stiff ode solver {\tt ode15s}. To determine the dynamics of the solution we then start with a random set of initial conditions and find the solution of the dynamical system starting from these. We then take a large enough time interval to allow the solution to converge onto its $\Omega-$limit set. To record this set we then plot the maximum and minimum values of $F$ on the asymptotic orbit. We choose to plot $F$ as this then allows us to see how the Omega limit set of the solution interacts with the discontinuity surface. By doing this for a set of values of $d$ we can determine the complete bifurcation picture for the solutions. 

\vspace{0.1in}

\begin{figure}[htbp]
    \centering
	\includegraphics[width=0.6\linewidth]{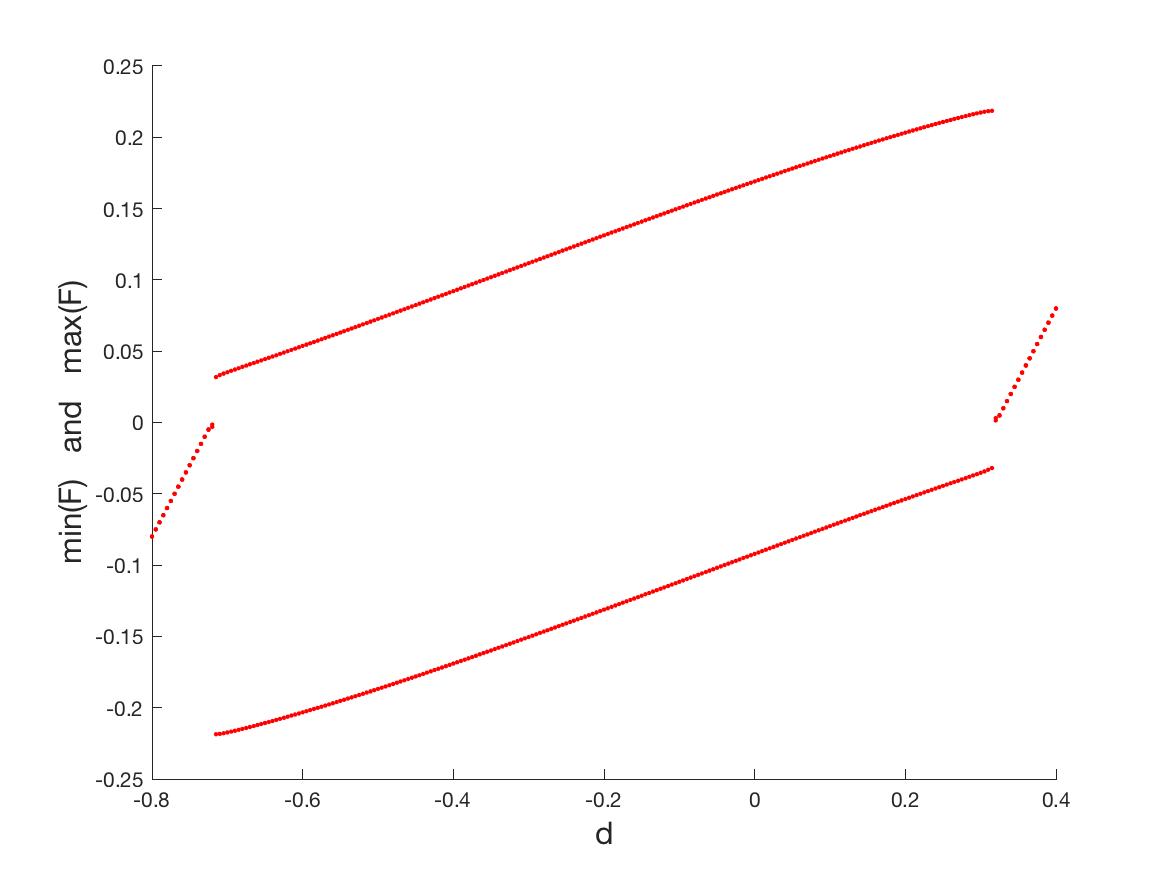}	\caption{The bifurcations of the solutions when $\eta = 1600$ showing the two border-collision bifurcations at $d = -0.72$ and $d = 0.32$. }
	\label{fig:Hopfbifur1}	
\end{figure}

\noindent If $\eta = 1600$ then $H_{\eta}(z)$ is a very good approximation to the Heavyside function, and we expect the dynamics of the smoothed system to be very close to that of the
Filippov system describing the PP04 model. The numerically computed bifurcation picture of the asymptotic behaviour of the solution as a function of $d$ is given in Figure \ref{fig:Hopfbifur1}. 

\noindent In this figure, as $d$ increases, we see the value of $F(Z^-)$ increasing
linearly with $d$ until the BCB when $F(Z^-) = 0$. The fixed point is then
immediately replaced by a periodic orbit with non-zero amplitude, which is in turn
destroyed at the second BCB when $F(Z^+) = 0$. 

\vspace{0.1in}

\noindent In a second figure we consider the bifurcation diagram, close to the rightmost bifurcation point, of the solutions as a function of $d$, when $\eta = 400, 800$ and $1600$.

\begin{figure}[htbp]
    \centering
	\includegraphics[width=0.7\linewidth]{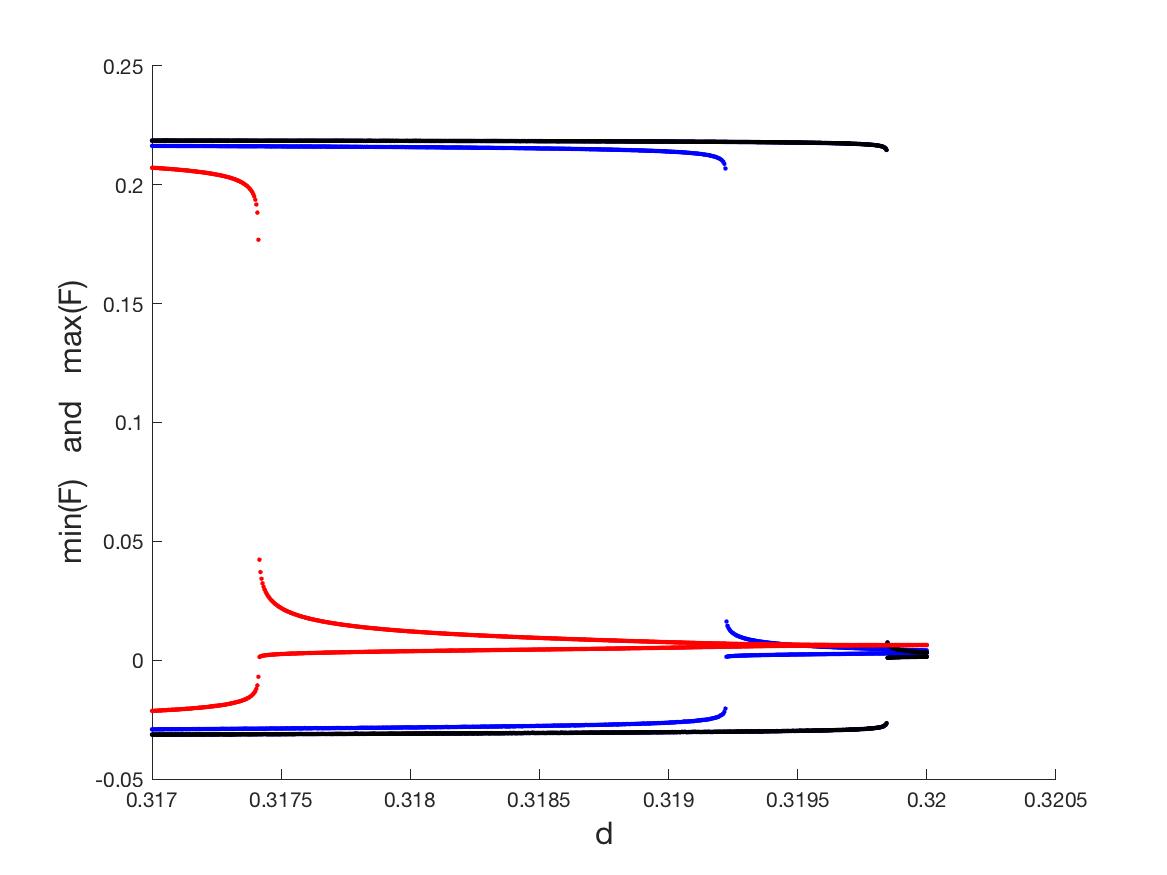}
	\caption{The bifurcation diagram of the solutions close to $d = 0.32$ when (from left to right)  $\eta = 400,800,1600$. These figure exhibit a smooth Hopf bifurcation from the fixed points followed by a rapid increase in size of the periodic orbit at a cyclic fold.}
	\label{fig:Hopfbifur}	
\end{figure}
\begin{figure}[htbp]
    \centering
	\includegraphics[width=0.7\linewidth]{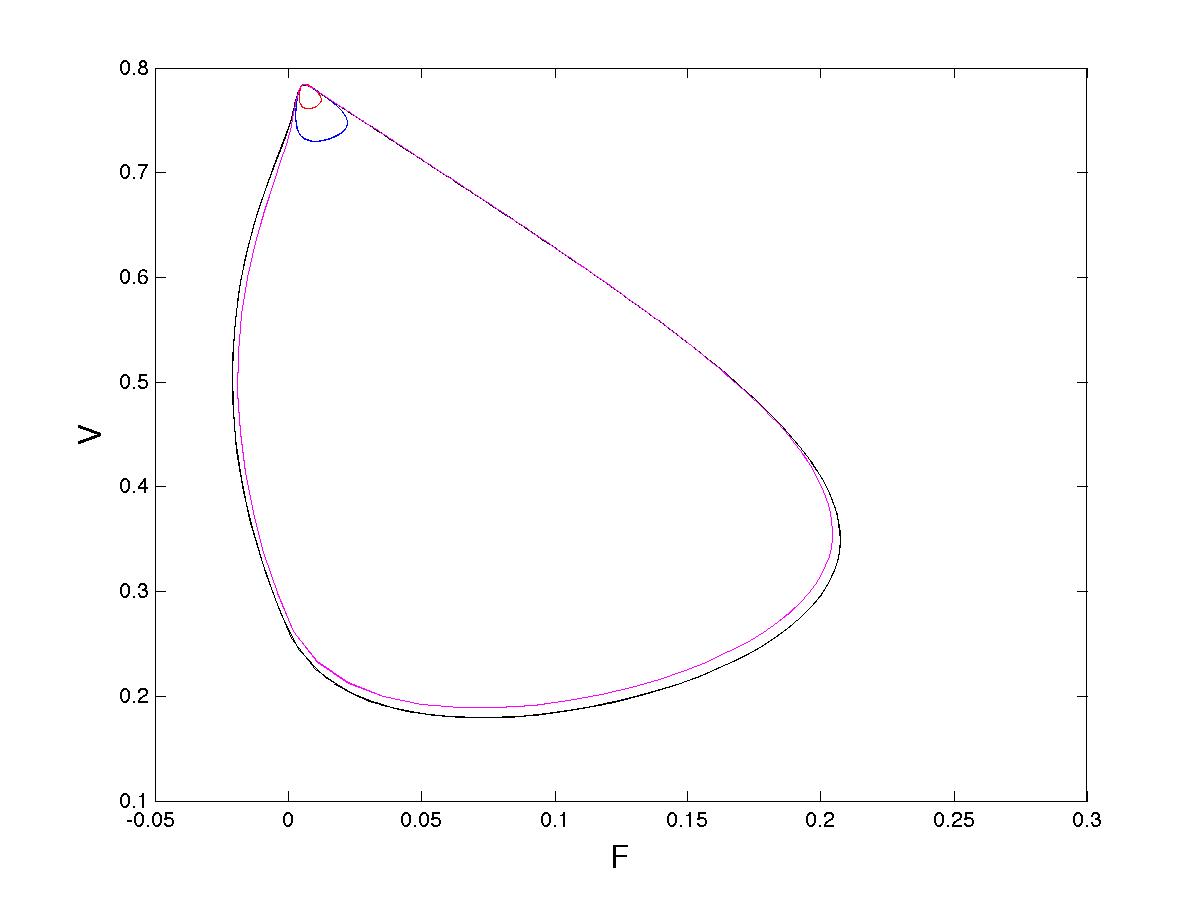}
	\caption{The $(F,V)$ phase plane of the solution as $d$ is increased from 0.317 (black)  to 0.318 (red) showing the rapid decrease in the size of the periodic orbit. }
	\label{FV}	
\end{figure}

 
\noindent In this figure we see that in all cases the fixed point $Z^+$ loses stability, as $d$ is decreased. The stability is lost to a periodic solution in what appears to be a super-critical Hopf bifurcation at a value of $d$ close to, but slightly smaller than, the BCB value of $d = 0.32$. Initially the periodic orbit is close to the fixed point. However, as $d$ is decreased further there appears to be a cyclic fold bifurcation at which point the periodic orbit expands rapidly in size, to approach the orbit to the discontinuous system. This behaviour was also observed in \cite{crucifix2012oscillators}. The sudden increase in the size of the periodic orbit as $d$ is decreased for the case of $\eta = 400$ is shown in Figure \ref{FV} in which we plot the trajectories in the $(F,V)$ phase plane for $d=0.317,d=0.31725,d=0.3175,d=0.318.$

\vspace{0.1in}

\noindent We see from these calculations that as $\eta$ increases the dynamics of the smooth system rapidly approximates the dynamics predicted by the analysis of the non-smooth system, with the (relatively simple) border-collision bifurcation in the non-smooth limit being replaced by a nearby, and more complex bifurcation structure in the smooth system. In the next section we will look at how the (apparently unique) periodic solution derived above changes when a periodic insolation forcing term is added.

\section{The analytic dynamics of the periodically forced PP04 model.}

\subsection{Overview} 

\noindent We now consider the solutions of the PP04 model when the insolation forcing has a single periodic mode. Clearly this form of forcing is unrealistic from a physical point of view. However, studying such systems allows us to gain insight into the more general case of quasi-periodic forcing, especially when one frequency is dominant in the insolation forcing. Indeed we will give evidence at the end of this paper that the behaviour of the quasi-periodically forced system is a simple perturbation of the periodically forced case.

\vspace{0.1in}

\noindent Accordingly, in this section we suppose that the insolation forcing has the form
\begin{equation}
    I_{65}(t) = \mu \sin(\omega t),
    \label{insol1}
\end{equation}
so that the period of the forcing is given by
\begin{equation}
    T = \frac{2 \pi}{\omega}.
    \label{insol2}
\end{equation}

\noindent In general we might expect to see the following types of solution behaviour for the PP04 system:

\begin{itemize}

\item [(a)] Synchronised periodic solutions (both stable and unstable)  of period $P = nT = 2\pi n/\omega$ with $n=1,2,3 ..$ which have precisely one glacial and one inter-glacial period (one glacial cycle) between repeats.
We define these to be $(1,n)$ periodic orbits. 
\item [(b)] Synchronised periodic solutions with several (for example $m$) different glacial cycles between repeats, of 'average' period $P = nT/m$ with $m=1,2,3, ..$. We define these to be $(m,n)$ periodic orbits. 
\item [(c)] Quasi-periodic solutions showing at least two distinct frequencies.
\item [(d)] Chaotic solutions.

\end{itemize}
 
\noindent In practice, for appropriate choices of parameters, we see all of these types of solutions, possibly
co-existing. Some of these  solutions arise through smooth bifurcations and others (as we have seen in the previous section) from non-smooth bifurcations as we vary parameters such as $\mu$ and $\omega$. 

\vspace{0.1in}

\noindent In this section we will consider those $(m.n)$ solutions which are initially small perturbations of the periodic orbit of the unforced system
constructed in the last section, given by taking the insolation forcing amplitude $\mu$ to be small.  The underlying periodic orbit has a well defined frequency $\omega^*$ and which (as we have shown) intersects the surface $\Sigma$ 
transversely. It follows from \cite{bernardo2008piecewise} that close to this orbit, the Poincar\'e return map defined by $P_S: {\mathbf X}(t) \to {\mathbf X}(t + 2\pi/\omega)$ is smooth. Thus we may apply the theory of {\em Arnold Tongues} \cite{pikovsky2003synchronization,simpson2010bifurcations}  to predict the existence of 'tongues' which are curves of (say) $(\mu,\omega)$ which define the boundaries of the existence regions for synchronised periodic solutions for the $(m,n)$ orbits when $\omega \approx n \omega^*/m$. 
Such tongues will be expected to have $|\omega -  n \omega^*/m|$ varying proportionally to $\mu^m$.

\subsection{Necessary algebraic conditions for the existence of the $(1,n)$ periodic solutions.}

\noindent It is relatively easy to construct algebraic conditions the satisfaction of which is necessary for the existence of the $(1,n)$  periodic orbits. Suppose that we have a periodic solution ${\mathbf X}(t)$ of period $P = 2n \pi/\omega$, and for which $F({\mathbf X}(t_i)) = 0$. For such a periodic orbit we will assume that exactly one glacial cycle exists for $t$ in the range $t \in [t_0,t_2 = t_0 + P].$ For this cycle we assume that the solution is {\em glacial} if $t \in [t_0,t_1]$, with $F(t) > 0$ and ${\mathbf X}(t) \equiv {\mathbf X}^+(t)$. Similarly the solution will be 
{\em inter-glacial} if $t \in [t_1,t_2]$, with $F(t) < 0$ and ${\mathbf X}(t) \equiv {\mathbf X}^-(t)$.   We define the set of points ${\mathbf X}_i = {\mathbf X}(t_i) $. It then follows that a periodic orbit must satisfy the conditions:
\begin{equation}
{
{\mathbf X}_0 = {\mathbf X}_2 \quad \mbox{and} \quad F(t_i) \equiv {\mathbf c}^T{\mathbf X}_i + d = 0 \quad i=0,1.
}
\label{c9}
\end{equation}
The differential equations satisfied by this system in the two regions are then given by:
\begin{equation}
{
\dot{{\mathbf X}}^{\pm} = L {\mathbf X^{\pm}} + {\mathbf b}^{\pm} + \mu \; {\mathbf e} \; \sin(\omega t). 
}
\label{c10}
\end{equation}
A particular integral of this system is given by
\begin{equation}
{\mathbf X}^{\pm}_{PI}(t) = {\mathbf Z}^{\pm} + \mu \; {\mathbf p} \; \cos(\omega t) + \mu \; {\mathbf q} \; \sin(\omega t) \equiv {\mathbf Z}^{\pm} + \mu \; {\mathbf r}(t).
\label{c11}
\end{equation}
Where (as before)
\begin{equation}
{\mathbf Z}^{\pm} = -L^{-1} {\mathbf b}^{\pm}, \quad {\mathbf p} = -(L^2 + \omega^2 I)^{-1} \; \omega \; {\mathbf e}, 
\quad {\mathbf q} = -(L^2 + \omega^2 I)^{-1} \; L \; {\mathbf e}.
\label{c12}
\end{equation}
We can then integrate the whole system to give
\begin{equation}
{
{\mathbf X}_1 = e^{L \Delta_1} \left( {\mathbf X}_0 - {\mathbf X}^{+}_{PI}(t_0) \right) +  {\mathbf X}^{+}_{PI}(t_1),
}
\label{c13}
\end{equation}
\noindent and similarly
\begin{equation}
{\mathbf X}_2 = e^{L \Delta_2} \left( {\mathbf X}_1 - {\mathbf X}^{-}_{PI}(t_1) \right) +  {\mathbf X}^{-}_{PI}(t_2).
\label{c14}
\end{equation}
Here we set
\begin{equation}
{
\Delta_1 = t_1 - t_0, \quad \Delta_2 = t_2 - t_1 = P - \Delta_1.
}
\label{c15}
\end{equation}

\noindent For a given period $P$, the problem of existence of a periodic solution is then reduced to finding the 5 unknowns comprising ${\mathbf X}_0$, together with the initial phase $t_0$  and the time of the transition between the glacial and inter-glacial states at $t_1$, so that the five equations in  (\ref{c9}) hold. (Alternatively we can take $t_0$ as given and find the period $P$ as part of the solution.) This nonlinear system may or may not have algebraic solutions, and we will consider this in the next sub-section. Furthermore the algebraic solutions, if they exist may or may not lead to physically relevant climate trajectories ${\mathbf X}(t)$, defined as follows: 

\vspace{0.2in}

\noindent {\bf Definition} We define a $(1,n)$ periodic solution to be {\em physical} if 
\begin{equation}
V(t) > 0, \; A(t) > 0, \; C(t) > 0 \quad \mbox{for all}  \quad t \in [t_0,t_2]
\label{c16a}
\end{equation}
and
\begin{equation}
F({\mathbf X}^+(t) ) > 0, \quad t_0 < t < t_1, \quad \mbox{and} \quad F({\mathbf X}^-(t)) < 0 \quad t_1 < t < t_2.
\label{c16}
\end{equation}

\vspace{0.2in}

\noindent Typically solutions lose algebraic existence through smooth (saddle-node or period-doubling) bifurcations, and lose physicality
through non-smooth (grazing) bifurcations, where we expect to see a dramatic change in the solution as indicated in Chapter 7 of \cite{bernardo2008piecewise}. We will return to this situation later. 

\subsection{Small $\mu$, synchronised, $(1,n)$ periodic solutions.}

\noindent We consider first the question of the existence of the $(1,n)$ periodic solutions. To do this we use perturbation theory, and look for explicit representations of the periodic solutions of the forced system, which are perturbations of the periodic solution of the unforced system when $\mu$ is {\em small}. We have shown in the previous section that if $\mu = 0$ (the unforced system) there is a periodic solution with one glacial cycle of frequency $\omega^*$ and period $P^* = 2\pi/\omega^*$ and which takes values
$X^* = (V^*,A^*,C^*)$ at the start of the glacial cycle. Extensive numerical experiments strongly indicate that this solution is also unique (up to the arbitrary starting time $t_0$) and attracting. Accordingly, for forcing at frequency $\omega$ with small $\mu$ we might expect to see a {\em synchronised} $(1,n)$ periodic orbit of period $P$ provided that $P \approx P^*$ so that
$$P = \frac{2 \pi n}{\omega} \approx P^* = \frac{2 \pi}{\omega^*}.$$
For the remainder of this section we will consider the periodic orbit that arises for the tabulated values of the parameters (so that for example $d=0.27$ for which we have found that 
$$\omega^* = 0.0429.$$
\noindent It follows that
\begin{equation}
{
\omega \approx n \;\omega^*.
}
\label{c17}
\end{equation}

\noindent We propose that for small $\mu$ that there is a range of $\omega$ values with $|\omega - n \; \omega^*| = {\cal O}(\mu)$ such that {\em two synchronised periodic solutions} exist within this range. Over the interval the phase $t_0$ of each such solution (defined as the phase of the forcing at the start of the glacial cycle) is well defined and varies over the whole range $[0,2\pi/\omega^*]$. Both solutions are perturbations of the unforced solution which is given when $\mu = 0$, and which has an arbitrary phase. Hence, both solutions are physical provided that $\mu$ is sufficiently small and the parameter $d$ is not close to the value at which a border collision occurs for the free system. The boundaries of the regions of
existence of both solutions are determined by the existence of saddle-node bifurcations
where the solutions coalesce.

\vspace{0.1in}

\noindent This result is an immediate consequence of the following:

\vspace{0.2in}

\noindent {\bf Lemma 6.1} {\em {\em (i)} For each $n$, if $\mu$ is {\em small} then there is a set of solutions $(\omega, {\mathbf x})$ to the algebraic system, which is parameterised by $t_0$.

\vspace{0.1in}

\noindent {\em (ii)} if $\mu$ is small then the curves $(\omega, V(\omega))$ of the $(1,n)$ orbits form ellipses which have the point $(n \omega^*,V^*)$ at the centre.

\vspace{0.1in}

\noindent {\em (iii)} The size (for example the semi-major axis) of the ellipses is (for sufficiently small $\mu$) directly 
proportional to $\mu$.

\vspace{0.1in}

\noindent {\em (iv)} As we go once around the small elliptical curves, the phase  $t_0$ increases by a factor of $2 \pi/(n \omega^*).$
}

\vspace{0.2in}

\noindent {\bf Corollary 6.2} {\em If $\mu \ll 1$ then for each value of $n$, the synchronised periodic solutions of the PP04 model exhibit saddle node (SN)  bifurcations at points $(\omega_{1,n},\omega_{2,n})$. At which points they change to quasi-periodic orbits. 
Synchronised periodic solutions exist in the interval $\omega \in (\omega_{1,n},\omega_{2,n})$. We have that
$$ |\omega_{i,n} - n\omega^*| = {\cal O}(\mu), \quad i=1,2.$$}

\vspace{0.2in}

\begin{figure}[htbp]
  \centering
  \includegraphics[width=0.7\textwidth]{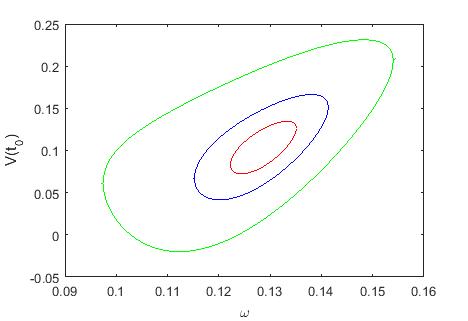}
\caption{The variation of $V(t_0)$ with $\omega$ for $\mu = 0.05$ (red), $\mu = 0.1$ (blue) and $\mu = 0.2$ (green) for $n = 3$ showing the (elliptical) curve of the solutions and the two saddle-node bifurcation points. It is clear that the size of the closed curve increases in proportion to $\mu$, and that it has a true elliptical shape for the smaller values of $\mu$.}
\label{fig:2}
\end{figure}

\begin{figure}[htbp]
  \centering
  \includegraphics[width=0.75\textwidth]{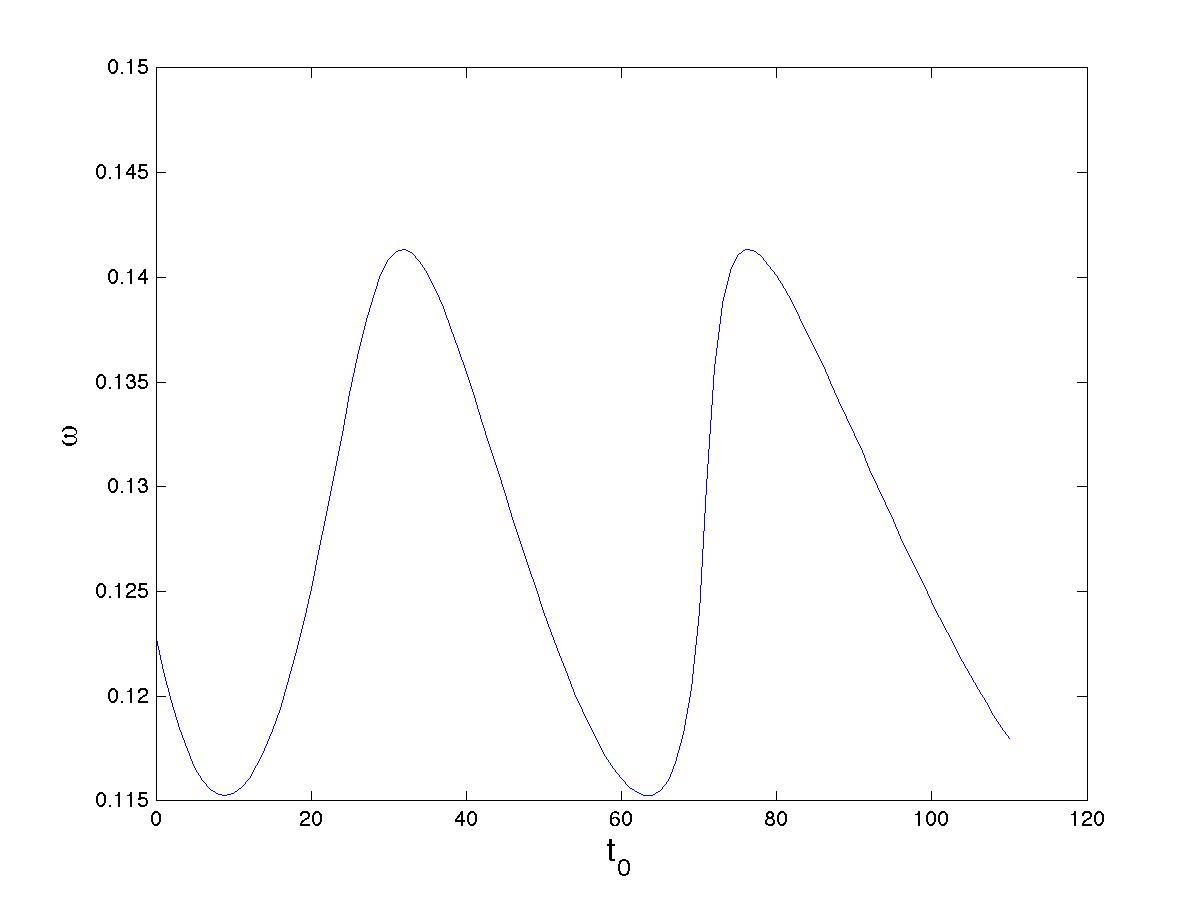}
\caption{The variation of $\omega$ with $t_0$ for $\mu = 0.1$ ($n = 3$) showing that $t_0$ increases by approximately $2 \pi/(3 \omega^*) = 49.0874$ over this cycle.}
\label{fig:2a}
\end{figure}

\begin{figure}[htbp]
  \centering
  \includegraphics[width=0.8\textwidth]{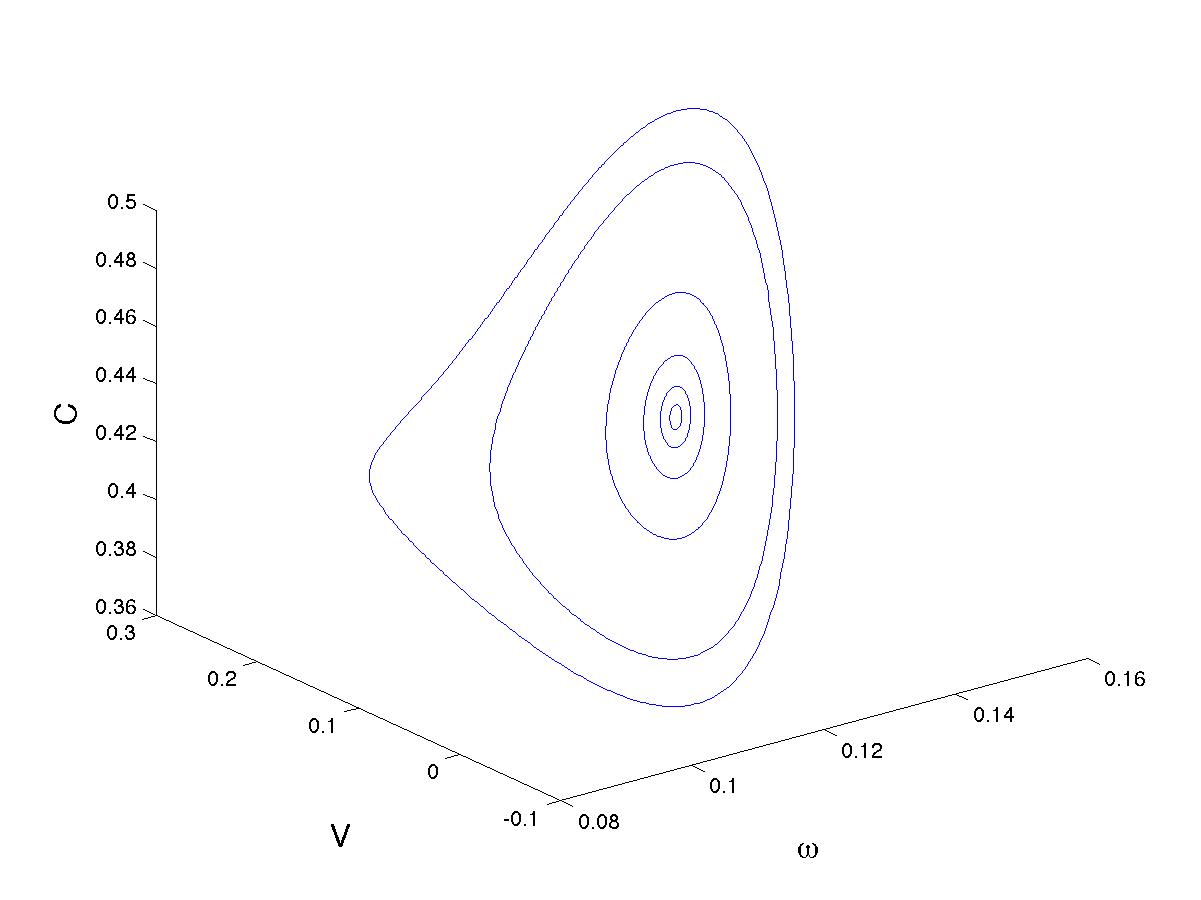}
\caption{The variation of $V(t_0)$ and $C(t_0)$ with $\omega$ when $n=3$ with $\mu$ increasing
from 0.01 to 0.24.}
\label{p5a}
\end{figure}

\noindent We illustrate the conclusions of Lemma 6.1 and Corollary 6.2 in Figure \ref{fig:2}, in which we plot the ellipses corresponding to the solutions $(\omega,V(t_0))$ for the parameter values $\mu=0.05, 0.1,0.2$ and $n = 3$. 
\noindent In Figure \ref{fig:2a} we plot $\omega$ as a function of $t_0$ for the case of the $(1,3)$ orbits which arise when $\mu = 0.1$, $n = 3$  as we go around the ellipse. 
For the problem considered we have $n \omega^* = 0.128$ We can see that $t_0$ increases by approximately $2\pi/(n \omega^*) = 49.0874$ over this cycle.
In Figure \ref{p5a} we plot the closed curves $(\omega(t_0),V(t_0),C(t_0))$ together for $\mu$ increasing from $0.01$ to $0.24$. We see that the closed elliptic curves exist for these parameter values in this extended space.
\vspace{0.1in}

\noindent {\em Proof} If $\mu = 0$ we have a periodic solution $X^*(t)$  of the autonomous system.
This can have an {\em arbitrary time} $t_0^*$ at the start of the glacial cycle for which $F(X^*(t_0^*)) = 0$.
We have a well defined set of time differences $\Delta_1^*$ and $\Delta_2^*$ so that the glacial period is in the interval $[t_0^*,t_0^* + \Delta_1^*]$ and the inter-glacial period in the time interval $[t_0^* + \Delta_1^*, t_0^* + \Delta_1^* + \Delta_2^*]$ with $\Delta_1^* + \Delta_2^* = 2\pi/\omega^*$.  In the forced case, with $\mu > 0$, the system will exhibit {\em phase-locking}, so that we expect to see a well defined start time $t_0$ in this case. Finding the form of the solution and the frequency $\omega$, in terms of $\mu$ and $t_0$ will be part of the solution process.
Suppose that we consider a periodic orbit ${\mathbf X}(t)$ of period $T = 2\pi n/\omega \approx 2 \pi/\omega^*$,
so that $\omega \approx n \omega^*$, with a glacial state in the interval $t \in (t_0,t_1)$ and an inter-glacial
state for $t \in (t_1,t_2)$ so that 
$$t_2 = t_0 + \frac{2 \pi n}{\omega}.$$
We define ${\mathbf X}_i = {\mathbf X}(t_i)$. Thus to have a periodic solution we must satisfy the following three conditions
\begin{equation}
{
{\mathbf X}_0 = {\mathbf X}_2 \quad \mbox{and} \quad F({\mathbf X}_i ) = {\mathbf c}^T{\mathbf X}_i + d = 0. 
}
\label{c5}
\end{equation}

\noindent We now consider a solution which is a perturbation of the unforced case so that to order ${\cal O}(\mu)$ we have 
$$\omega = n \omega^* + \mu\alpha, \quad  \Delta_1= \Delta^* + \mu \delta, \quad {\mathbf X}_0 = {\mathbf X}_0^* + \mu {\mathbf x}.$$

\vspace{0.1in}

\noindent To find the leading order form of the perturbed solution we then determine
$$\alpha, \delta, {\mathbf x}$$
as functions of the phase $t_0$. 

\vspace{0.1in}

\noindent It follows immediately that
$$\Delta_2 = \frac{2 \pi n}{\omega} - \Delta_1 = \Delta_2^* - \mu \delta - \mu \frac{2 \pi \alpha }{(\omega^*)^2}.$$
It then follows from (\ref{c11}) that 
$${\mathbf X}_1 = e^{L(\Delta_1 + \mu \delta)}({\mathbf X^*}_0 + \mu {\mathbf x} - {\mathbf Z}^{+} + \mu {\mathbf r}(t_0) ) +{\mathbf Z}^+ + \mu {\mathbf r}(t_1).$$
Hence, after some manipulation
$${\mathbf X}_1 = {\mathbf X}^*_1 + \mu\left ( e^{L \Delta_1^*} \left(\delta L {\mathbf X}^*_0 + {\mathbf x} - {\mathbf r}(t_0)\right)  + {\mathbf r}(t_1) \right) + {\cal O}(\mu^2).$$
Thus 
$${\mathbf X}_1 = {\mathbf X}_1^* + \mu {\mathbf y} + {\cal O}(\mu^2),$$
where
$${\mathbf y} = e^{L \Delta_1^*} \left(\delta L {\mathbf X}^*_0 + {\mathbf x} - {\mathbf r}(t_0)\right)  + {\mathbf r}(t_1) $$
Similarly, 
$${\mathbf X}_2 = {\mathbf X}_2^* + \mu {\mathbf z} + {\cal O}(\mu^2),$$
where
$${\mathbf z} = e^{L \Delta_2^*} \left(-(\delta - 2\pi \alpha/(\omega^*)^2)) L {\mathbf X}^*_1 + {\mathbf y} - {\mathbf r}(t_1)\right)  + {\mathbf r}(t_0).$$
The conditions $F(X_i) = 0$ are then satisfied provided that
$${\mathbf x} = {\mathbf z} \quad \mbox{and} \quad {\mathbf c}^T{\mathbf x} = 0, \quad {\mathbf c}^T{\mathbf y} = 0.$$
\noindent We define the linear operators $A_1$ and $A_2$ by:
$$e^{L \Delta_1} = A_1, \quad e^{L \Delta_2} = A_2.$$
It follows that
$${\mathbf y} = A_1(\delta L X_0^* + {\mathbf x} - {\mathbf r}(t_0)) + {\mathbf r}(t_1), \quad z = A_2(-(\delta + \gamma \alpha)L X_1 + {\mathbf y} - {\mathbf r}(t_1)) + {\mathbf r}(t_0).$$
Hence, as ${\mathbf z} = {\mathbf x}$ we have
\begin{equation}
{
{\mathbf x} = A_2(-(\delta + \gamma \alpha)L {\mathbf X}_1 + A_1(\delta L X_0 + {\mathbf x} - {\mathbf r}(t_0)) + r(t_0),
}
\label{c111}
\end{equation}
\begin{equation}
{
{\mathbf c}^T{\mathbf x} = 0, \quad {\mathbf c}^T{\mathbf y} = 0.
}
\label{c112}
\end{equation}

\noindent Now we look at the structure of the equations (\ref{c111},\ref{c112}). We note that to leading order, as $t_1 = t_0 + \Delta_1^*$ that
there are vectors $p_0,q_0,p_1,q_1$ so that 
$${\mathbf r}(t_0) = {\mathbf p}_0 \cos(n \; \omega^* t_0) + {\mathbf q}_0 \sin(n \; \omega^* t_0), \quad {\mathbf r}(t_1) =
{\mathbf p}_1 \cos(n \; \omega^* t_0) + {\mathbf q}_1 \sin(n \; \omega^* t_0).$$
It follows that there is a linear operator $M$ and vectors ${\mathbf a}$ and ${\mathbf b}$ so that (\ref{c111},\ref{c112}) can be put into the form
\begin{equation}
{
M \left[ 
\begin{array}{c} 
{\mathbf x} \\
\delta \\
\alpha \\
\end{array}
\right]
+ {\mathbf a} \cos(n \; \omega^* t_0) + {\mathbf b} \sin(n \; \omega^* t_0) = {\mathbf 0}.
}
\label{c113}
\end{equation}
\noindent The linear operator $M$ and the vectors ${\mathbf a},{\mathbf b}$ can all be constructed explicitly. 
We will make the assumption that $M$ is invertible. Numerical evidence clearly indicates that this is always the case. Under this assumption, for each value of $t_0$ the system
(\ref{c113}) can be solved uniquely to give the values of ${\mathbf x}, \delta$ and $\alpha$.
These then take the form
$$\left[ 
\begin{array}{c} 
{\mathbf x} \\
\delta \\
\alpha \\
\end{array}
\right]
=
{\mathbf f} \cos(n \; \omega^* t_0) + {\mathbf g} \sin(n \; \omega^* t_0)
$$
for appropriate (constant) vectors ${\mathbf f}$ and ${\mathbf g}$.
 In particular there will be unique values $f_5$ and $g_5$ so that
$$\alpha = f_5 \cos(n \; \omega^* t_0) + g_5 \sin(n \; \omega^* t_0).$$
As $t_0$ varies over the whole range of $[0,2\pi/(n \; \omega^*)]$ so $\alpha$ will range over
the interval.
\begin{equation}
\alpha \in \left[ -\sqrt{f_5^2 + g_5^2},\sqrt{f_5^2+g_5^2} \right].
\label{c114}
\end{equation}
This interval sets the limits of existence of the solutions of (\ref{c111},\ref{c112}) and hence the width of the tongues over which we will see synchronised periodic solutions.
Clearly if $W_{\alpha} = \sqrt{f_5^2 + g_5^2}$ then there is a phase $\phi_{\alpha}$ so that
\begin{equation}
\alpha = W_{\alpha} \cos(n \; \omega^*t_0 - \phi_{\alpha}).
\label{p1}
\end{equation}
\noindent If we set $[v a c ] = {\mathbf x}^T$, then an identical argument  implies that there are
amplitudes $W_V, W_A$ and $W_C$, and phases $\phi_V,\phi_A$ and $\phi_C$ so that
\begin{equation}
v = W_V  \cos(n \; \omega^*t_0 - \phi_V), \quad  a = W_A \cos(n \; \omega^*t_0 - \phi_A), \quad c  = W_C \cos(\omega^*t_0 - \phi_c).
\label{p2}
\end{equation}
It follows immediately that the curves $(\alpha,v), (\alpha,a)$ and $(\alpha,c)$ are all ellipses centred on the origin.    \qed

\subsubsection{The nature of the small $\mu$ solution ellipses.}

\begin{figure}[htbp]
  \centering
  \includegraphics[width=0.7\textwidth]{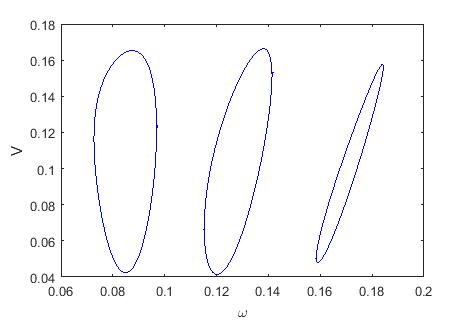}
\caption{The variation of $V(t_0)$ with $\omega$ for $\mu = 0.1$ for (from left to right) the
three cases of $n = 2,3,4$.}
\label{p3}
\end{figure}

\noindent The values of the coefficients of the vectors ${\mathbf f}$ and ${\mathbf g}$ are determined explicitly by the calculation above, but are hard to estimate from this.  However, the basic calculation of the $(1,n)$ periodic orbits is identical for all values
of $n = 1,2,3,4, ..$ although the precise values of the coefficients will change in each case. In particular, for small $\mu$ we expect
to see small ellipses in each case, the size of which is directly proportional to $\mu$. In Figure \ref{p3}
we plot the resulting ellipses when $\mu = 0.1$ for $n= 1, 2,3,4$. These ellipses are computed by numerically solving the nonlinear equations for $V,A,C,\omega$ and $t_1$.
As these solutions are parameterised by the initial time $t_0$, it is convenient in this calculation to use $t_0$ as the path following variable.  Each of these ellipses are centred on the values
of $\omega_1 = 0.0426$ $\omega_2 = 0.0853$, $\omega_3 = 0.128$ and $ \omega_4 = 0.1706$ respectively, corresponding to the integer multiples of the frequency of the periodic solution to the unforced problem. 
We note that as $n$ increases the size of the minor axis of the ellipse appears to decrease, although the size of the major axis stays approximately constant.
The value of $t_0$  varies over the interval $[0,2\pi/(n \; \omega^*)]$ as we travel 
around the ellipse. In particular, it follows from (\ref{c114}) that the value of $t_0$ changes by $\pi/(n \; \omega^*)$ between the 
two saddle node bifurcation points. This is of interest as it demonstrates that the phase of the response $(V,A,C)$
to the insolation forcing, whilst locked to it for a particular periodic orbit, differs from it. This phenomenon has
been observed in the record of the ice ages, in which the Milankovitch cycles are not always seen to 
be in phase with the cooling and warming periods. 

\vspace{0.1in}

\begin{figure}[htbp]
  \centering
  \includegraphics[width=0.6\textwidth]{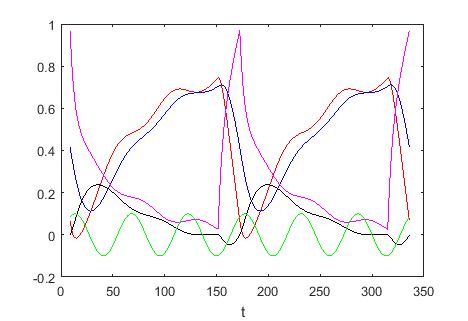} \quad
\includegraphics[width=0.6\textwidth]{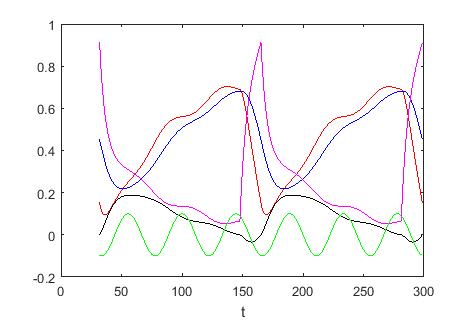}
\includegraphics[width=0.6\textwidth]{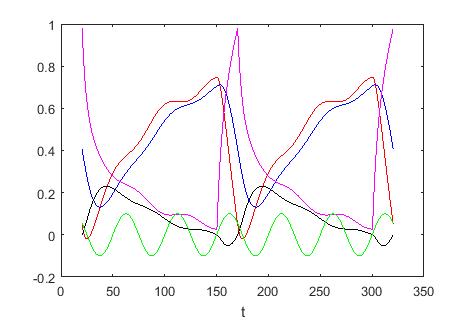} 
\caption{Two cycles of the periodic solutions when $n = 3$. In this plot we see $V$ (blue), $A$ (red), $C$ (maroon),
$F$ (black), and the insolation (green). We have Left: $t_0 = 8.79$, Right: $t_0 = 31.882$ and Bottom: $t_0 = 20.3156$.}
\label{p35}
\end{figure}

\noindent As an example we take the case $n = 3$ and $\mu = 0.1$. A numerical calculation in this case shows that
solutions exist for $\omega \in [0.1152,0.1413]$, $ 2\pi/\omega \in [ 44.46, 54.52],$ $T = 6\pi/\omega = [ 133.37, 163.57 ]$ and $t_0 \in [8.7965,31.882].$
In Figure \ref{p35} we plot two cycles of the resulting periodic orbits for the three cases $t_0= 8.7965$, $t_0 =  20.3156$
and $t_0 = 31.882$ representing the left and right limits and the middle of the range of values for which we see a solution.

\subsubsection{The regions of existence of the $(1,n)$ orbits for small $\mu$.}

\noindent The previous analysis has shown that $\omega^* = 0.0429.$
Further numerical studies lead to the following approximations for small $\mu$ of the regions of existence
of the $(1,n)$ orbits.
\begin{equation}
\begin{array}{l l}

\omega_{1,1} = \;  \omega^*-0.0905\mu    & \; \omega_{2,1} =  \omega^* +  0.0905\mu    \\
\omega_{1,2} = 2\omega^*-0.1228\mu    &  \omega_{2,2} = 2\omega^* + 0.1228\mu    \\
\omega_{1,3} = 3\omega^*-0.1348\mu    &  \omega_{2,3} = 3\omega^* + 0.1348\mu     \\
\omega_{1,4} = 4\omega^*-0.1296\mu    &  \omega_{2,4} = 4\omega^* + 0.1296\mu     \\

\end{array}
\end{equation}

\begin{figure}[htbp]
  \centering
  \includegraphics[width=0.6\textwidth]{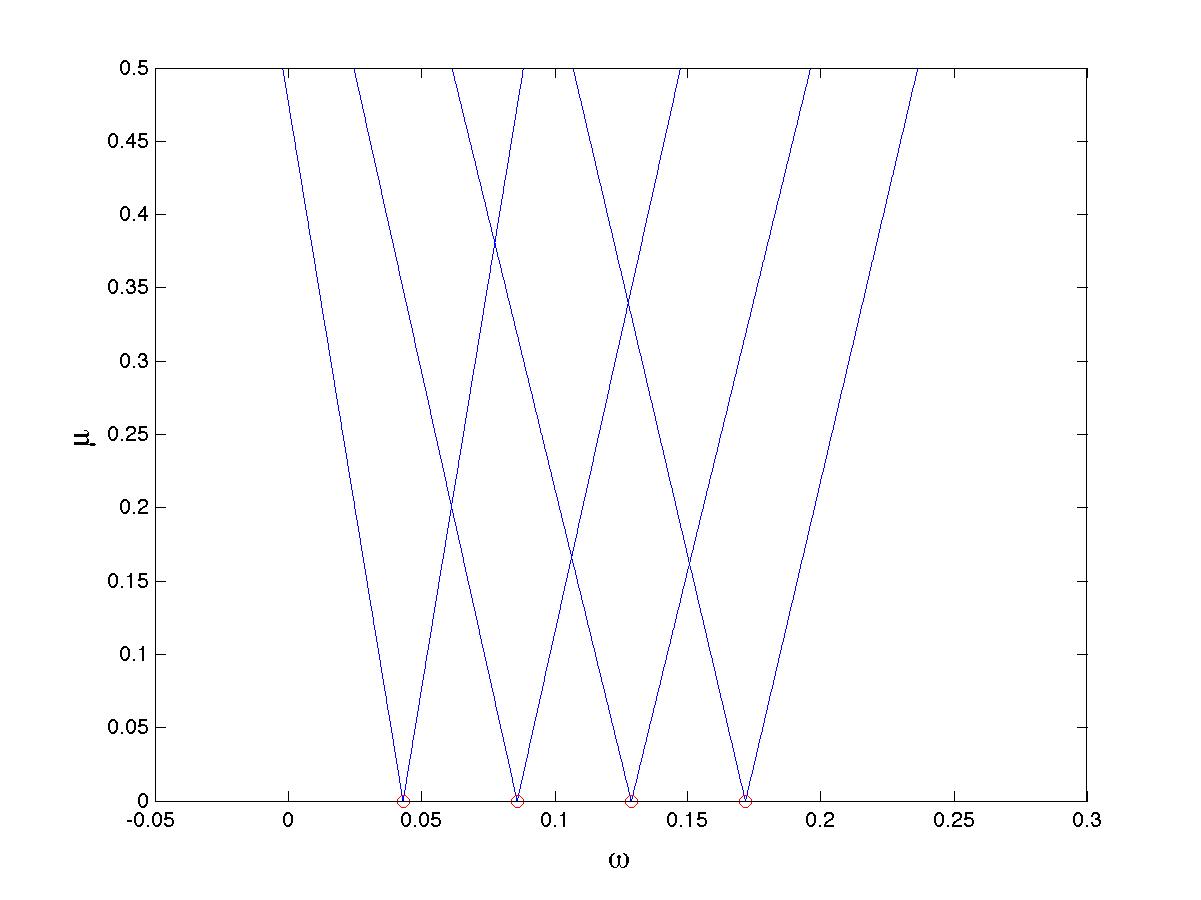}
\caption{The existence regions for the periodic solutions of the linearised problem.}
\label{p33}
\end{figure}

\noindent In Figure \ref{p33} we give the graph of the regions of existence of the periodic solutions for
$n=1,2,3,4$ for the linearised problem as described above. We can see that the regions of existence for this linear problem 
start to overlap if $\mu > 0.15$. For $\mu > 0.15$ we will expect to see (as we in fact do see) the co-existence
of periodic solutions with different values of $n$ and hence of different periods $T = 2n \pi/\omega$. In fact,
as we shall see, the original (nonlinear) problem has rather larger regions of overlap of the existence regions.

\subsection{Larger values of $\mu$.}

\noindent The above calculation has given only a small $\mu$ analysis, showing that for small $\mu$ the width of the existence tongues and the associated ellipses of the solutions of the algebraic system (\ref{c9}) increases in direct proportion to $\mu$ for all values of $n$. Similar results for other systems are given in \cite{schilderandpeckham2007}.

\vspace{0.1in}

\noindent For larger values of $\mu$ nonlinear effects become important, and the ellipses determined above will form part of the complex surface of the solutions of (\ref{c9}). In this scenario, as we shall see, the ellipses calculated above become distorted, and then can break up and expand as they coalesce with other curves of solutions. However, we note that (unlike the small $\mu$ case) many of the solutions of the algebraic equations (\ref{c9}) for larger values of $\mu$  will not represent physical climate states. For example this may be a trajectory starting from an initial state at $t0$ and ${\mathbf X}_0$ calculated as a solution of (\ref{c9}) on the assumption that it remains in $S^+$ for $t_0 < t < t_1$ which may, in fact, cross $\Sigma$ at a time $t_0 < t^* < t_1$. 

\vspace{0.1in}

\begin{figure}[htbp]
  \centering
  \includegraphics[width=\textwidth]{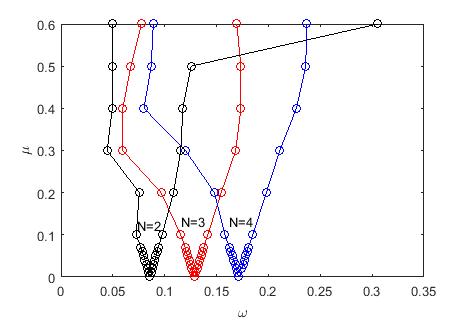}
  \caption{ A set of graphs showing the regions of existence of the $(1,n)$ orbits when both $ \mu$ and $\omega$ are varied. The red case $n = 2,3,4$ are illustrated.}
\label{fig:35}
\end{figure}

\noindent In Figure \ref{fig:35} we plot computed regions of existence of the $(1,n)$ periodic solutions. These regions are determined by first fixing the value of $\mu$ and solving the full algebraic system numerically for a set of values of $\omega = n\omega^* \pm \delta$ increasing $\delta$ from 0. The calculation was done using the Matlab solver {\tt fsolve} with an initial guess given by ${\mathbf X}_0 = (5,5,5,5,5,5,100,200)$. We then plot the first values of $\omega_{new}$ against $\mu$ for which the algebraic solver breaks down. 
As can be seen, the regions of existence are linear (as predicted) for small values of $\mu$. They then expand significantly as $\mu$ increases. This is due to a coaelescence of the  small $\mu$ ellipses with other solution curves as described above.

\vspace{0.1in}

\noindent We note that the physically interesting case of $(\mu,\omega) = (0.467,0.1532)$ (see \ref{muandomega}) lies in the region where there is only a $(1,3)$ periodic solution, and we will return to this observation later.

\vspace{0.1in}

\begin{figure}[htbp]
  \centering
  \includegraphics[width=\textwidth]{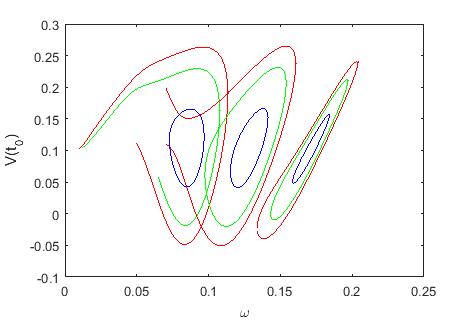}
\caption{The variation of $V(t_0)$ with $\omega$ when $n=2,3,4$ (from left to right respectively) for $\mu = 0.1$ (blue), $\mu = 0.2$ (green) and
$\mu = 0.25$ (red) showing the break up of the elliptical curves when they coalesce with other solution curves as $\mu$ increases. Here $n=2,3,4$}
\label{p5}
\end{figure}

\noindent In Figure \ref{p5} we see the set of elliptical curves for the cases of $n=2,3,4$ taking larger values of $\mu$ than before.  For $n=2,3$ we see a coalesecence of the ellipses with other solution curves at $\mu = 0.25$. In the case of $n = 4$ the coalesecence occurs for a larger value of $\mu$.  Indeed, we observe in general, that the coalescence of the ellipses with other solution curves occurs for smaller values of $\mu$ as $n$ decreases. We note further that if $\mu = 0.25$ then (as expected from the linear analysis) the regions of existence of the $n = 2$ and $n = 3$ periodic orbits overlap. As a consequence we might expect to see both $n = 2$ and $n = 3$ orbits in this case, with related domains of attraction for the initial data.

\begin{figure}[htbp]
  \centering
  \includegraphics[width=0.8\textwidth]{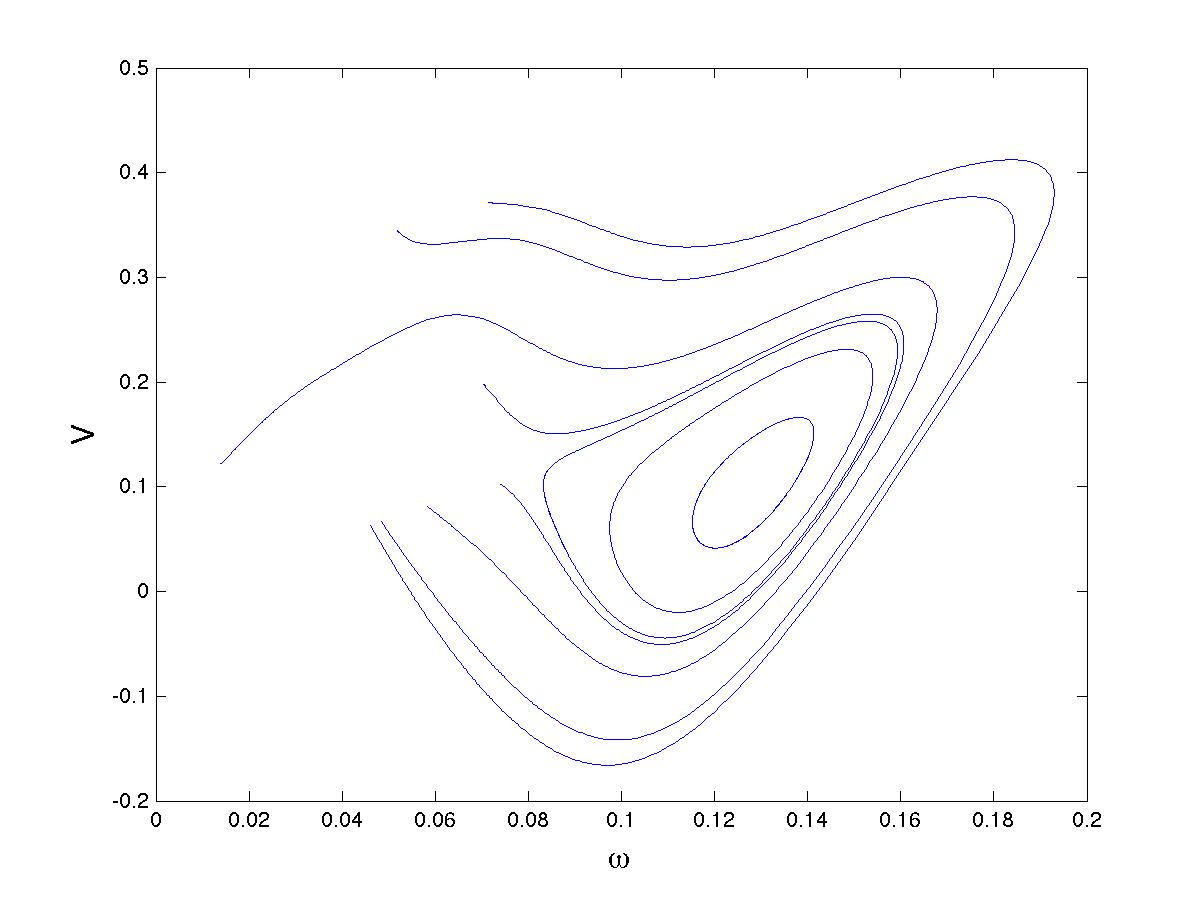}
\caption{The variation of $V(t_0)$ with $\omega$ for the $(1,3)$ orbit when $\mu$ increases from $\mu = 0.1$ to $\mu = 0.467$ showing the break up of the elliptical curve at $\mu = 0.25$}
\label{fig:3}
\end{figure}

\begin{figure}[htbp]
  \centering
  \includegraphics[width=0.6\textwidth]{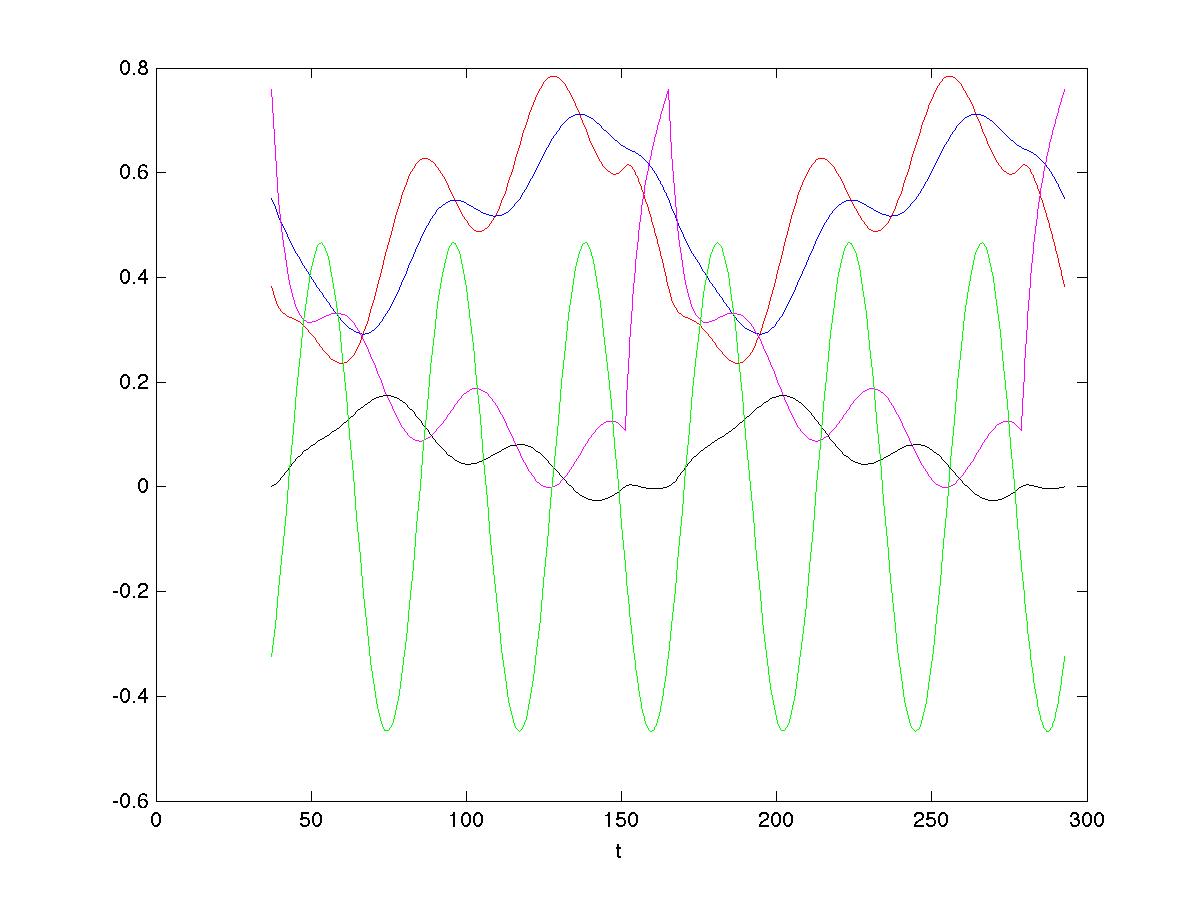}
\includegraphics[width=0.6\textwidth]{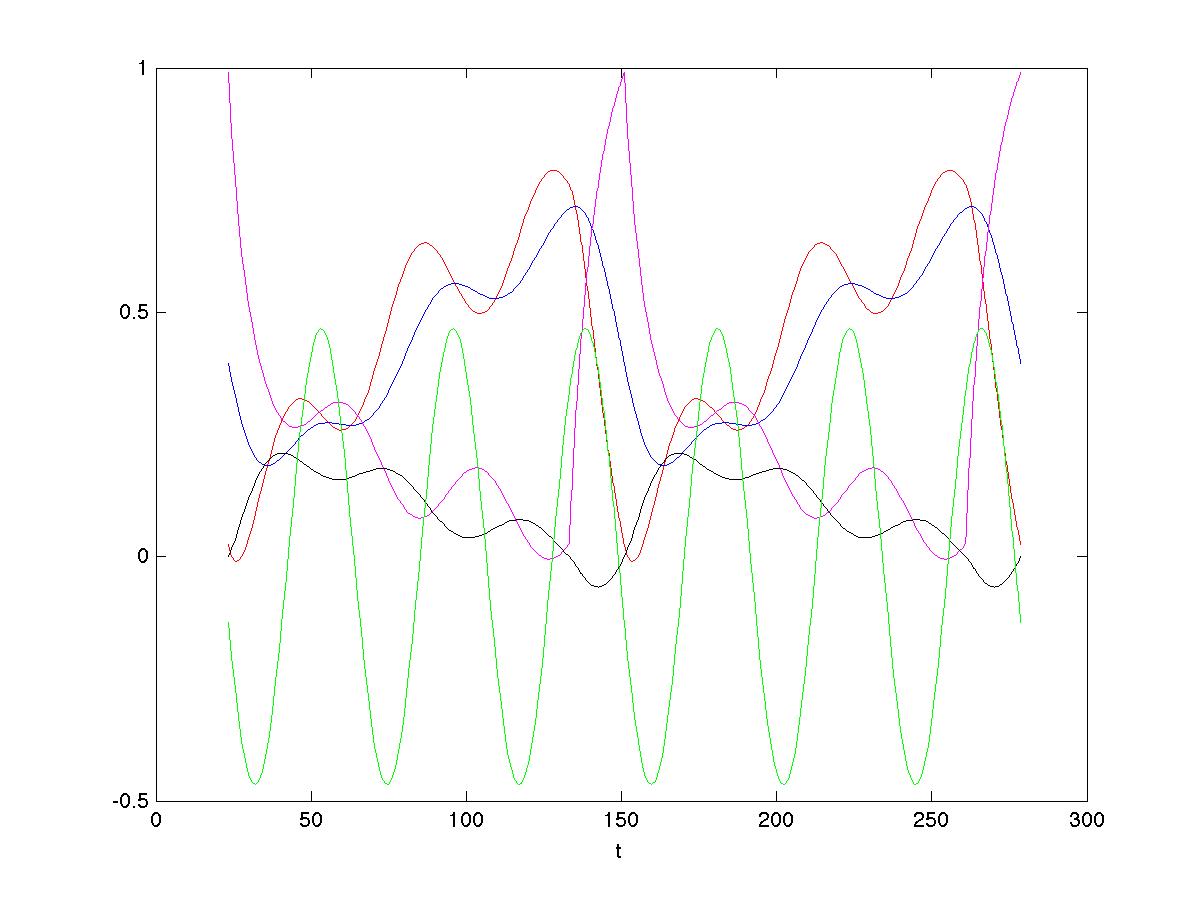}
\caption{The solutions on the upper branch ($S_1$ top) and the lower branch ($S_2$ bottom). In these graphs we plot (as functions of time) $V$ (red), $A$ (blue), $C$ (purple), $F$ (black) and the insolation forcing (green). The solution $S_1$ is not physical as $F(t)$ changes sign within the glacial cycle. }
\label{fig:3d}
\end{figure}

\noindent If we take the larger, and physically relevant, value of $\mu = 0.467$ then we see a more complicated curve, and the range of existence of the solutions in this case is more difficult to predict. In Figure \ref{fig:3} we show the curves of the $(1,3)$ orbit for a range of values of $\mu$ increasing from $\mu = 0.1$ to $\mu = 0.467$. In this figure we observe solution existence ellipses for $\mu < 0.245$. These then break up at around $\mu = 0.25$ and enlarge as $\mu$ increases. When $\mu = 0.467$ we see that the maximum value of $\omega = 0.194$. There is no minimum value shown on this graph, however we note as described earlier, that not all of the solutions of the algebraic system (\ref{c9}) are physical over this range. 
To see this we take $\mu = 0.467$ and consider the physically relevant value of $\omega = 0.1532$. For this value of $\omega$ it is apparent from Figure \ref{fig:3} that there are (at least) two solutions, $S_{1,2} \equiv [V(t_0),A(t_0),C(t_0),t_0,\Delta]$ to the algebraic equations, with $S_1$ on the upper side of the curve of solutions and $S_2$ on the lower. A careful calculation shows that these solutions are given by
$$S_1 = [0.393,0.5541, 0.7508, 35.726,  113.8284],$$
and
$$ S_2 = [0.059, 0.4113, 0.9674, 23.2854, 106.0578].$$
The corresponding functions $(V(t),A(t),C(t),F(t))$ are plotted in Figure \ref{fig:3d}, along with the insolation forcing. It is clear from this figure that only the solution $S_2$ can be physical. This is because 
when we consider the solution $S_1$ we can see from the graph that the function $F(t)$ does not
keep a constant sign during either the glacial or the inter-glacial cycles. 
\vspace{0.1in}

\begin{figure}[htbp]
  \centering
  \includegraphics[width=0.45\textwidth]{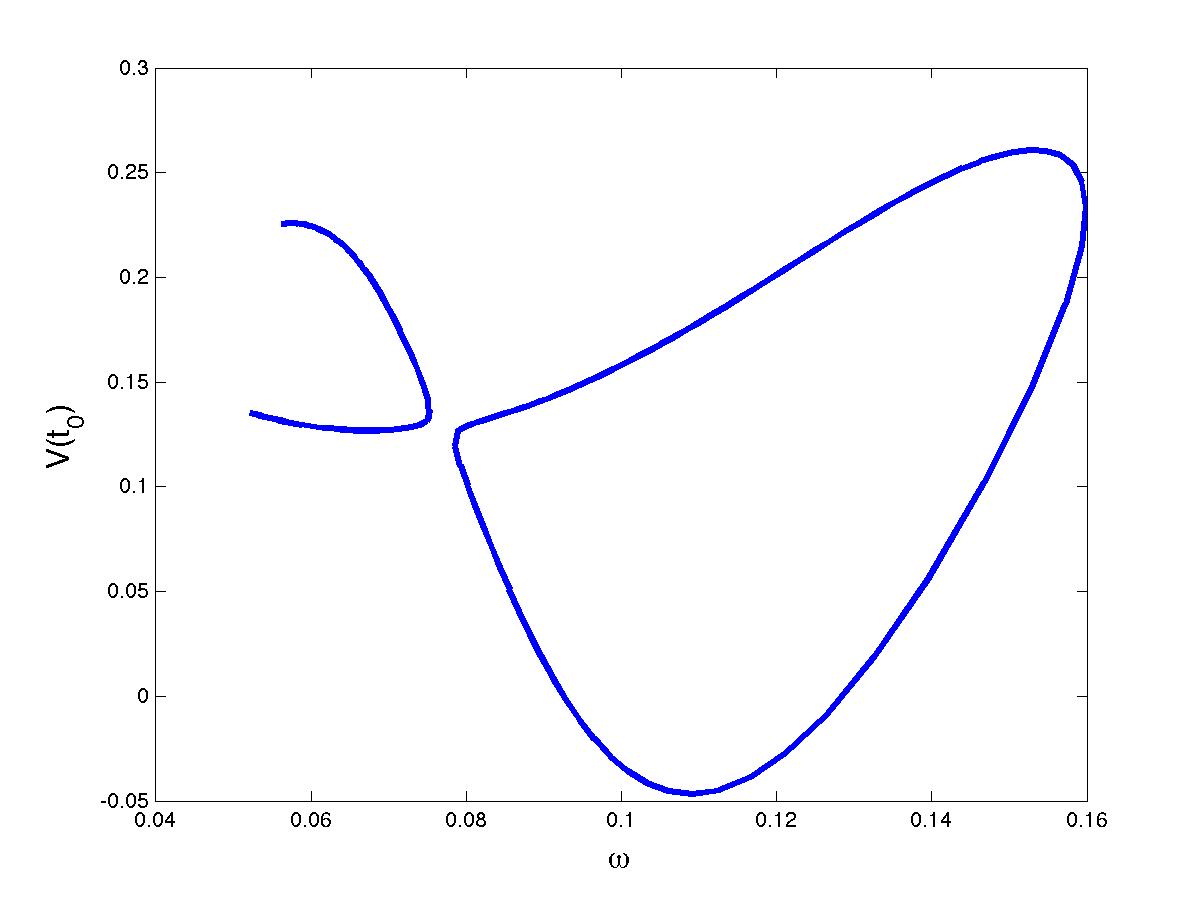}
  \includegraphics[width=0.45\textwidth]{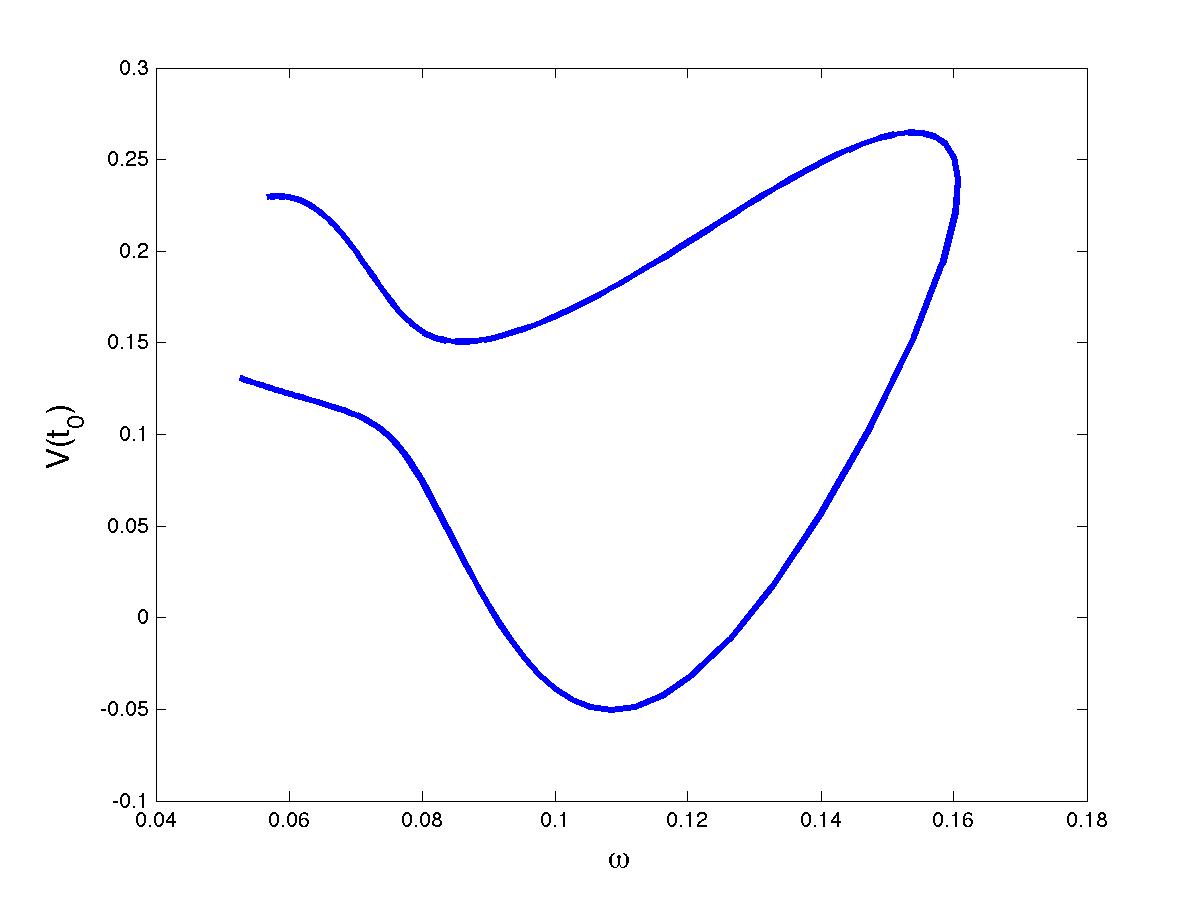}
\caption{The variation of $V(t_0)$ with $\omega$ for the $(1,3)$ orbit for $\mu = 0.244$ (left) and  $\mu = 0.25$ (right). Here we can see the coalescence of two solution curves when $\mu=0.244$ leading to an expansion of the solution ellipse when $\mu=0.25.$}
\label{fig:3aa}
\end{figure}

\begin{figure}[htbp]
\centering
\includegraphics[width=0.5\textwidth]{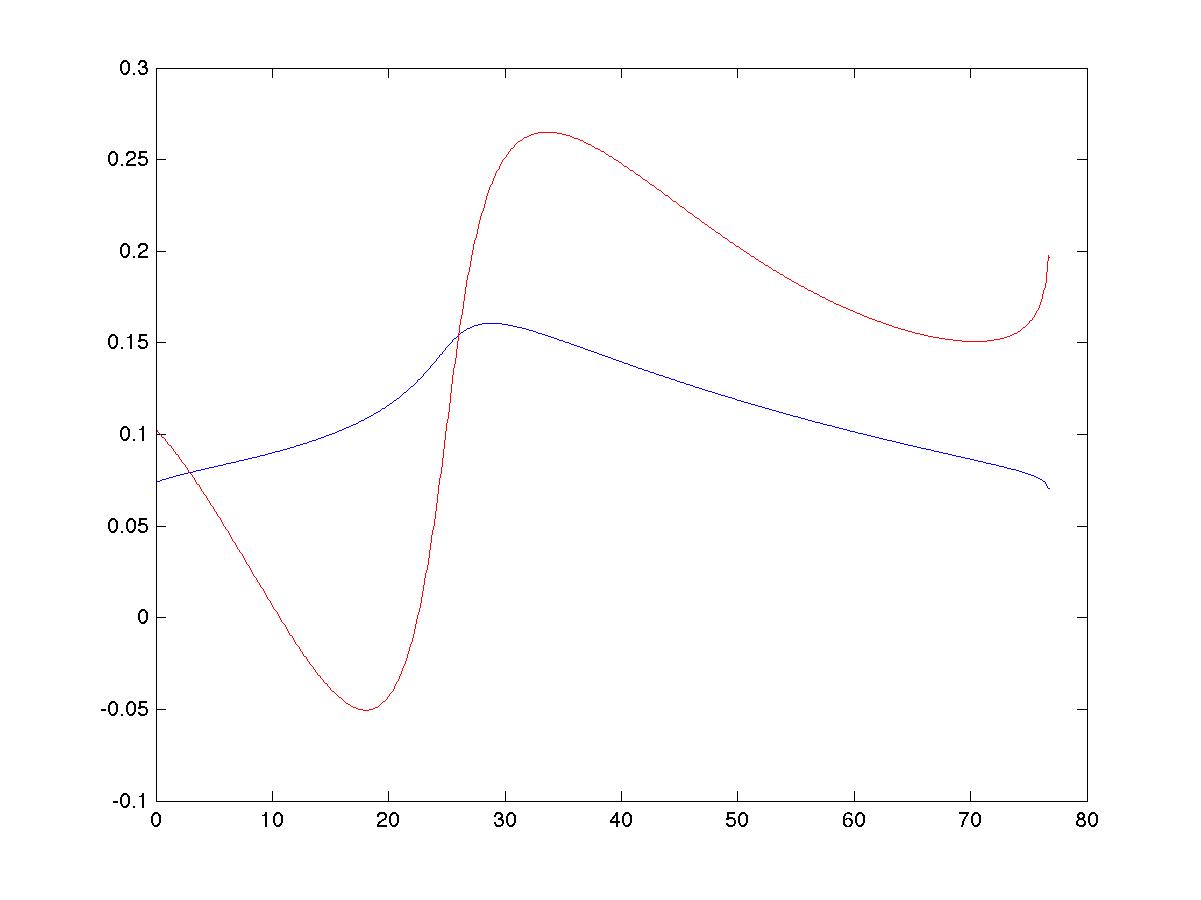}
\caption{The variation of $V(t_0)$ (red), and of $\omega$ (blue) with $t_0$ for the $(1,3)$ orbit when $\mu = 0.25.$ We can see evidence for a limit point at $t_0=0.78.$}
\label{fig:3ab}
\end{figure}

\noindent The solutions close to  $\mu = 0.25$ are of theoretical interest as here we see the reason for the break up of the closed elliptical curves. In Figure \ref{fig:3aa} we show the solution existence curves for $\mu = 0.244$ (left) and for $\mu = 0.245$ (right). The curve for $\mu = 0.244$ shows two separated solution branches, one of which is a distorted ellipse. As predicted earlier, these two branches then coalesce close to $\mu = 0.25$, leading to a sudden expansion of the rightmost elliptical curve.
A plot of the curve of $(t_0,\omega)$ and of $(t_0,V(t_0))$ for the case of $\mu = 0.25$ is given in Figure \ref{fig:3ab}. We see that unlike the case of small $\mu$ when $t_0$ could take arbitrary values, in this case we have an upper limit of $t_0 < 78$. We note, however, that the solutions on these curves are not necessarily physical as $V$

\subsection{Stability and physicality}

\noindent As we have seen, not all of the orbits on the computed curves are physical, in the sense that the function $F$ on a solution trajectory can change sign at an intermediate point $t_0 < t^* < t_1$ during a glacial period, or similarly during an inter-glacial period. 

\vspace{0.1in}

\noindent Also of significant interest is the stability of the resulting orbits. The right extremes of the $(\omega,V)$ solution curves are in all cases marked by saddle-node bifurcations. In general such bifurcations are associated with changes in the {\em stability} of the solutions. It is difficult to determine
the stability algebraically. However a large number of numerical experiments demonstrate clearly that it is the {\em lower branch of the curves} which is (in general) stable, and the {\em upper branch} is unstable.

\vspace{0.1in}

\noindent We will see later that as a parameter such as $\omega$ is varied, the solutions can also lose stability at {\em period-doubling bifurcations},
where a $(m,n)$ orbit is replaced by a $(2m,2n)$ orbit. A further loss of stability is associated with a {\em grazing bifurcation}, which is the first value of the parameter at which a solution loses physicality with the trajectory grazing the discontinuity surface $\Sigma$. (Such events are known to be highly destabilising \cite{bernardo2008piecewise}.)

\subsection{More general $(m,n)$ periodic orbits} 

\noindent  A similar analysis can be applied to the more general $(m,n)$ orbits. In such orbits we see  $m$ glacial cycles of warming and cooling, in a period of $2\pi n/\omega$. To construct, and analyse these, we  introduce a series of $m$ intervals $\Delta_{i,1}$ and $\Delta_{i,2}$ with $i = 1 \dots m-1$, summing in total to $2\pi n/\omega$, being the times between successive glacial and inter-glacial periods. Each such interval will start at a time $t_{i,1}$ or $t_{i,2}$, with $i=0 \ldots m-1$. Here
each such $t_{i,1,2}$ can be computed from the initial time $t_{i,1}$ of the first glacial cycle by adding up the appropriate time periods $\Delta_{i,1,2}$. For small $\mu$ Each $\Delta_{i,1}$, and $\Delta_{i,2}$ is then a perturbation, $\delta_{i,1}$ or
$\delta_{i,2}$, of the respective times of the glacial and inter-glacial periods of the periodic solution of the unforced problem. Similarly, we let ${\mathbf X}_{i,1}$ and
${\mathbf X}_{i,2}$ be the initial conditions at the start of the respective glacial and inter-glacial periods. For small $\mu$ these will be perturbations $x_{i,1}$ and $x_{i,2}$ of the related values for the periodic orbit of the unperturbed system. The algebraic equations for a $(1,n)$ orbit then extend to the following system for $i=0 \ldots m-1$:
\begin{eqnarray}
    {\mathbf X}_{i,2} &=& {\mathbf E}(t_{i,1},\Delta_{i,1},{\mathbf X}_{i,1}), 
     \\ \label{caug0}
    {\mathbf X}_{i+1,1} &=& {\mathbf E}(t_{i,2},\Delta_{i,2},{\mathbf X}_{i,2})  \\ 
    F({\mathbf X}_{i,1}) &=& 0,  \\
    F({\mathbf X}_{i,2}) &=& 0, \\
    \sum_{i=0}^{m-1} \Delta_{i,1} + \Delta_{i,2} &=& \frac{2\pi n}{\omega}, \\
    {\mathbf X}_{0,1} &=& {\mathbf X}_{m,1}.
    \label{caug1}
\end{eqnarray}
Here ${\mathbf E}(t_{i,1},\Delta_{i,1},{\mathbf X}_{i,1})$ is the evolutionary operator which we have constructed explicitly. If we specify the start time $t_{0,1}$ and the amplitude $\mu$ then the system (\ref{caug1}) constitutes $8m + 1$ equations for the $8m+1$ unknowns
${\mathbf X}_{i,1}, {\mathbf X}_{i,2}, \Delta_{i,1},
\Delta_{i,2},$ and $\omega.$

\vspace{0.1in}

\noindent As before, the complete system (\ref{caug1}) can be linearised about the periodic solution when $\mu = 0$. In this case we take $\omega$ = $n \omega^*/m + \delta \omega$ with $|\delta \omega| \ll 1.$ 
The resulting system will be identical in form to that given in equation (\ref{c113}) with a corresponding linear operator $M$ in this case.
However, from the earlier discussion of the general rules for the asymptotic behaviour of the Arnold tongues, we expect that $\delta \omega = {\cal O}(\mu^m)$ in this case.

\vspace{0.1in}

\begin{figure}[htbp]
  \centering
  \includegraphics[width=0.45\textwidth]{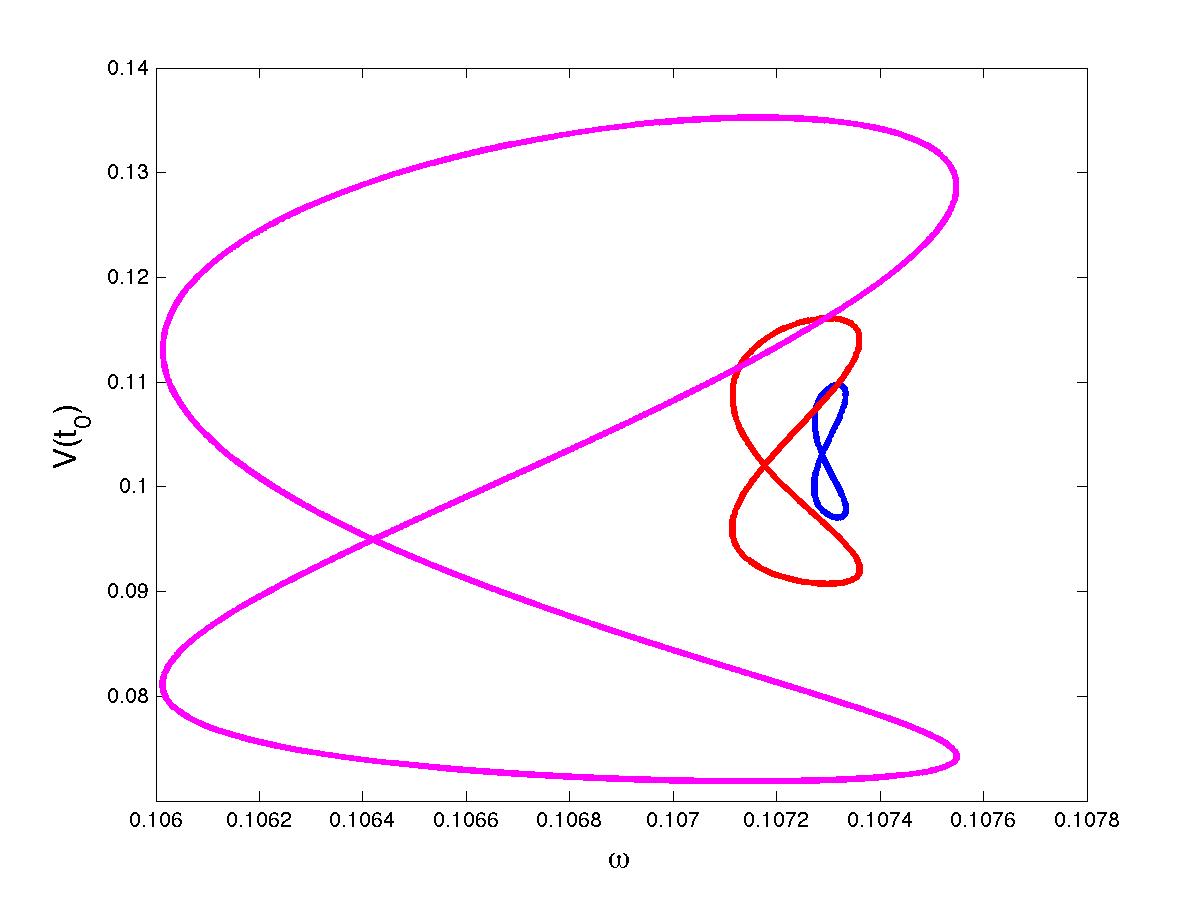}
  \includegraphics[width=0.45\textwidth]{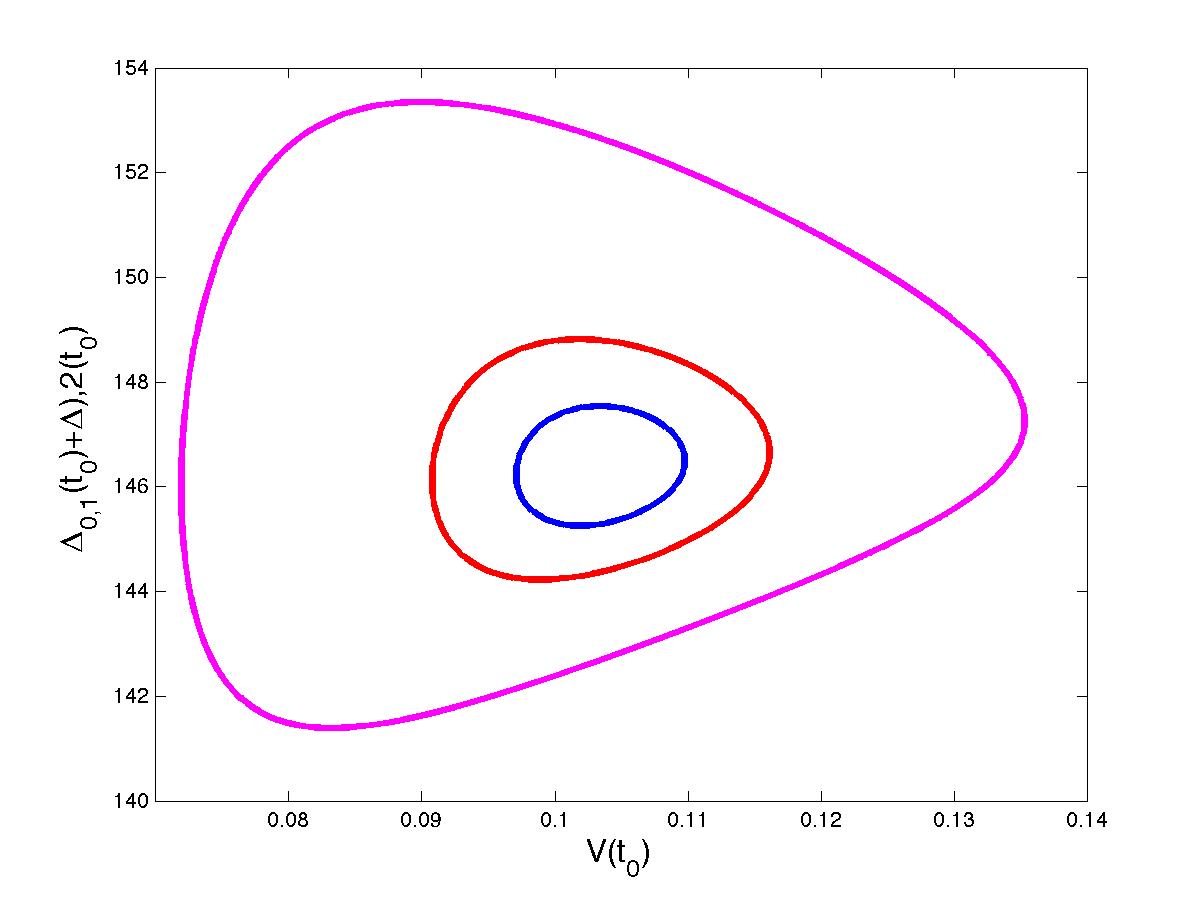}
\caption{The computed variation of the $(2,5)$ orbit with $mu = 0.01,0.02,0.05$. On the left we see $V(t_0)$ as a function of $\omega$ showing linear dependence in $\mu$ for the perturbations of  $V(t_0)$ and quadratic dependence of $\delta \omega$. On the right the variation in the period of the first full glacial cycle with $V(t_0)$ showing linear dependence in $\mu$ for both.}
\label{fig:3aac}
\end{figure}

\noindent We present in Figure \ref{fig:3aac} an example calculation of solving this algebraic system numerically for the case of a periodic solution with $(m,n) = (2,5)$ with $\mu=0.01,0.02$ and $0.05$. In this figure on the left we plot $V(t_0)$ as a function of $\omega$, and on the right we plot
the period of the first full glacial cycle $P_G = \Delta_{0,1} + \Delta_{0,2}$ as a function of $V(t_0)$. It is clear from these figures that $V(t_0)$ and $P_G$ have a single 'cycle' as $t_0$ varies over one period, and the perturbation form the unforced value scale linearly with $\mu$, with the $(V(t_0),P_g)$ curve being a perturbed ellipse. In contrast $\delta \omega$ scales quadratically with $\mu$ and has a double cycle (in the form of a figure of eight) in this period. (A plot of the same curves for the $(3,5)$ orbit shows, as expected, similar behaviour for $V(t_0)$ and $P_G$ and a triple cycle for $\delta \omega$ which scales as ${\cal O}(\mu^3).$ 

\vspace{0.1in}

\noindent An excellent account of the computation of Arnold tongues for general circle maps is given in \cite{schilderandpeckham2007}, with general surfaces for the solutions obtained for varying parameters. The surfaces determined above (for example the ellipses and figure of eight
can be also found in the examples computed in \cite{schilderandpeckham2007}.

\section{More general dynamics of the PP04 model}

\noindent The previous sections have allowed us to gain an analytical insight into the general behaviour of the periodic solutions of the PP04 model for small periodic forcing, but give less information about the general behaviour of the system. Of course this is of most interest in a general discussion of how well the model applies to climate dynamics for which the periodic insolation forcing $\mu \sin(\omega t)$ takes larger values. We now make a systematic numerical study of this case which both confirms the predictions of the previous section for small $\mu$, and also which allows us to explore the rich dynamics of the forced PP04 system for the case of larger values of the insolation forcing.

\subsection{Poincar\'e sections and Mont\'e-Carlo plots}

\noindent A natural tool for analysing the PP04 climate model under periodic forcing is the {\em stroboscopic Poincare map} $P_S$ mentioned in the last section.  This map is defined as follows

\vspace{0.1in}

\noindent {\bf Definition} Let the PP04 model be forced by the insolation function $\sin(\omega t)$, with state vector ${\mathbf X}(t)$ then
\begin{equation}
P_s \; {\mathbf X}(t) \equiv {\mathbf X}(t + 2\pi/\omega).
\label{eq:g1}
\end{equation}

\vspace{0.1in}

\noindent Using this map we can construct a set of points ${\mathbf x}_m$ defined by the iteration
\begin{equation}
{\mathbf X}_{m+1} = P_S \; {\mathbf X}_m.
\label{eq:g2}
\end{equation}

\noindent A $(m,n)$ periodic orbit, as constructed above, then corresponds to an orbit which is an $n-$cycle
$({\mathbf X}_0, {\mathbf X}_{1}, \ldots , {\mathbf X}_{n-1})$ of $P_S$ for which 
$${\mathbf X}_0 = P_S \; {\mathbf X}_{n-1}.$$
Such an orbit crosses the discontinuity manifold $2m$ times.
\noindent The nature of such Poincar\'e maps for Filippov flows has been studied some detail in \cite{bernardo2008piecewise} Chapter 7.
In general, it follows from the theory presented in  \cite{bernardo2008piecewise} that the map $P_S$ will be smooth if the intersection between the solution trajectory and $\Sigma$ is transversal. However it will lose smoothness if there is a grazing
event in the interval $[t,t+2\pi/\omega]$ leading to a non transversal intersection. As the vector field is continuous across $\Sigma$ but has a derivative discontinuity, then the map will typically have a square root type behaviour close to the grazing point. We will explore the impact that this has on the dynamics of the PP04 model in more detail in a forthcoming paper. 

\vspace{0.1in}

\noindent The general dynamics of the PP04 system can now be studied by considering the iterations of the map $P_S$. To do this we use a Monte-Carlo approach in which, for a given parameter,  we take a random set of initial data (typically for computations this set will have 5 members) and iterate the solution starting from points in this set forward. To do this calculation we take the smoothed system with $\eta = 1000$ in the approximation of the Heavyside function, and solve forward in time it using the Matlab code {\tt ode15s} (with tolerance set to $1e-10$) for a period of 6000 kyr.  The Omega limit set given by the displaying the final set of iterations of the map. By choosing a {\em set} of random initial data, we obtain a Mont\'e-Carlo plot of (hopefully) all of the possible Omega-limit sets. This gives significant insight into the overall dynamics of the system. It is convenient to represent the state of the whole system by plotting the values of the single variable $F({\mathbf X_i})$ at the points $X_i$. The advantage of this approach over the path-following methods used, for example, in the AUTO code \cite{AUTO}, is that it can locate  Omega-limit sets which are disjoint from the main solution branch. The disadvantage is that it can only find asymptotically stable sets.

\subsubsection{Varying $\omega$.}

\begin{figure}[htbp]
  \centering
  \includegraphics[width=0.8\textwidth]{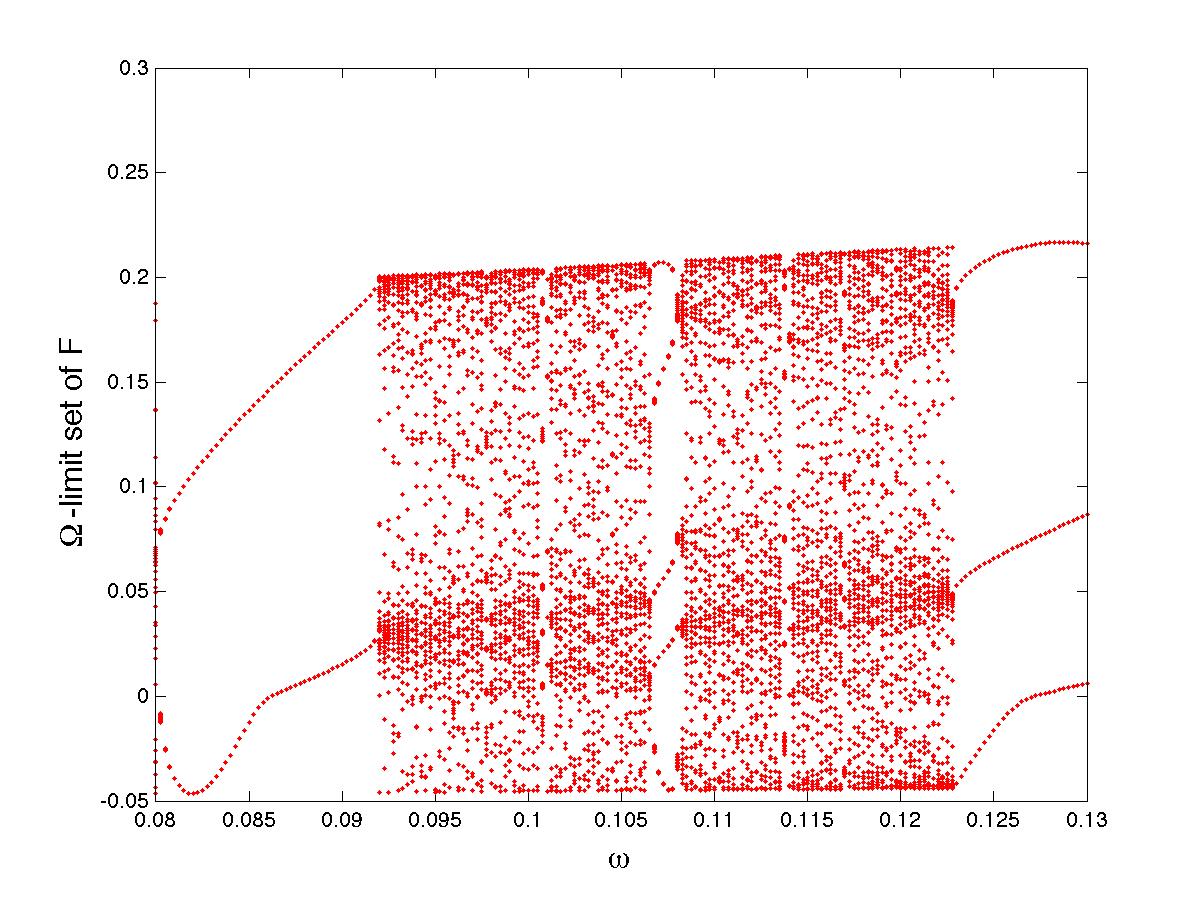}
\caption{The Poincar\'e section points of $F$ on the omega limit set, as a function of $\omega$ with $\mu = 0.05$ showing (as $\omega$ increases), a large window of existence for the  $(1,2)$ periodic orbit, a much smaller window of existence for the $(2,5)$ periodic orbit close to $\omega = 0.107$, and then another large window of existence for the $(1,3)$ periodic orbit. We can clearly see the transition from quasi-periodic motion when $\omega < 0.1245$ to the
period $(1,3)$ motion at a saddle-node bifurcation. All of the windows are separated by intervals of quasi periodic motion}
\label{fig:40}
\end{figure}

\begin{figure}[htbp]
  \centering
  \includegraphics[width=0.8\textwidth]{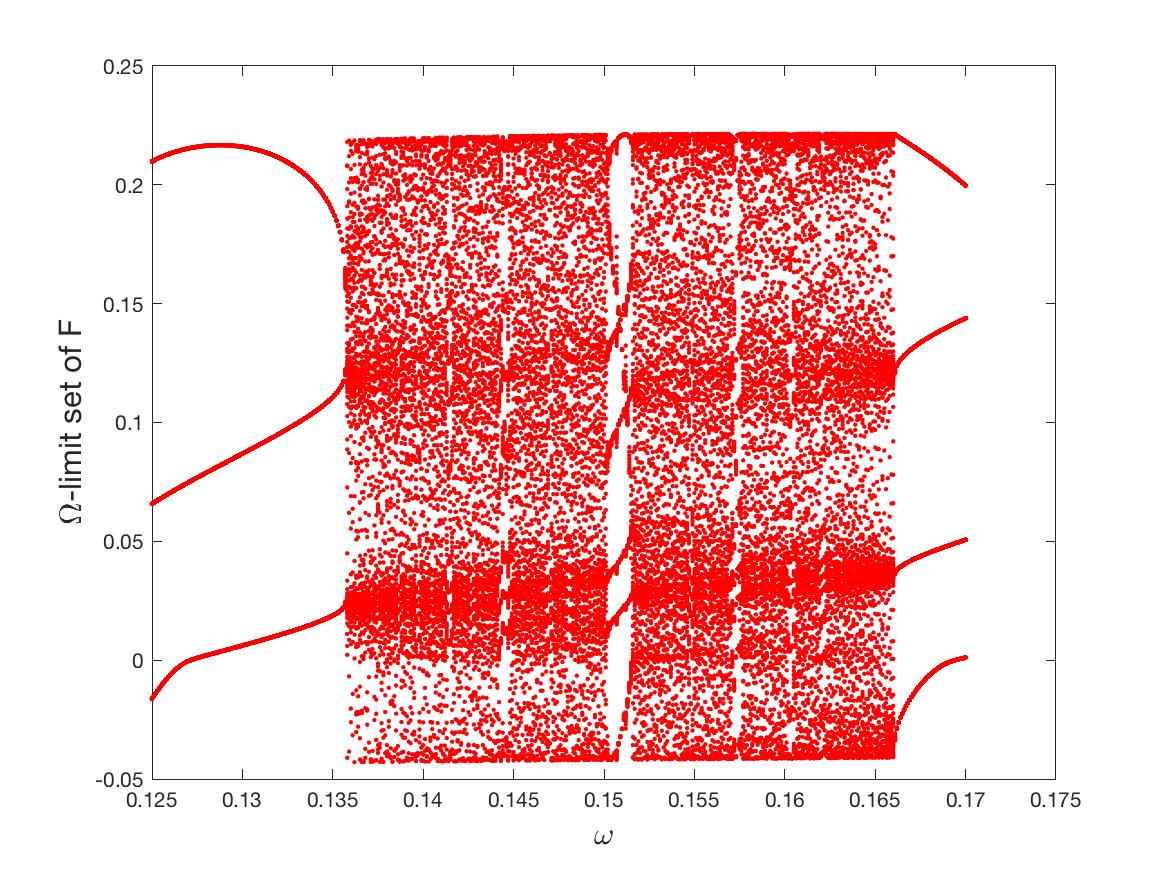}
\caption{The Poincar\'e section of $F$  on the Omega limit set, as a function of $\omega$ with $\mu = 0.05$ showing period $(1,3)$ and period $(1,4)$ motions separated by an interval of quasi periodic motion, containing a small window with a period $(2,7)$ orbit, and evidence of other periodic orbits.}
\label{fig:34}
\end{figure}

\noindent Initially we take fixed small values of $\mu$ (consistent with the earlier analysis) and vary the value of $\omega$. In Figure \ref{fig:40} we take $\mu = 0.05$ and increase $\omega$ from $0.08$ to $0.13$, plotting the omega-limit set of the resulting orbit in each case. It is convenient to represent these orbits by plotting the values of the function $F$. In this figure we can see a clear $(1,2)$ orbit for smaller values of $\omega$ and an equally clear $(1,3)$ orbit for the larger values. For $\omega \approx 0.107$ there is a small window of existence for the $(2,5)$ periodic orbit, and there is some evidence of windows of existence for more complex period motions. Away from these values we observe quasi-periodic behaviour. 
In Figure \ref{fig:34}  we see (again for $\mu = 0.05$) the $(1,3)$ orbit changing to a quasi-periodic orbit
when $\omega = 0.135$ followed by an interval of quasi-periodic motion, which then turns into a $(1,4)$ orbit 
when $\omega = 0.165$.  There is a thin window of existence for a $(2,7)$ orbit between the $(1,3)$ and $(1,4)$ orbits, and evidence of other periodic orbits.

\begin{figure}[htbp]
  \centering
  \includegraphics[width=0.8\textwidth]{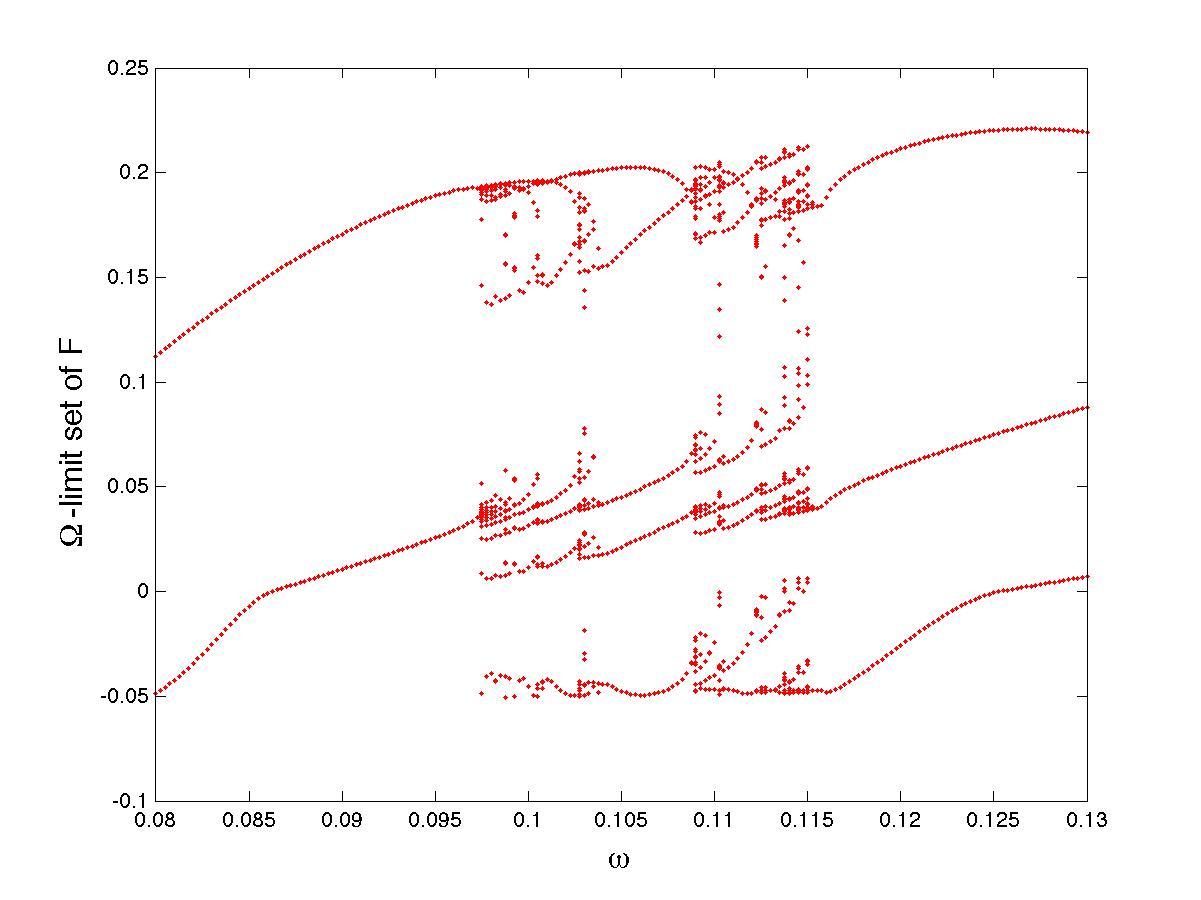}
  \caption{The Poincar\'e section points of $F$ on the Omega limit set, as a function of $\omega$, with $\mu = 0.1$, showing the $(1,2)$, $(2,5)$ and $(1,3)$ periodic solutions
  and a variety of other types of dynamics.}
\label{fig:38}
\end{figure}

\noindent In Figure \ref{fig:38}  we take the larger value of the forcing $\mu = 0.1$. Again we see the $(1,2)$, $(2,5)$ and $(1,3)$ orbits with larger regions of existence, together with other types of more complex dynamics, but less evidence of a full quasi-periodic attractor.

\begin{figure}[htbp]
  \centering
  \includegraphics[width=0.8\textwidth]{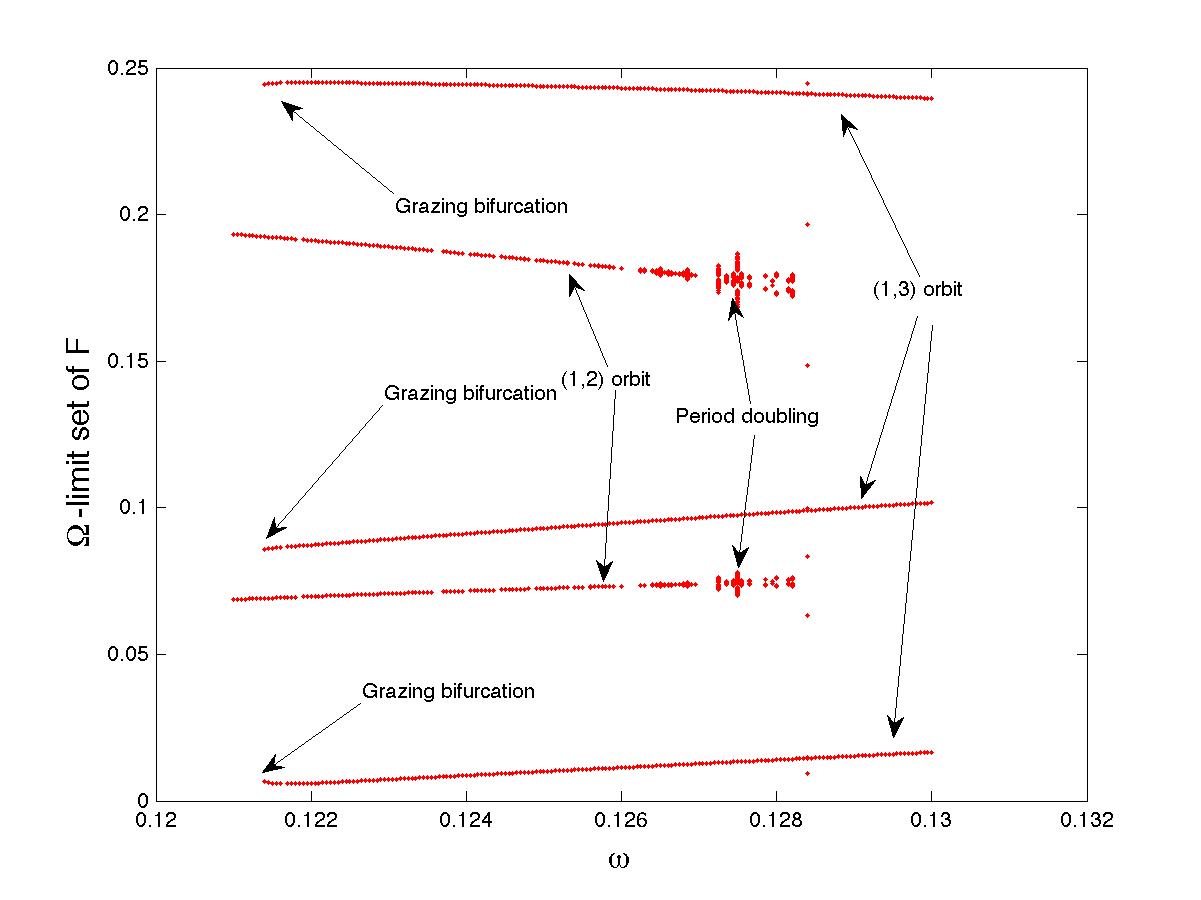}
  \caption{A Mont\'e-Carlo plot of the Omega-limit set of  $F$ for $\mu = 0.467$. To the left we observe a $(1,2)$ orbit and to the right a $(1,3)$ orbit. Various transitions between these orbits can also be observed. These include a grazing bifurcation of the $(1,3)$ orbit at $\omega = 0.122$ and an expansion of the $(1,2)$ orbit at $\omega = 0.1275$ followed by a period-doubling cascade.}
\label{fig:1aa}
\end{figure}

\noindent 
\noindent For a final calculation we take the physically relevant value of $\mu = 0.467$ (see Section 2 for a motivation of this value) and vary $\omega$ from $0.121$ to $0.129$. The results of this calculation are presented in Figure \ref{fig:1aa}. 
As we would expect from the previous results for the smaller values of $\omega$, we see a $(1,2)$ periodic solution and, for the larger values of $\omega$, a $(1,3)$ periodic solution. There is no quasi-periodic behaviour in this case. Indeed, for a wide range of values of $\omega$ the $(1,2)$ and $(1,3)$ solutions co-exist. Two interesting transitions can be observed in this figure as $\omega$ increases. At $\omega = 0.122$ the $(1,3)$ solution abruptly appears. The reason for this can be seen from studying the values of $F$. In particular, at the bifurcation point, there is a value of $t$ strictly within the glacial period, at which $F(t)= 0$. This is an example of a (non-smooth) grazing bifurcation (as mentioned above) at which the $(1,3)$ orbit suddenly starts to become physical. We will study this transition in more detail in a future paper.  A (smooth) super-critical period-doubling bifurcation can also be seen at $\omega = 0.1275$. At this point the $(1,2)$ solution loses stability to a nearby $(2,4)$ orbit as $\omega$ increases. There is evidence of a period-doubling cascade close to this value. 
\subsubsection{Varying $\mu$}

\begin{figure}[htbp]
  \centering
  \includegraphics[width=0.8\textwidth]{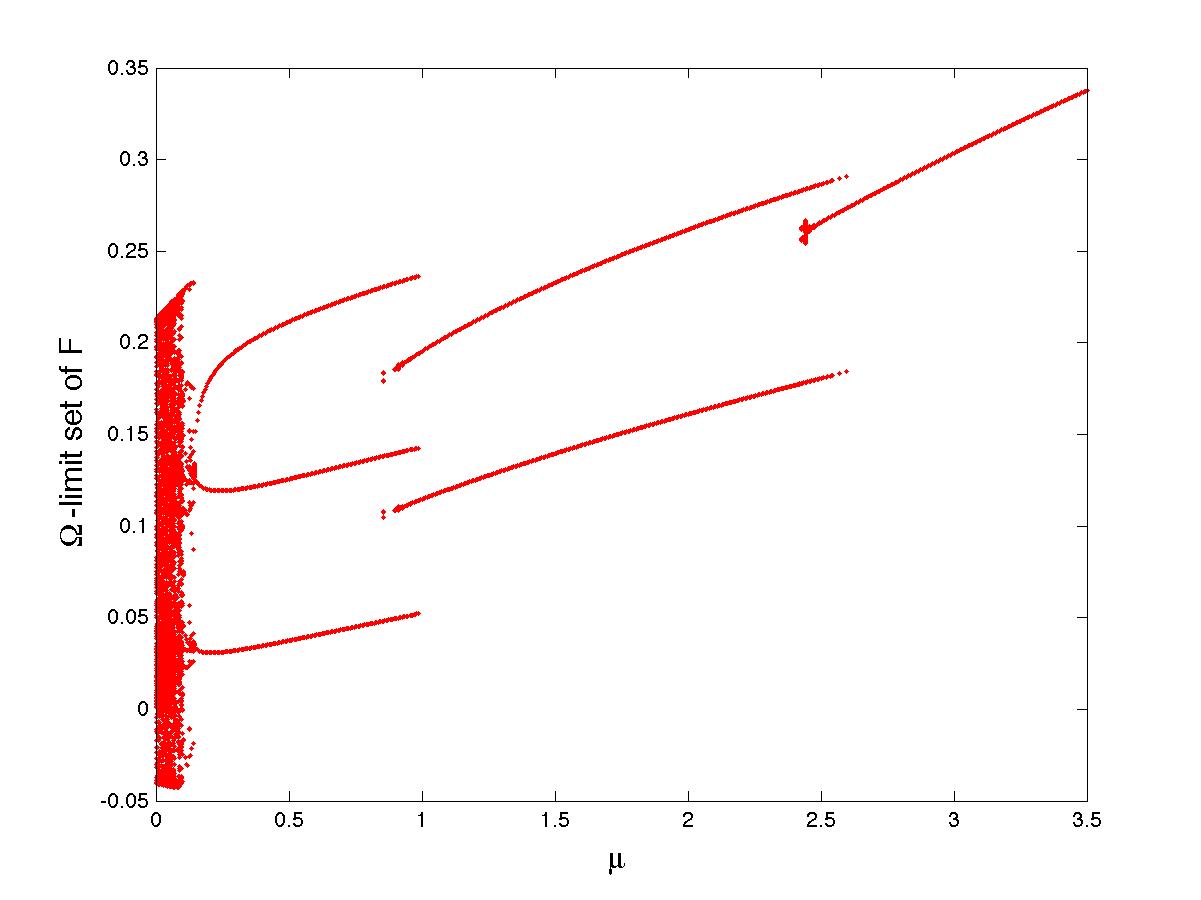}
  \caption{The Poincar\'e section points on the Omega limit set of $F$, as a function of $\mu$ with fixed $\omega = 0.1532$. This shows the different types of solutions as $\mu$ increases from quasi-periodic, to  $(1,3)$,$(1,2)$ periodic orbits and then a $(1,1)$ periodic solution, with regions of co-existence. The $(1,3)$ orbit starts at a saddle-node bifurcation and terminates at a grazing bifurcation. The $(1,1)$ periodic solution shows evidence of a period-doubling bifurcation at $\mu \approx 2.5$.}
\label{fig:37}
\end{figure}

\noindent As a second calculation, we fix $\omega$ at the physically relevant value of $\omega = 0.1532$ (see \ref{muandomega}) and increase $\mu$ from zero. The resulting Mont\'e-Carlo calculation is presented in Figure 
\ref{fig:37}. In this figure we see quasi-periodic behaviour for small values of $\mu$.
The $(1,3)$ orbit arises at a saddle-node bifurcation at around $\mu = 0.15$ and persists until it is destroyed at a grazing bifurcation at $\mu \approx 1$. For a short interval of values of $\mu$ there are coexisting $(1,2)$ and $(1,3)$ orbits. The $(1,2)$ orbit then persists until it too is destroyed at a grazing bifurcation when $\mu \approx 2.6$. It co-exists with a $(1,1)$ orbit which loses stability at a period-doubling bifurcation when $\mu \approx 2.5$. For  larger values of $\mu$ we see only the $(1,1)$ periodic orbits, completely locked to the forcing. At the physically interesting value of $\mu = 0.467$ (see Section 2) we see only a $(1,3)$ periodic orbit. 

\subsection{Domains of attraction}

\begin{figure}[htbp]
  \centering
  \includegraphics[width=0.45\textwidth]{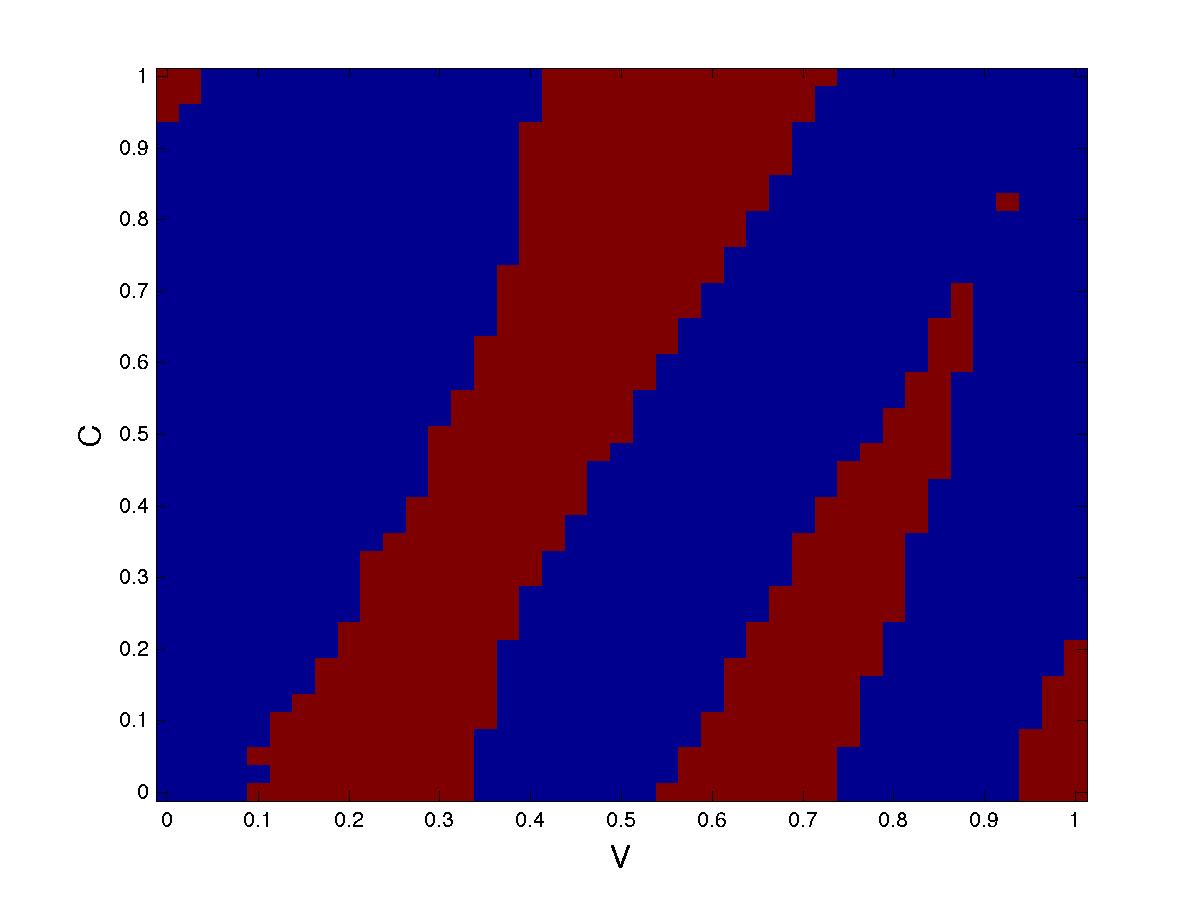}
  \includegraphics[width=0.45\textwidth]{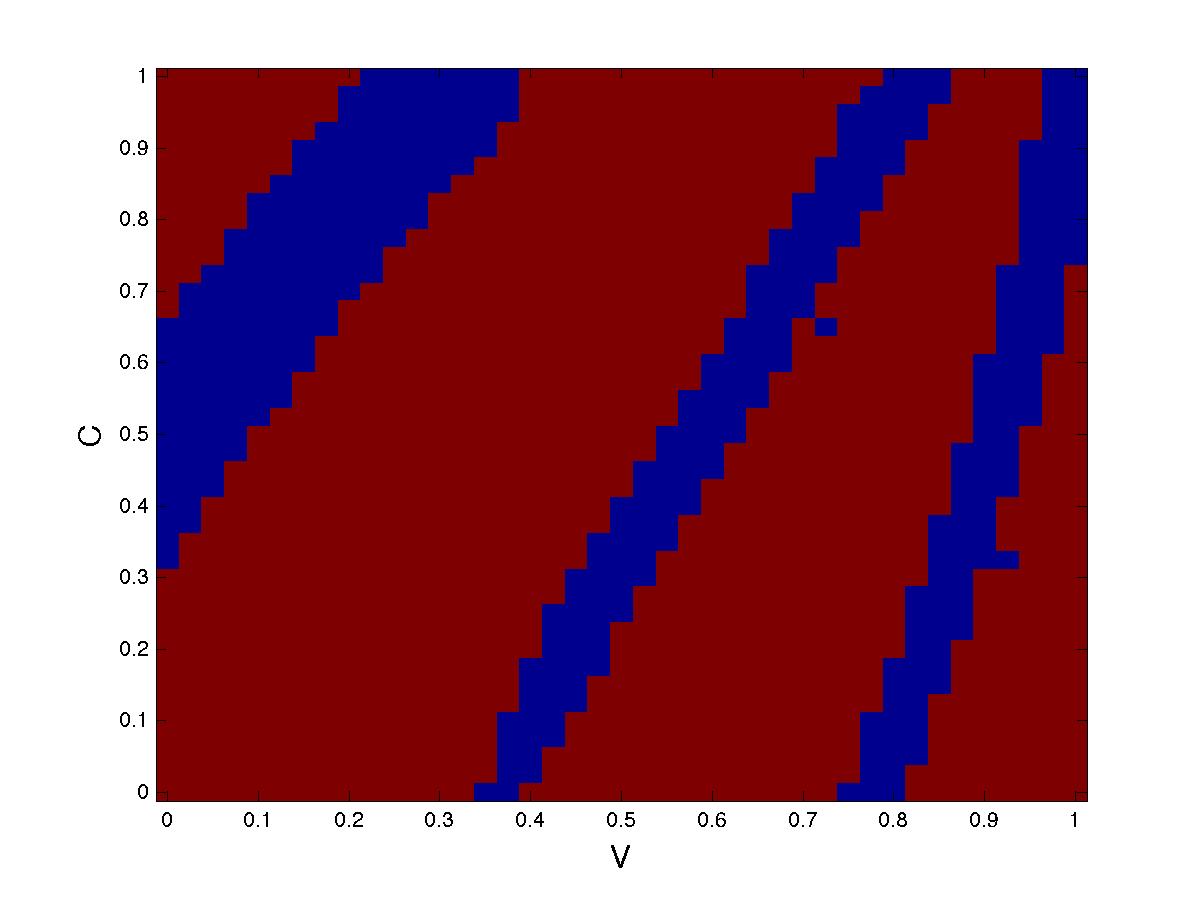}
  \caption{ A cross-section of the domains of attractions of the periodic solutions of the periodically forced system with $t_{initial} = 0, A=0.55$, when  $\omega = 0.124$ (left) and $\omega = 0.128$ (right) with $\mu = 0.467$.  Here the red regions represents the domain of attraction of the $(1,3)$ periodic solution, and the blue regions the domain of attraction of the $(1,2)$ (left) or $(2,4)$ (right) periodic solution. The very small blue regions are the domains of attraction for other solutions.}
\label{fig:41}
\end{figure}

\noindent The co-existence, for example, of the $(1,2)$ and $(1,3)$ solutions when $\omega = 0.124$ and $\mu = 0.467$, and the $(2,4)$ and $(1,3)$ orbits when $\omega = 0.128$, leads to the possibility of seeing both types of behaviour in the solution of the PP04 system, depending upon the initial conditions.
Furthermore we may also expect to see, for certain initial conditions, an evolution from behaviour which is close to one type of periodic motion to behavior close to the other. To investigate this phenomenon we calculate the domains of attraction for the periodic orbits above. These domains are the subsets of the three dimension phase space $(V,A,C)$ such that the omega-limit set of the iterations of the map $P_S$ is either the $(1,2)$, $(2,4)$ or the $(1,3)$ orbit. It is problematic to find the full three dimensional sets, so for convenience we find a two-dimensional projection by fixing $t_{initial} = 0, A = 0.55$. The resulting two-dimensional cross-sections of the domains of attraction are given in Figure \ref{fig:41}. In these figures we see a rapid increase in the domain of attraction of the $(1,3)$ orbit as $\omega$ increases from 0.124 to 0.128.

\vspace{0.1in}

\begin{figure}[htbp]
  \centering
  \includegraphics[width=0.7\textwidth]{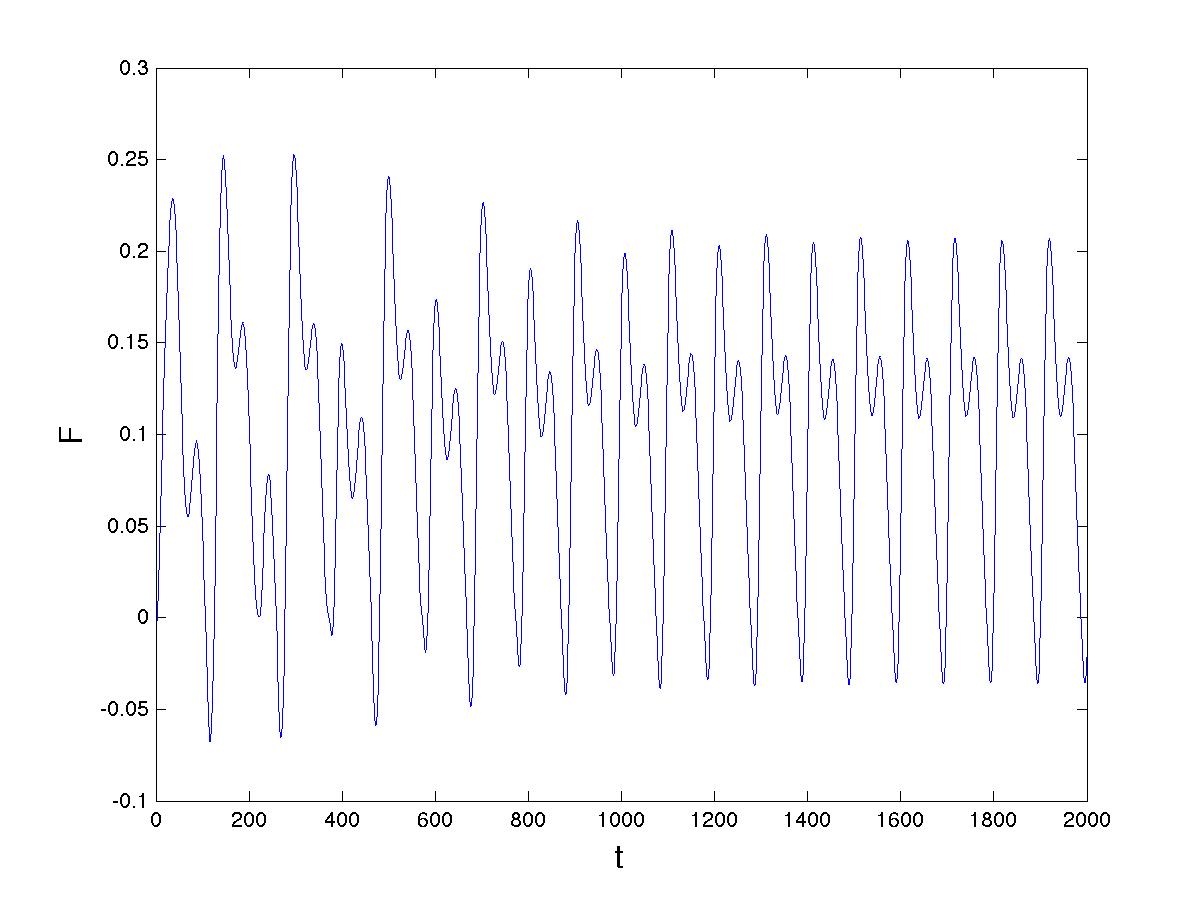}
  \caption{The time evolution of $F(t)$ for the system started close to the boundary of the domain of attraction. This figure shows the slow evolution from a $(1,3)$ periodic orbit to a $(1,2)$ periodic orbit. Here $\omega = 0.124$ and  $\mu = 0.467$ and the initial conditions are $(V(0),A(0),C(0))=(0.341,0.55,0.6)$.}
\label{fig:39a}
\end{figure}

\noindent Motivated by this figure we now explore the time evolution of the solutions from a variety of initial conditions. In Figure \ref{fig:39a} we take $\omega = 0.124$ and plot $(t,F)$ for a solution in which we take initial conditions in the green region but close to the red boundary with $(V(0),A(0),C(0))=(0.341,0.55,0.6)$. We  observe an initial transient with dynamics close to that of the $(1,3)$ periodic orbit, which then ultimately evolves to a $(1,2)$ orbit. We note that there is a dramatic change in the behaviour of the system when $t \approx 600 kyr$. This occurs when there is a local minimum at which $F < 0$ which occurs for the first time in a 'glacial region'. The resulting instability is the result of a {\em grazing transition} \cite{bernardo2008piecewise}.  

\noindent As a separate calculation we take $\mu = 0.467$ and $\omega = 0.128$, which is just greater than the period-doubling value. We now take as initial conditions $(V(0),A(0),C(0)) = (0.13,0.55,0.6).$ In the resulting intermittent dynamics we see a $(2,4)$ orbit
evolve into a larger amplitude $(1,3)$ orbit in a manner which qualitatively resembles that at the mid-Pleistocene transition. The sudden expansion in the solution amplitude (and the consequent change in period) again seems to occur just after the function $F$ grazes zero. We will return to this in the forthcoming paper on grazing transitions in the PP04 model. 
\begin{figure}[htbp]
  \centering
  \includegraphics[width=0.7\textwidth]{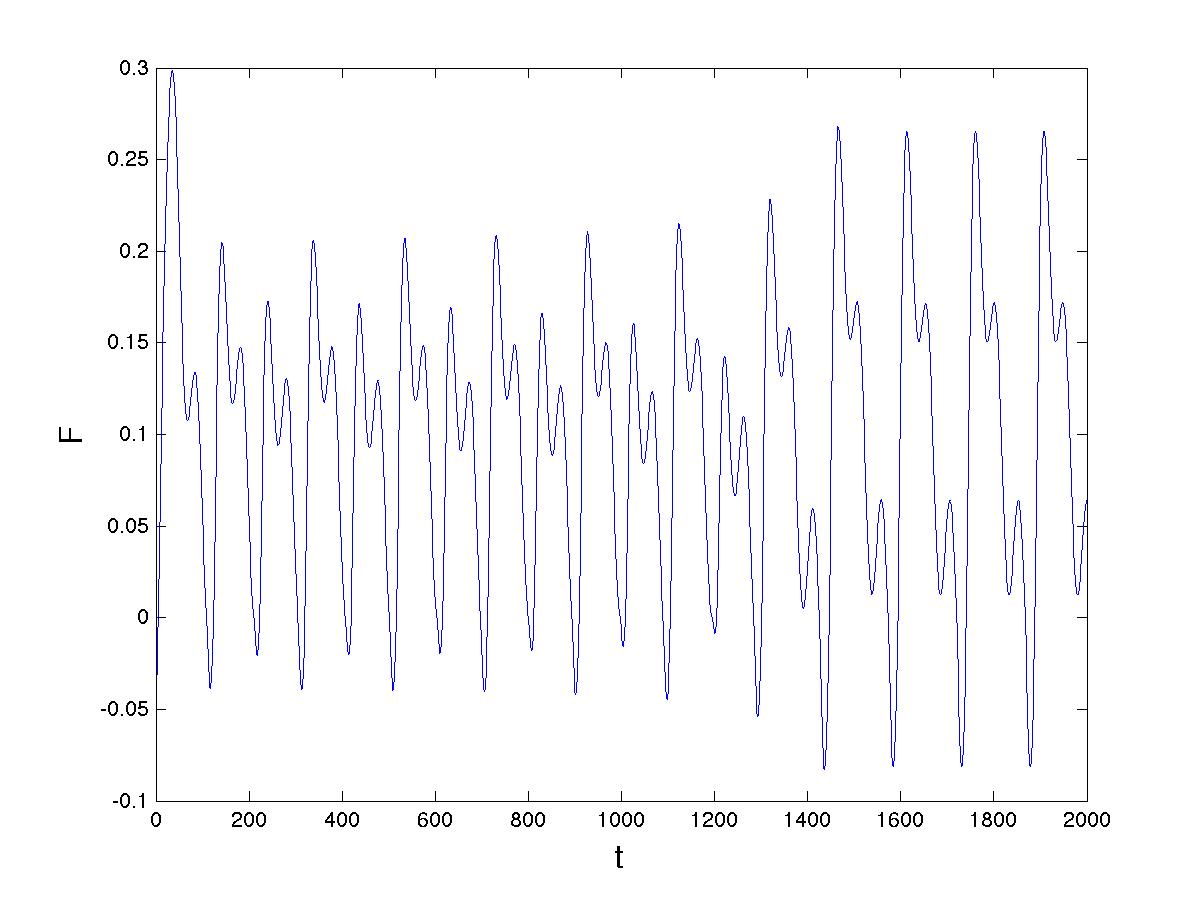}
  \caption{The time evolution of $F(t)$ when $\omega = 0.128$ and  $\mu = 0.467$ with initial conditions $(V(0),A(0),C(0)) = (0.13,0.55,0.6).$ In this case we see a slightly unstable, low amplitude, $(2,4)$ orbit evolve into a larger amplitude $(1,3)$ orbit.}
\label{fig:39}
\end{figure}

\section{The implications of these results for climate modelling.}

\subsection{The unforced system}

\noindent From the results that we have obtained for the PP04 model, we have shown that if there is no insolation forcing on the system and $-0.72 < d < 0.32$, then there is a periodic orbit of period of about 140 kyr. Numerically this orbit appears to be both stable and unique. The existence of this orbit suggests that the Earth's climate, if left alone  without the contribution of insolation forcing, will have periodic glacial cycles. In these it will spend most  of its time, say about 120 kyr, in the glacial state and less time in the inter-glacial state. On the other hand, if the $d$  is greater than $0.32$ or less than $-0.72$, we have stable equilibria and the climate can get locked into either a glacial state or an inter-glacial state. 

\subsection{The existence and persistence of the $(1,3)$ orbit under changes to the 
insolation forcing. }

\begin{figure}[htbp]
  \centering
  \includegraphics[width=0.8\textwidth]{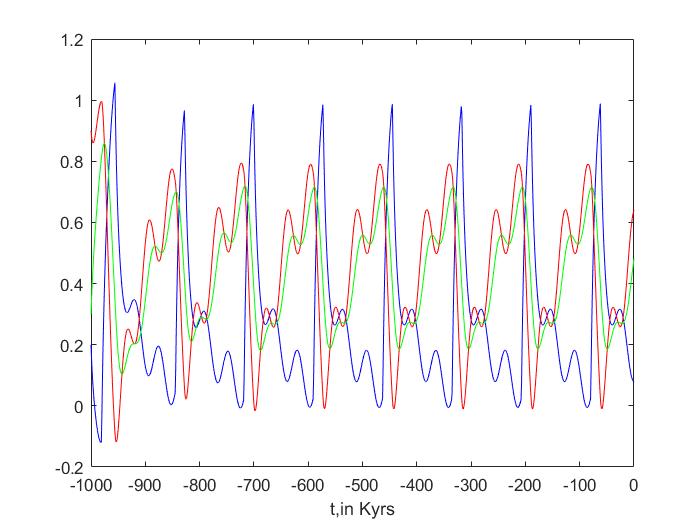}
  \caption{The evolution of the solution when $(\mu,\omega) = (0.467,0.1532)$. This shows a
  rapid evolution towards the $(1,3)$ periodic solution. In this figure $V$ is shown in red, $A$ in green and $C$ in blue.}
\label{fig:4a}
\end{figure}

\noindent When purely periodic insolation forcing is introduced, and we consider the  physically relevant values of $(\mu,\omega) = (0.467,0.1532)$ we see only a stable $(1,3)$ periodic orbit. This orbit has period $6\pi/\omega = 123$ kyr, which is slightly longer than the observed period of $100$ kyr.
(We note that $100$ kyr is very close to the period of the $(2,5)$ periodic orbit. However, we have not seen any evidence of this orbit existing close to the realistic parameter values). From extensive numerical experiments, for these parameter values, the $(1,3)$ orbit appears to be unique, globally stable, and indeed strongly attracting, for all physical initial states. The resulting orbit and a short transient is shown in Figure \ref{fig:4a}

\vspace{0.1in}

\begin{figure}[htbp]
  \centering
  \includegraphics[width=0.8\textwidth]{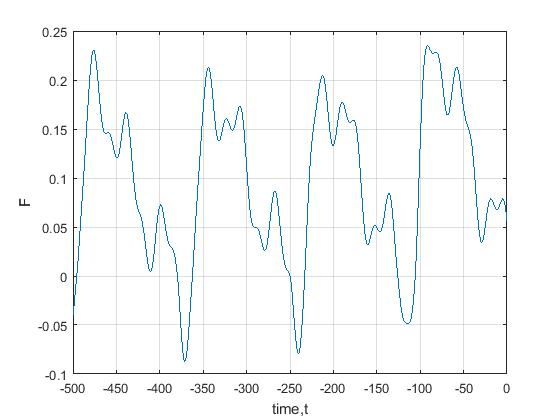}
  \caption{The time solution of the system showing the quasi-periodic forced solution with the $(1,3)$ periodic solution evident for $\mu_1 =0.467$, $\omega_2 =0.1476$,$\mu_2 =0.5$ and $\omega_2 =0.331$. This figure demonstrates that the $(1,3)$ periodic solution is perturbed to a more complex orbit on a torus. However, the basic form of the $(1,3)$ orbit remains.}
\label{fig:qp}
\end{figure}

\noindent Of course this analysis has only been made for the case of {\em periodic forcing}. In practice the Milankovich cycles lead to quasi-periodic insolation forcing. The structural stability of the $(1,3)$ orbit constructed above means that this orbit persists, appropriately perturbed to an invariant torus, when quasi-periodic forcing is introduced, with a small additional forcing. We demonstrate this by considering an insolation forcing of the form
$$I(t) = \mu_1 \sin(\omega_1 t) + \mu_2 \sin(\omega_2 t).$$ 
Provided that $\mu_2$ is not too large,
the $(1,3)$ orbit in this case is replaced by a quasi-periodic orbit on a torus in the phase space close to the original periodic curve. This is illustrated in Figure \ref{fig:qp} which we compare with the above figure Figure \ref{fig:4a}. 
 The study in more detail of the quasi-periodic forced  PP04 model will be given later in a later paper, where we consider the break up of the tori for larger forcing $\mu_2$. Similar results for quasi-periodic forcing of the PP04 model (and other similar reduced climate models) are described in the paper by Ashwin et. al \cite{ashwin2018chaotic} (see also \cite{crucifix2013}) in which apparently chaotic behaviour of the solutions was observed for certain types of quasi-periodic forcing.

\subsection{Transitions}

\noindent When $\mu = 0.467$ and $\omega = 0.1532$ the only observed solution is the stable $(1,3)$ periodic orbit. However, for values of $\omega$ close to $0.128$ we also see stable $(1,2)$ solutions and even stable $(2,4)$ solutions. Note that if $\omega = 0.128$ then the period of the $(1,2)$ orbit is 98.17 kyr and of the $(1,3)$ orbit is 147 kyr. If $\omega$ is fixed and the initial data is taken close to the boundary of the domains of attraction of these orbits, the we see transitions, for values of $\omega$ close to 0.128, both from $(1,3)$ orbits to $(1,2)$ orbits and from (period-doubled) $(2,4)$ orbits to $(1,3)$ orbits. The latter transition, in particular, has some resemblance to the qualitative changes in the behaviour of the climate at the MPT. During such transitions there is a long transient motion close to one form of periodic orbit, before the solution converges on the other. Such examples of transitions raise the hope of understanding the MPT through a bifurcation type of analysis. However, much more work needs to be done on this to explore the various transitions possible given the large number of parameters that can be varied in the PP04 model. We will be discussing, in particular, sudden transitions due to grazing bifurcations in a forthcoming paper.

\section{Conclusions}

\noindent In this paper we have made a first mathematical study using the theory of non-smooth dynamical systems of the (periodically forced)  PP04 model for climate change. This has revealed the existence of stable and unstable periodic orbits, with subtle domains of attraction and transitions between them. 
The stable orbits calculated for physically realistic values of the parameters persist under small additional quasi-periodic forcing and have a similar form to those of the observed glacial cycles. The results make an interesting comparison to those of descriptions of the glacial cycle using smooth dynamics systems models, for  example \cite{KaperVo}.

\vspace{0.1in}

\noindent Much more work needs to be done on the PP04 model to  understand fully both the transitions in the whole of the parameter space and also the effect of additional larger terms in the quasi periodic forcing. Both of these will be the subject of further work, which will look in more detail at the effect of grazing bifurcations on the stability of the orbits in the PP04 model and how these (grazing) transitions change when the insolation forcing is quasi-periodic.  Furthermore additional work is  needed to understand better the effect of including additional climatic terms into the PP04 model.  However, we conclude that the PP04 model both has a rich structure as a discontinuous dynamical system, and is a plausible explanation of the glacial cycles. As such it deserves much further study.

\section*{Acknowledgement}
\noindent This research was funded in part by an award from the Botswana International University of Science and Technology (BIUST). We would like to thank Prof. Rachel Kuske (Georgia Tech) and Prof. Paul Glendinning (University of Manchester) for many stimulating conversations related to this work, and the anonymous referees for their very insightful comments on an earlier version of this work.

\section*{Bibliography}

\bibliography{references}

\newcommand{\etalchar}[1]{$^{#1}$}
\providecommand{\bysame}{\leavevmode\hbox to3em{\hrulefill}\thinspace}
\providecommand{\MR}{\relax\ifhmode\unskip\space\fi MR }
\providecommand{\MRhref}[2]{%
  \href{http://www.ams.org/mathscinet-getitem?mr=#1}{#2}
}
\providecommand{\href}[2]{#2}
\begin{thebibliography}{ADCvdH18}

\bibitem[AD15]{ashwin2015middle}
Peter Ashwin and Peter Ditlevsen, \emph{The middle pleistocene transition as a
  generic bifurcation on a slow manifold}, Climate dynamics \textbf{45} (2015),
  no.~9-10, 2683--2695.

\bibitem[ADCvdH18]{ashwin2018chaotic}
Peter Ashwin, Charles David~Camp, and Anna~S von~der Heydt, \emph{Chaotic and
  non-chaotic response to quasiperiodic forcing: Limits to predictability of
  ice ages paced by milankovitch forcing}, Dynamics and Statistics of the
  Climate System \textbf{3} (2018), no.~1, 1--20.

\bibitem[AFO05]{awrejcewicz2005continuous}
Jan Awrejcewicz, Michal Fe{\v{c}}kan, and Pawel Olejnik, \emph{On continuous
  approximation of discontinuous systems}, Nonlinear Analysis: Theory, Methods
  \& Applications \textbf{62} (2005), no.~7, 1317--1331.

\bibitem[BBCK08]{bernardo2008piecewise}
Mario Bernardo, Chris Budd, Alan~Richard Champneys, and Piotr Kowalczyk,
  \emph{Piecewise-smooth dynamical systems: theory and applications}, vol. 163,
  Springer Science \& Business Media, 2008.

\bibitem[CD10]{colombo2010discontinuity}
Alessandro Colombo and Fabio Dercole, \emph{Discontinuity induced bifurcations
  of nonhyperbolic cycles in nonsmooth systems}, SIAM Journal on Applied
  Dynamical Systems \textbf{9} (2010), no.~1, 62--83.

\bibitem[CDBHJ12]{colombo2012bifurcations}
Alessandro Colombo, M~Di~Bernardo, SJ~Hogan, and MR~Jeffrey, \emph{Bifurcations
  of piecewise smooth flows: Perspectives, methodologies and open problems},
  Physica D: Nonlinear Phenomena \textbf{241} (2012), no.~22, 1845--1860.

\bibitem[Cor08]{cortes2008discontinuous}
Jorge Cortes, \emph{Discontinuous dynamical systems}, IEEE Control systems
  magazine \textbf{28} (2008), no.~3, 36--73.

\bibitem[Cru12]{crucifix2012oscillators}
Michel Crucifix, \emph{Oscillators and relaxation phenomena in pleistocene
  climate theory}, Philosophical Transactions of the Royal Society A:
  Mathematical, Physical and Engineering Sciences \textbf{370} (2012),
  no.~1962, 1140--1165.

\bibitem[Cru13]{crucifix2013}
\bysame, \emph{Why could the ice ages be unpredictable}, Clim. Past \textbf{9}
  (2013), 2253--2267.

\bibitem[DBBC{\etalchar{+}}08]{di2008bifurcations}
Mario Di~Bernardo, Chris~J Budd, Alan~R Champneys, Piotr Kowalczyk, Arne~B
  Nordmark, Gerard~Olivar Tost, and Petri~T Piiroinen, \emph{Bifurcations in
  nonsmooth dynamical systems}, SIAM review \textbf{50} (2008), no.~4,
  629--701.

\bibitem[DBH10]{di2010discontinuity}
M~Di~Bernardo and SJ~Hogan, \emph{Discontinuity-induced bifurcations of
  piecewise smooth dynamical systems}, Philosophical Transactions of the Royal
  Society A: Mathematical, Physical and Engineering Sciences \textbf{368}
  (2010), no.~1930, 4915--4935.

\bibitem[DCF{\etalchar{+}}98]{AUTO}
Eusebius Doedel, Alan Champneys, Thomas Fairgrieve, Bjorn Sandstedte, and
  Xainjun Wang, \emph{Auto97:c (continuation and bifurcation software for
  ordinary differential equation, with homcont)}, Concordia University,
  Technical Report (1998).

\bibitem[Dij13]{dijkstra2013nonlinear}
Henk~A Dijkstra, \emph{Nonlinear climate dynamics}, Cambridge University Press,
  2013.

\bibitem[DSCW13]{de2013astronomical}
Bernard De~Saedeleer, Michel Crucifix, and Sebastian Wieczorek, \emph{Is the
  astronomical forcing a reliable and unique pacemaker for climate? a
  conceptual model study}, Climate Dynamics \textbf{40} (2013), no.~1-2,
  273--294.

\bibitem[EKKV17]{KaperVo}
Hans Engler, Hans Kaper, Tasso Kaper, and Theodore Vo, \emph{Dynamical systems
  analysis of the maasch-saltzman model for glacial cycles}, Physica D:
  Nonlinear Phenomena \textbf{359} (2017), 1--20.

\bibitem[Gle16]{glendinning2016classification}
Paul Glendinning, \emph{Classification of boundary equilibrium bifurcations in
  planar filippov systems}, Chaos: An Interdisciplinary Journal of Nonlinear
  Science \textbf{26} (2016), no.~1, 013108.

\bibitem[GOH13]{garcia2013simulation}
Antonio Garc{\'\i}a-Olivares and Carmen Herrero, \emph{Simulation of
  glacial-interglacial cycles by simple relaxation models: consistency with
  observational results}, Climate dynamics \textbf{41} (2013), no.~5-6,
  1307--1331.

\bibitem[GST08]{guardia2008topological}
Marcel Guardia, TM~Seara, and MA~Teixeira, \emph{Topological equivalences for
  planar filippov systems}, Talk during ‘‘Problems in Nonsmooth Dynamical
  Systems’’, University of Bristol (2008), 28--29.

\bibitem[GT00]{gildor2000sea}
Hezi Gildor and Eli Tziperman, \emph{Sea ice as the glacial cycles’ climate
  switch: Role of seasonal and orbital forcing}, Paleoceanography and
  Paleoclimatology \textbf{15} (2000), no.~6, 605--615.

\bibitem[Hel82]{held1982climate}
Isaac~M Held, \emph{Climate models and the astronomical theory of the ice
  ages}, Icarus \textbf{50} (1982), no.~2-3, 449--461.

\bibitem[HIS{\etalchar{+}}76]{hays1976variations}
James~D Hays, John Imbrie, Nicholas~J Shackleton, et~al., \emph{Variations in
  the earth’s orbit: pacemaker of the ice ages}, Science \textbf{194} (1976),
  no.~4270, 1121--1132.

\bibitem[IBB{\etalchar{+}}93]{imbrie1993structure}
John Imbrie, Andr{\'e} Berger, EA~Boyle, SC~Clemens, A~Duffy, WR~Howard,
  G~Kukla, J~Kutzbach, DG~Martinson, A~McIntyre, et~al., \emph{On the structure
  and origin of major glaciation cycles 2. the 100,000-year cycle},
  Paleoceanography \textbf{8} (1993), no.~6, 699--735.

\bibitem[JLP{\etalchar{+}}87]{jouzel1987vostok}
Jean Jouzel, Cl~Lorius, JR~Petit, C~Genthon, NI~Barkov, VM~Kotlyakov, and
  VM~Petrov, \emph{Vostok ice core: a continuous isotope temperature record
  over the last climatic cycle (160,000 years)}, Nature \textbf{329} (1987),
  no.~6138, 403.

\bibitem[KE13]{kaper2013mathematics}
Hans Kaper and Hans Engler, \emph{Mathematics and climate}, vol. 131, Siam,
  2013.

\bibitem[KJea18]{hadgem3}
Till Kuhlbrodt, Colin Jones, and et. al., \emph{The low‐resolution version of
  hadgem3 gc3.1: Development and evaluation for global climate}, Journal of
  advances in modelling earth systems \textbf{10} (2018), 2865--2888.

\bibitem[MA14]{mitsui2014dynamics}
Takahito Mitsui and Kazuyuki Aihara, \emph{Dynamics between order and chaos in
  conceptual models of glacial cycles}, Climate dynamics \textbf{42} (2014),
  no.~11-12, 3087--3099.

\bibitem[MCA15]{mitsui2015bifurcations}
Takahito Mitsui, Michel Crucifix, and Kazuyuki Aihara, \emph{Bifurcations and
  strange nonchaotic attractors in a phase oscillator model of
  glacial--interglacial cycles}, Physica D: Nonlinear Phenomena \textbf{306}
  (2015), 25--33.

\bibitem[Pai98]{paillard1998timing}
Didier Paillard, \emph{The timing of pleistocene glaciations from a simple
  multiple-state climate model}, Nature \textbf{391} (1998), no.~6665, 378.

\bibitem[Pai01]{paillard2001glacial}
\bysame, \emph{Glacial cycles: toward a new paradigm}, Reviews of Geophysics
  \textbf{39} (2001), no.~3, 325--346.

\bibitem[Pai17]{paillard2017climate}
\bysame, \emph{Climate science: Predictable ice ages on a chaotic planet},
  Nature \textbf{542} (2017), no.~7642, 419.

\bibitem[PJR{\etalchar{+}}99]{petit1999climate}
Jean-Robert Petit, Jean Jouzel, Dominique Raynaud, Narcisse~I Barkov, J-M
  Barnola, Isabelle Basile, Michael Bender, J~Chappellaz, M~Davis, G~Delaygue,
  et~al., \emph{Climate and atmospheric history of the past 420,000 years from
  the vostok ice core, antarctica}, Nature \textbf{399} (1999), no.~6735, 429.

\bibitem[PK08]{piiroinen2008event}
Petri~T Piiroinen and Yuri~A Kuznetsov, \emph{An event-driven method to
  simulate filippov systems with accurate computing of sliding motions}, ACM
  Transactions on Mathematical Software (TOMS) \textbf{34} (2008), no.~3, 13.

\bibitem[PP04]{paillard2004antarctic}
Didier Paillard and Fr{\'e}d{\'e}ric Parrenin, \emph{The antarctic ice sheet
  and the triggering of deglaciations}, Earth and Planetary Science Letters
  \textbf{227} (2004), no.~3-4, 263--271.

\bibitem[PRKK03]{pikovsky2003synchronization}
Arkady Pikovsky, Michael Rosenblum, Jurgen Kurths, and J{\"u}rgen Kurths,
  \emph{Synchronization: a universal concept in nonlinear sciences}, vol.~12,
  Cambridge university press, 2003.

\bibitem[Sim10]{simpson2010bifurcations}
David John~Warwick Simpson, \emph{Bifurcations in piecewise-smooth continuous
  systems}, vol.~70, World Scientific, 2010.

\bibitem[SM90]{saltzman1990first}
Barry Saltzman and Kirk~A Maasch, \emph{A first-order global model of late
  cenozoic climatic change}, Earth and Environmental Science Transactions of
  the Royal Society of Edinburgh \textbf{81} (1990), no.~4, 315--325.

\bibitem[SM91]{saltzman1991first}
\bysame, \emph{A first-order global model of late cenozoic climatic change ii.
  further analysis based on a simplification of co 2 dynamics}, Climate
  Dynamics \textbf{5} (1991), no.~4, 201--210.

\bibitem[SP07]{schilderandpeckham2007}
Frank Schilder and Bruce Peckham, \emph{Computing arnold tongue scenarios}, J.
  Comp. Phys. \textbf{220} (2007), 932--951.

\bibitem[WWHM16]{EW}
James Walsh, Esther Widiasih, Jonathan Hahn, and Richard McGehee,
  \emph{Periodic orbits for a discontinuous vector field arising from a
  conceptual model of glacial cycles}, Nonlinearity \textbf{29} (2016), 1843.

\end{thebibliography}
\bibliographystyle{amsalpha}

\end{document}